\documentclass[prb,reprint,superscriptaddress,twocolumn]{revtex4-2}

\usepackage{amsmath,amssymb,bm,braket,mathrsfs,mathtools}
\usepackage{multirow}
\usepackage{hyperref}
\usepackage[dvipdfmx]{graphicx}
\usepackage{color}

\usepackage{ulem}

\DeclareMathOperator{\trace}{Tr}

\DeclareMathOperator{\pf}{Pf}

\begin{document}
\title{Fragile topological insulators protected by rotation symmetry without spin-orbit coupling
}
\author {Shingo Kobayashi}
\affiliation{RIKEN Center for Emergent Matter Science, Wako, Saitama, 351-0198, Japan}
\author{Akira Furusaki}
\affiliation{RIKEN Center for Emergent Matter Science, Wako, Saitama, 351-0198, Japan}
\affiliation{Condensed Matter Theory Laboratory, RIKEN, Wako, Saitama, 351-0198, Japan}


\begin{abstract}
We present a series of models of three-dimensional rotation-symmetric fragile topological insulators in class AI (time-reversal symmetric and spin-orbit-free systems), which have gapless surface states protected by time-reversal ($T$) and $n$-fold rotation ($C_n$) symmetries ($n=2,4,6$).
Our models are generalizations of Fu's model of a spinless topological crystalline insulator, in which orbital degrees of freedom play the role of pseudo-spins.
We consider minimal surface Hamiltonian with $C_n$ symmetry in class AI and discuss possible symmetry-protected gapless surface states,
i.e., a quadratic band touching and multiple Dirac cones with linear dispersion.
We characterize topological structure of bulk wave functions in terms of two kinds of topological invariants obtained from Wilson loops:
$\mathbb{Z}_2$ invariants protected by $C_n$ ($n=4,6$) and time-reversal symmetries, and $C_2T$-symmetry-protected $\mathbb{Z}$ invariants (the Euler class) when the number of occupied bands is two.
Accordingly, our models realize two kinds of fragile topological insulators.
One is a fragile $\mathbb{Z}$ topological insulator whose only nontrivial topological index is the Euler class that specifies the number of surface Dirac cones.
The other is a fragile $\mathbb{Z}_2$ topological insulator having gapless surface states with either a quadratic band touching or four (six) Dirac cones, which are protected by time-reversal and $C_4$ ($C_6$) symmetries.
Finally, we discuss the instability of gapless surface states against the addition of $s$-orbital bands and demonstrate that surface states are gapped out through hybridization with surface-localized $s$-orbital bands. 
\end{abstract}
\maketitle

\section{Introduction}
Over the past decade, considerable progress has been made in understanding topological phases of matter.
The classification theory of gapped free-fermion systems with on-site symmetry and/or crystalline symmetry has been developed using the K-theory approach~\cite{Schnyder08,Kitaev09,Schnyder09,Ryu10,Morimoto13,Chiu13,Freed2013,Shiozaki14,Shiozaki15,Shiozaki16,Chiu16,Shiozaki17,CFang17,Shiozaki18atiyah,Cornfeld19,Shiozaki19classification,Okuma19,Song19topological}. The K-theoretical classification is valid for systems with an arbitrary number of energy bands and ensures the existence of gapless boundary states as long as the relevant symmetry is preserved.
Such topological phases are called stable topological phases, examples of which include Chern insulators in class A and $\mathbb{Z}_2$ topological insulators in class AII in the tenfold-way classification~\cite{Schnyder08,Kitaev09,Schnyder09,Ryu10}. 
In addition, the comprehensive classification schemes of topological quantum chemistry~\cite{Bradlyn17topological} and symmetry-based indicators~\cite{Kruthoff17,Po17,ZSong18quantitative,Khalaf18,Elcoro2020magnetic,Po2020symmetry} have strengthened the search for a variety of topological phases protected by crystalline symmetry.
In particular, the approach incorporating these schemes with materials database and first-principles calculations has discovered several thousands of topological material candidates~\cite{Zhang19catalogue,Vergniory19complete,Tang19sciadv,Tang2019efficient,DWang19,Xu2020high}.

The search for topological materials beyond the K-theoretical classification has found topological insulators that do not belong to any of stable topological phase.
Examples of these unstable topological phases are Hopf insulators~\cite{Schnyder08,Moore2008,Deng2013,Kennedy2016,Liu2017,Schuster2019,Schuster2021}, fragile insulators~\cite{Po2018fragile,Wieder2018axion,Kooi2019,Hwang2019,Song2020fragile}, and  delicate insulators~\cite{Nelson2021}.
Among these, fragile topological insulators have unstable topological properties (obstructions) that can be nullified by adding a trivial band to the valence band~\footnote{There are various definitions of unstable/fragile topological insulators in the literature.
An initial example of unstable topological insulators is a Hopf insulator~\cite{Schnyder08,Moore2008} of a three-dimensional two-band insulator. The term {"fragile topology"} was introduced in the  
analysis using the symmetry-based indicators~\cite{Po2018fragile}
for energy bands that can be mathematically written as a \textit{difference} between trivial (atomic) bands.
Since atomic insulators do not host gapless surface states, the same holds true for these fragile topological bands.
In contrast, recent studies~\cite{Bouhon2020,Alexandradinata2020} have found fragile topological insulators that are not captured by the symmetry-based indicators but often have gapless surface states.
In this paper, we adopt the broad definition of fragile topological insulators that includes the latter cases. }.
In particular, crystalline-symmetry-protected fragile topological insulators without spin-orbit coupling~\cite{Fu2011,Alexandradinata2014spin} and the Hopf insulators
can be realized in specific band representations such as pseudo-spin states and two-band representations, respectively.
They exhibit topologically-protected surface states with a weaker type of robustness, i.e., a representation-dependent stability~\cite{Fu2011,Alexandradinata2020}.

It has been known from the K-theoretical classification that band insulators in class AI, i.e., 
gapped free-fermion systems with time-reversal (TR) symmetry and SU(2) spin rotation symmetry,
have no stable symmetry-protected topological phase in one, two, and three spatial dimensions~\cite{Schnyder08,Kitaev09,Schnyder09,Ryu10}, even when additional crystal symmetry is taken into account~\cite{Shiozaki19classification,Song19topological}. 
In contrast, several topological phases have been proposed and observed experimentally in photonic crystals and metamaterials, some of which have recently been identified as fragile topological phases in class AI~\cite{Alexandradinata2020}.
This implies that band insulators of class AI have the potential of bringing a variety of fragile topological phases,
and systematic approach for exploring them is called for.

In this paper, we present a series of models for fragile topological insulators protected by TR symmetry ($T; T^2=+1$) and $n$-fold rotation (C$_n$) symmetry in class AI ($n=2,4,6$) and discuss their topological properties.
Our findings are summarized as follows.

First, we take the model of a fragile insulator introduced by Fu\cite{Fu2011} and its variants with C$_n$ symmetry, and develop a theory for gapless surface states which are protected by TR and C$_n$ symmetries and represented in terms of orbital pseudo-spin basis with angular momentum $l$.
We systematically examine the stability of symmetry-protected gapless surface states for minimal $2\times2$ surface Hamiltonian, and find that $n$ surface Dirac cones can emerge when $2l =0 \mod n$ $(n=2,4,6)$, where each Dirac cone is locally protected by $C_2T$ symmetry\cite{CFang15}, the combination of $C_2$ and TR symmetries.
In particular, we find that the Fu model, which originally has C$_4$ symmetry with $l=1$, hosts two surface Dirac cones when the C$_4$ symmetry is broken down to C$_2$ symmetry.
When $2l\ne0$ mod $n$ ($n=2,4,6$), the minimial $2\times2$ surface Hamiltonian has gapless surface states with a quadratic band touching at a high-symmetry point.
When the minimal Hamiltonian is doubled to $4\times4$ Hamiltonian, this band touching can be either gapped or changed into multiple Dirac cones.
We define two kinds of surface topological numbers protecting gapless surface states.

Second, we study topological properties of electronic states in the bulk using minimal $4\times4$ bulk Hamiltonian and its doubled ($8\times8$) bulk Hamiltonian.
We calculate two kinds of topological invariants from Wilson loops:
$\mathbb{Z}_2$ topological invariants protected by TR and C$_4$ rotation~\cite{Fu2011} (or C$_6$ rotation~\cite{Alexandradinata2016Berry}) symmetries, 
and $C_2T$ symmetry-protected $\mathbb{Z}$ topological invariants, known as the Euler class~\cite{Ahn2018band,Bouhon2019,Ahn2019,Song2019all,Ahn2019failure,Bouhon2020,Unal2020}.
The Euler class ($\mathbb{Z}$) is a well-defined topological invariant as long as $C_2T$ symmetry is intact and the number of occupied bands is two, whereas the $\mathbb{Z}_2$ invariants are well defined under C$_n$ ($n=4,6$) and TR symmetries.
We find from the combined analysis of surface and bulk states that our models are categorized into two classes of fragile topological insulators depending on the orbital angular momentum $l$.
\begin{itemize}
\item[(i)]
Our minimal models with $2l=0$ mod $n$ realize fragile topological insulators that are characterized by the Euler class and have multiple of $n$ surface Dirac cones, provided that the number of occupied bands in the bulk equals two.
When more than two bands are occupied~\cite{Ahn2018band,Ahn2019}, the $\mathbb{Z}$-valued Euler class is reduced to the $\mathbb{Z}_2$-valued second Stiefel-Whitney class, thereby yielding a trivial insulator.
\item[(ii)]
Our minimal models with $2l\ne0$ mod $n$ ($n=4,6$) realize fragile topological insulators protected by C$_n$ ($n=4,6$) and TR symmetries.
These fragile topological insulators have gapless surface states with either a quadratic band touching at a high-symmetry point or multiple Dirac cones at generic momenta.
The former gapless states can exist for the minimal Hamiltonian with a nontrivial Euler class as well as a $\mathbb{Z}_2$ index, and the Fu model is an example of this case.
The latter multiple Dirac cones can be
obtained for doubled Hamiltonian with a proper choice of combined C$_n$ symmetry.
\end{itemize}

Finally, we examine the instability of surface states when an $s$-orbital band is added to the occupied band.
It turns out that the addition of an $s$-orbital band to the valence band does not necessarily gap out the surface states, even when it resolves the Wannier obstruction of the occupied states in the bulk Wilson-loop spectra.
We demonstrate that the surface states are gapped out through hybridization with an $s$-orbital band localized at the surface.

Throughout this paper we treat electrons as spinless fermions because we ignore spin-orbit coupling.
Instead, we introduce pseudo-spins that correspond to orbital degrees of freedom.
We assume that the C$_n$ rotation axis is along the $z$ axis.

This paper is organized as follows.
In Sec.~\ref{sec:surface_theory}, generalizing the Fu's argument, we develop a surface theory and classify possible gapless surface states including $n$ surface Dirac cones under C$_n$ and TR symmetries.
In Sec.~\ref{sec:bulk_topology}, we study the bulk band topology using the Wilson loop approach and compare it with the surface energy spectra.
In Sec.~\ref{sec:fragile_topology}, we calculate the surface states and Wilson loop spectra for the Fu model coupled with additional $s$-orbital bands and demonstrate the instability of surface states.
In Sec.~\ref{sec:other_cases}, we present C$_n$-symmetric lattice models having gapless surface states and show the correspondence between surface Dirac cones and relevant topological invariants of the valence bands.
We summarize our results in Sec.~\ref{sec:conclude}.
Some mathematical details are presented in appendices.

\section{Surface theory}
\label{sec:surface_theory}

According to the K-theory classification of topological insulators in which the number of filled bands is arbitrary,
there is no stable topological phase in class AI in one, two, and three spatial dimensions.
However, it is known that a fixed number of filled bands can host a fragile topological insulator phase with gapless surface states.
A representative example is the Fu model of a 3D spinless insulator that has surface states with quadratic band touching. 
Note that the symmetry-based indicators under $C_4$ and TR symmetries are not able to capture the nontrivial fragile topology of the Fu model~\cite{Song2020fragile}.
Key ingredients of the model are C$_4$ rotation symmetry and pseudo-spin basis consisting of two orbital states with finite angular momenta.
In this basis, the $C_4$ operator has a two-dimensional (2D) representation under TR symmetry, and the energy band structure has two-fold degeneracy, similar to the Kramers degeneracy, at the $C_4T$ invariant points in the Brillouin zone (BZ).
In the following subsections we first review the Fu model and its surface theory in Sec.~\ref{sec:revew_Fumodel}. 
We generalize the model to the cases with $C_n$ symmetry ($n=2, 4, 6$) and develop the surface theory in terms of arbitrary orbital pseudo-spin states in Sec.~\ref{sec:surftheory_pseudo}.
We then show the emergence of $n$ surface Dirac cones under $C_n$ and TR symmetries in Sec.~\ref{sec:surf_Dirac}. 
Finally, we define two surface topological numbers protecting gapless surface states and discuss surface anomaly in Sec.~\ref{sec:surface_anomaly}.

\subsection{The Fu model}
\label{sec:revew_Fumodel}

\begin{figure}[tbp]
\centering
 \includegraphics[width=8cm]{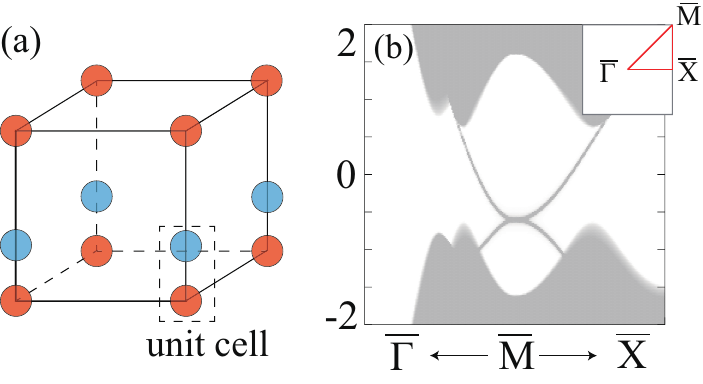}
 \caption{(a) Tetragonal lattice structure with atoms A and B in the unit cell. (b) Surface density of states of the Fu model (\ref{eq:Fumodel}) on the $(001)$ surface.
 The parameters are $(t_0^A,t_1^A,t_2^A,t_0^B,t_1^B,t_2^B,t_1',t_2',t_z')=(0.4,1,0.5,-0.4,-1,-0.5,2.5,0.5,2)$. The quadratic band touching appears at the $\bar{\mathrm{M}}=(\pi,\pi)$ point.}\label{fig:Fumodelsurf}
\end{figure}

We consider a tetragonal lattice with atoms A and B in the unit cell as shown in Fig.~\ref{fig:Fumodelsurf} (a). The two atoms align along $c$ axis in the unit cell, and each type of atoms forms a square lattice in the $ab$ plane. We consider electrons in $(p_x,p_y)$ orbitals on each atom and assume that there is no spin-orbit interaction. The tight-binding Hamiltonian of the Fu model is written in the momentum space $\bm{k}=(k_x,k_y,k_z)$ as~\cite{Fu2011} 
\begin{align}
 &\mathcal{H}_{\rm Fu}(\bm{k}) = \begin{pmatrix} h^A(\bm{k}) & h^{AB}(\bm{k}) \\ h^{AB \, \dagger}(\bm{k}) & h^B(\bm{k}) \end{pmatrix}, \label{eq:Fumodel}\\
 &h^a(\bm{k}) = t_0^a\bm{1}_2 +2 t_1^a \begin{pmatrix} \cos (k_x) & 0 \\ 0 & \cos (k_y)\end{pmatrix} \notag \\
 &  \qquad \qquad + 2t_2^a \begin{pmatrix} \cos (k_x) \cos (k_y) & \sin (k_x) \sin (k_y) \\ \sin (k_x) \sin (k_y) & \cos (k_x) \cos (k_y) \end{pmatrix}, \notag \\
 &h^{AB} (\bm{k})= \{t_1' + 2 t_2' [\cos (k_x) + \cos (k_y)] +t_z' e^{i k_z}\} \bm{1}_{2}, \notag
\end{align}
where $t_0^a$, $t_1^a$, and $t_2^a$ in $h^a$ ($a=A,B$) are the on-site potential, the nearest-neighbor, and the next-nearest-neighbor hoppings in the $ab$ plane; $t_1'$, $t_2'$, and $t_z'$ in $h^{AB}$ are the hopping matrix elements between atoms A and B.
The Hamiltonian has TR symmetry
\begin{equation}
\mathcal{T}\mathcal{H}_{\rm Fu}(\bm{k})\mathcal{T}^{\dagger} = \mathcal{H}_{\rm Fu}(-\bm{k}),
\quad
\mathcal{T}=\bm{1}_4K,
\end{equation}
and C$_4$ rotation symmetry around the $z$ axis
\begin{equation}
\mathcal{C}_4 \mathcal{H}_{\rm Fu}(k_x,k_y,k_z)\mathcal{C}_4^{\dagger} = \mathcal{H}_{\rm Fu}(-k_y,k_x,k_z), 
\quad
\mathcal{C}_{4} = i\sigma_y \otimes \bm{1}_{2},
\end{equation}
where $\sigma_i$ are the Pauli matrices in the orbital space with the basis $(p_x,p_y)$, $\bm{1}_{n}$ is the $n \times n$ identity matrix, and $K$ is the complex conjugation operator.
Here we have used caligraphic fonts for the Hamiltonian and symmetry operators in the 3D momentum space.
Imposing the semi-infinite boundary condition along the $z$ direction, 
we obtain the surface density of states in the $(001)$ plane shown in Fig.~\ref{fig:Fumodelsurf} (b). Here, the surface density of states is defined by $\rho(\bm{k}_{\parallel},E) = - {\rm Im}[G_s(\bm{k}_{\parallel},E+i\eta)]/\pi$ and the surface Green's function $G_s(\bm{k}_{\parallel},E+i\eta)$ is calculated by the method of Ref.~\onlinecite{Umerski1997}, where $\bm{k}_{\parallel}=(k_x,k_y)$, $E$ is the energy, and $\eta=10^{-5}$ is the smearing factor.
In Fig.~\ref{fig:Fumodelsurf} (b), we observe the surface energy bands with quadratic dispersion touching at $\bar{M}=(\pi,\pi)$.

Let us discuss the stability of gapless surface states in the pseudo-spin basis $(p_x, p_y)$ under the fourfold rotation ($C_4=i\sigma_y$) and TR ($T=\bm{1}_2K$) symmetries.
For spinless fermions (class AI) on the surface, the symmetry operators satisfy $T^2=1$ and $(C_4)^4=1$, while their combination satisfies $(C_4T)^2=-\bm{1}_2$, similar to the TR operator for spin-$\frac12$ electrons. These symmetries impose constraints on the surface Hamiltonian such that 
\begin{subequations}
\begin{align}
&
C_4 H(k_x,k_y)C_4^{\dagger} = H(-k_y,k_x),\\
&
TH(\bm{k}_{\parallel})T^{-1} = H(-\bm{k}_{\parallel}),
\end{align}
\end{subequations}
where $\bm{k}_\parallel$ is measured from the $\bar{M}$ point.
We note that we have used italic fonts for the Hamiltonian and symmetry operators in the 2D momentum space.
Thus, near the $\bar{M}$ point the surface Hamiltonian has the form
\begin{align}
H_{\rm Fu}(\bm{k}_{\parallel})  = v_1 (k_x^2-k_y^2) \sigma_z + 2v_2 k_x k_y \sigma_x , \label{eq:Fu_surface} 
\end{align}
where $v_{1,2}\in\mathbb{R}$, and we have dropped terms proportional to the identity matrix $\bm{1}_2$, which do not affect the stability analysis of gapless surface states.
Note that the form of Eq.~(\ref{eq:Fu_surface}) is not uniquely determined by $C_4$ and TR symmetries as we will discuss in Sec.~\ref{sec:surftheory_pseudo}.
In deriving Eq.~(\ref{eq:Fu_surface}), we have used the additional constraint $\sigma_x H(k_x,k_y)\sigma_x = H(k_y,k_x)$ that descends from mirror symmetry in $\mathcal{H}_\mathrm{Fu}(\bm{k})$.

The surface Hamiltonian (\ref{eq:Fu_surface}) tells us the following three important properties unique to class AI.
First, the quadratic band touching is a consequence of the presence of both TR and $C_4$ symmetries that admits neither a $k$-linear term nor a constant Dirac mass term.
Indeed, Eq.~(\ref{eq:Fu_surface}) reproduces the quadratic dispersion of the surface states around the $\bar{\mathrm{M}}$ point in Fig.~\ref{fig:Fumodelsurf}.
Second, the surface states are gapped by breaking either $T$ or $C_4$ symmetry.
The breaking of $T$ allows a Dirac mass term $M_1 \sigma_y$ to appear in the surface Hamiltonian (\ref{eq:Fu_surface}), which opens a gap at $\bm{k}_\parallel=0$. 
The breaking of $C_4$ gives rise to $M_2 \sigma_x$ and $M_3 k_y \sigma_y$ terms, which work together to gap out the surface states, as follows.
The $M_2\sigma_x$ term splits the quadratic band touching into two Dirac points, and they are gapped out by the {\it momentum-dependent mass term} $M_3 k_y\sigma_y$; see also Appendix~\ref{sec:massterm}.
Third, we can infer the topological structure of the model from the stacking of two surface Hamiltonians, $H_{\rm Fu} \oplus H_{\rm Fu}$, which we call a double Hamiltonian in this paper. Then, without breaking the symmetries, we can find a mass term $M_4 \sigma_y \otimes \tau_y$ gapping out the surface states, where $\tau_y$ is a Pauli matrix coupling the two surface Hamiltonians. Hence, the system has a $\mathbb{Z}_2$ topology, which is consistent with the bulk band topology~\cite{Fu2011}. 

\subsection{Surface theory in orbital pseudo-spin states}
\label{sec:surftheory_pseudo}

In this section we generalize the Fu model to spinless systems with TR and $C_n$ symmetries ($n=2,4,6$).
Here we consider surface states from a pair of orbital pseudo-spin states, which form a real 2D representation of the $C_n$ symmetry compatible with TR symmetry ($T=\bm{1}_2K$).
We start from gapless nonlinear dispersion at a high-symmetry point and show the emergence of surface Dirac cones, band crossing points with linear dispersion.
These Dirac points are located at generic points in the surface Brillouin zone and protected by $C_2T$ symmetry~\cite{CFang15}.
The form of symmetry-protected gapless structures is determined by the angular momenta of the pseudo-spin states. Then, we consider symmetry-allowed mass terms to examine the stability of gapless surface states. Using these complementary approaches, we determine possible topological structures of surface states.


Let us first consider the gapless structure at a high-symmetry point that is invariant under $C_n$ and TR operations, such as $\bar{\Gamma}=(0,0)$ and $\bar{M} =(\pi,\pi)$ etc. Here we choose the $\bar{\Gamma}$ point, but the discussion below does not depend on the choice of a high-symmetric point.
To begin with, we specify an orbital pseudo-spin basis by the orbital angular momentum $l$. For example, $l=0$ corresponds to a pair of $s$ orbitals, $l=1$ to $(p_x,p_y)$ or $(d_{zx},d_{yz})$ orbitals, and $l=2$ to $(d_{x^2-y^2},d_{xy})$ orbitals.
We assume the commutation relation $[C_{n},T]=0$, so that the representation of the $C_n$ operator is restricted to a real $2\times 2$ matrix. Hence, the matrix form of the $C_n$ operator with orbital angular momentum $l$ can be written as
\begin{align}
C_{n,l}=\exp \!\left(i \frac{2\pi l}{n} \sigma_y \right)
\label{eq:rot_op}
\end{align}
for $l=0,1,\ldots,n/2$,
where $\sigma_i$ ($i=x,y,z$) are the Pauli matrices  in the orbital space. 
It is clear from Eq.~(\ref{eq:rot_op}) that $C_{n,l}$ is proportional to the $2 \times 2$ identity matrix when $2l=0$ mod $n$; otherwise it has nonzero off-diagonal elements.

The surface Hamiltonian written in terms of the orbital pseudo-spin basis has a $2 \times 2$ Hermitian matrix form
\begin{align}
 H(\bm{k}_{\parallel}) = &f(\bm{k}_{\parallel}) \sigma_+ + [f(\bm{k}_{\parallel})]^{\ast} \sigma_- + g(\bm{k}_{\parallel}) \sigma_y, \label{eq:sym_analysis}
\end{align}
where $\sigma_{\pm} = (\sigma_z \pm i \sigma_x)/2$, and $f$ and $g$ are complex and real functions of $\bm{k}_{\parallel}=(k_x,k_y)$, respectively, due to the Hermiticity of $H(\bm{k}_\parallel)$. The form of the surface Hamiltonian~(\ref{eq:sym_analysis}) is constrained by $C_n$ and TR symmetries such that
\begin{align}
 &TH(\bm{k}_{\parallel})T^{-1} = H(-\bm{k}_{\parallel}), \label{eq:trs_surface} \\
 &C_{n,l} H(\bm{k}_{\parallel})C_{n,l}^{\dagger} = H(R_{n} \bm{k}_{\parallel}), \label{eq:cn_surface}
\end{align}
where $R_n$ represents the rotation around the $z$ axis in the momentum space:
\begin{equation}
R_{n} \bm{k} =
\begin{pmatrix} \cos(\frac{2\pi}{n}) & -\sin(\frac{2\pi}{n}) & 0 \\ \sin(\frac{2\pi}{n}) & \cos(\frac{2\pi}{n}) & 0 \\ 0 & 0 & 1 \end{pmatrix}
\begin{pmatrix} k_x \\ k_y \\ k_z \end{pmatrix}.
\end{equation}
From the TR symmetry constraint~(\ref{eq:trs_surface}), $f$ and $g$ must be even and odd functions of $\bm{k}_{\parallel}$, respectively.
Similarly, the $C_n$ symmetry constraint~(\ref{eq:cn_surface}) imposes the conditions
on $f$ and $g$ functions:
\begin{subequations}
\label{constraints}
\begin{align}
& e^{i 4 \pi l/n}f(k_+,k_-) = f(e^{ i 2\pi/n}k_+, e^{- i 2\pi/n}k_-),\\
& g(k_+,k_-) = g(e^{ i 2\pi/n}k_+,e^{- i 2\pi/n}k_-),
\end{align}
\end{subequations}
where $k_{\pm} \equiv k_x \pm i k_y$.
Since we are interested in the form of gapless structures at $\bm{k}_{\parallel}=0$, we expand $f$ and $g$ to leading order in $\bm{k}_{\parallel}$,
\begin{equation}
f(k_+,k_-)=v k_+^p k_-^q,
\quad
g(k_+,k_-)=v' k_+^r k_-^s + \mathrm{H.c.},
\label{f, g expanded}
\end{equation}
where $v,v'\in\mathbb{C}$ and $p,q,r,s$ are non-negative integers.
Substituting Eq.\ (\ref{f, g expanded}) into Eq.\ (\ref{constraints}) yields 
\begin{subequations}
\label{p-q, r-s}
\begin{align}
 &p-q = 2l \mod n, \label{eq:kp_result1}\\
 &r-s =0 \mod n. \label{eq:kp_result2}
\end{align}
\end{subequations}
Thus the leading power-law dependence of $f$ and $g$ on $\bm{k}_{\parallel}$ is determined by the $C_n$ symmetry and the orbital angular momentum $l$.
Furthermore, recalling that $g$ is an odd function of $\bm{k}_{\parallel}$, we find $g(\bm{k}_\parallel)=0$ when $n=2,4,6$.

Table~\ref{tab:kp_theory} summarizes the symmetry-allowed form of $f(\bm{k}_\parallel)$
for small $\bm{k}_\parallel$.
There are two distinct cases\footnote{It should be noted that $C_3$ symmetry allows another possibility that $f$ does not have a constant term and $g \propto k_+^3 + \mathrm{H.c.}$, which leaves a quadratic band touching intact. However, the surface quadratic band touching cannot be split into multiple Dirac cones due to the absence of the $C_2T$ symmetry.}
that are distinguished by the presence or absence of a constant term in $f(\bm{k}_\parallel)$.
\begin{itemize}
\item[(i)]
When $2l=0$ (mod $n$), the leading term of $f(\bm{k}_\parallel)$ is a constant.
In this case the quadratic band touching at the high-symmetry point $\bm{k}_\parallel=0$ is split into $n$ Dirac cones with linear dispersion at generic momenta $\bm{k}_0, R_n \bm{k}_0, \cdots, R_n^{n-1} \bm{k}_0$, which will be discussed further in Sec.~\ref{sec:single_hami}.
\item[(ii)]
When $2l\ne0$ (mod $n$), the leading term of $f(\bm{k}_\parallel)$ is proportional to $k_\parallel^2$, and
the surface states have quadratic dispersion at $\bm{k}_\parallel=0$, as in the Fu model.
This quadratic band touching is associated with a two-fold degeneracy enforced by 2D representations for $n=4,6$.
\end{itemize}

\begin{table}[tb]
\caption{
Summary of the classification of the surface Hamiltonian (\ref{eq:sym_analysis}) in the pseudo-spin basis under $C_n$ symmetry. The second and third columns represent the orbital angular momentum of pseudo-spin states and the leading term in the symmetry-allowed $f(\bm{k}_\parallel)$. The fourth and fifth columns show possible $C_n$ symmetry-protected surface states (SS) in $H(\bm{k}_{\parallel})$ with $C_{n,l}$ symmetry and $H(\bm{k}_{\parallel}) \oplus H(\bm{k}_{\parallel})$ with $C_{n,l}^-$ symmetry, respectively. The coefficients in $f(\bm{k}_\parallel)$ are complex numbers, $M(\bm{k}_\parallel) = m_0 + v_0 |\bm{k}_{\parallel}|^n $ with $m_0, v_0, v_\pm \in\mathbb{C}$.
}
\label{tab:kp_theory}
\begin{tabular}{ccccc}
\hline\hline
$n$ & $2l$ & $f(\bm{k}_{\parallel})$ & SS of $H$ & SS of $H\oplus H$ \\
\hline 
$2$ & $0$ & $M(\bm{k}_\parallel) \!+\! v_+ k_+^{2} \!+\! v_- k_-^2$ & 2 Dirac cones & Gapped \\
$4$ & $0$ & $M(\bm{k}_\parallel) \!+\! v_+ k_+^{4} \!+\! v_- k_-^4 $ & 4 Dirac cones & Gapped  \\
 4    & $2$ & $v_+ k_+^{2} + v_- k_-^2 $& Quadratic & 4 Dirac cones\\
$6$  & $0$ & $M(\bm{k}_\parallel) \!+\! v_+ k_+^{6} \!+\! v_- k_-^6 $ & 6 Dirac cones & Gapped \\
  6     & $2$ & $v_+ k_+^{2}$ & Quadratic & 6 Dirac cones \\
  6     & $4$ & $v_- k_-^{2}$ & Quadratic & 6 Dirac cones \\
\hline\hline
\end{tabular} 
\end{table}

These gapless points are unstable when either TR or $C_n$ symmetry is absent.
Without TR symmetry, the mass term $g\sigma_y$ with $g = \mathrm{const}.$\@ is allowed, and surface states are gapped out.
When $C_n$ symmetry is absent, the leading terms in $f$ and $g$ functions are $f=\mathrm{const.}$ and $g\propto k_\parallel$, which together gap out the surface states as we discussed for the Fu model in Sec.~\ref{sec:revew_Fumodel}.


Let us discuss the topological structure of gapless states for the cases (i) and (ii) listed above. To this end, we examine symmetry-allowed mass terms in the double Hamiltonian $H \oplus H$, for which $C_{n,l}$ is extended to two different representations,
\begin{align}
C_{n,l}^{\pm} = C_{n,l} \oplus (\pm C_{n,l}), \label{eq:doublerot}
\end{align}
where the sign difference originates from different choices of orbital angular momentum for basis states. For instance, $C_{4,1}^-=C_{4,1}\otimes\tau_z$ corresponds to the combination of $l=1$ and $l=3$ states. As we discuss below, the sign difference plays an important role in the stability of surface states.
We note that the combined symmetry of the type in Eq.\ (\ref{eq:doublerot}) is sufficient in our discussion here; the cases of other combinations of $l$'s are briefly discussed in Appendix~\ref{sec:doublehami}.

First, let us consider the representation $C_{n,l}^+=C_{n,l}\otimes\tau_0$, where $\tau_0=\bm{1}_2$. In this case, we have a symmetry-preserving mass term $M \sigma_y \otimes \tau_y$, which opens a gap in both cases (i) and (ii).

When we impose the $C_{n,l}^-=C_{n,l}\otimes\tau_z$ symmetry, the mass term $M \sigma_y \otimes \tau_y$ is forbidden.
Instead, momentum-dependent terms are allowed by the symmetry: e.g., $[a(k_x^2-k_y^2)+bk_xk_y]\sigma_y \otimes \tau_y$ with $a,b\in\mathbb{R}$ for $C_{4,l}^-$.
Such momentum-dependent mass terms can gap out the $n$ Dirac cones that exist at generic momenta in the case (i) $2l=0$ (mod $n$); see Appendix~\ref{sec:massterm} for further discussion.
We infer that the topological structure in this case is $\mathbb{Z}_2$.

On the other hand, in the case (ii) $2l\ne0$ (mod $n$) with the $C_{n,l}^-$ representation, the above momentum-dependent mass terms do not affect the quadratic band touching at the high-symmetry point $\bm{k}_\parallel=0$.
Thus, the quadratic band touching is stable in the double Hamiltonian $H\oplus H$ under $C_{n,l}^-$ symmetry.
Moreover, as we demonstrate in Sec.~\ref{sec:double_hami}, symmetry-preserving perturbations can turn the gapless (band touching) point into $n$ Dirac cones, which cannot be gapped out by symmetry-preserving perturbations as long as the number of surface bands is fixed. 
However, the gapless surface states are unstable in quadruple Hamiltonian $H\oplus H \oplus H \oplus H$ with representation $C_{n,l}^-\otimes\bm{1}_2$.
This suggests that the topological structure of the case (ii) should be
$\mathbb{Z}_2\times\mathbb{Z}_2$, where the double Hamiltonians $H \oplus H$ with $C_{n,l}^{+}$ and $C_{n,l}^{-}$ representations can be identified with $(2,0)$ and $(1,1) \in \mathbb{Z}_2 \times \mathbb{Z}_2$, respectively.


\subsection{Surface Dirac cones}
\label{sec:surf_Dirac}

In Sec.~\ref{sec:surftheory_pseudo} we have performed symmetry analysis of the general form of surface Hamiltonian for the
$2 \times 2$ Hamiltonian $H$ and the $4 \times 4$ double Hamiltonian $H \oplus H$. 
The former accommodates $n$ Dirac cones for the case (i), whereas the latter admits four and six Dirac cones for the case (ii) under the $C_{4,l}^-$ and $C_{6,l}^-$ symmetry, respectively.
In this section we will discuss the emergence of $n$ Dirac cones in more detail.

\subsubsection{Dirac cones in $2 \times 2$ Hamiltonians}
\label{sec:single_hami}

The $2 \times 2$ Hamiltonian of the case (i) $2l=0$ (mod $n$) has a constant term in $f$, which generates $n$ Dirac cones. To see this mechanism in a concrete example, we start from the surface Hamiltonian in Eq.~(\ref{eq:Fu_surface}) with $C_{4,1}$ symmetry and the pseudo spin states with $l=1$.
If the symmetry $C_{4,1}$ is reduced to $C_{2,1}$, the Hamiltonian satisfies the condition of the case (i). The Hamiltonian with such symmetry breaking perturbations is given by
\begin{align}
H_{\rm Fu}(\bm{k}_{\parallel}) + M_1 \sigma_z + M_2 \sigma_x,
\label{eq:Fu_surface+}
\end{align}
where $M_1$ and $M_2$ are real constants.
The energy spectrum of Eq.\ (\ref{eq:Fu_surface+}) is given by
\begin{equation}
E(\bm{k}_{\parallel}) =
 \pm \sqrt{[v_1k^2\cos(2 \theta)+M_1]^2+[v_2k^2\sin(2 \theta)+M_2]^2}, \label{eq:ene_Fusurf}
\end{equation}
which vanishes at $\bm{k}_\parallel=\pm(k_0\cos\theta_0, k_0\sin\theta_0)$,
where $k_0=[(M_1/v_1)^2+(M_2/v_2)^2]^{1/4}$,
$\cos(2\theta_0) = -M_1/v_1k_0^2$ and
$\sin(2\theta_0)=-M_2/v_2k_0^2$.
Around the zero-energy points, the energy spectrum disperses linearly, forming two Dirac cones. Therefore, the splitting of the quadratic band touching to two Dirac cones occurs under the perturbations that break the $C_{4,1}$ symmetry down to $C_{2,1}$.
See Fig.~\ref{fig:wilsonloop} (g) for the surface energy band structure.
The Dirac cones are located away from high-symmetry points and protected by the $C_2T$ symmetry.
When $|M_1/v_1|$ or $|M_2/v_2|$ is on the order of or larger than the inverse lattice spacing, the Dirac cones move far away from the $\bar{\Gamma}$ point, beyond the range of the low-energy theory.

The above mechanism for generating $n$ Dirac cones is generalized to the cases when $2l =0$ (mod $n$).
To see this, we choose a minimal form of the surface Hamiltonian as $f(\bm{k}_{\parallel}) = m_0+v_+k_+^{n}$ with  $v_+, m_0 \in \mathbb{C}$. The corresponding energy spectrum is $E= \pm |m_0+v_+k_+^n|$.  When $m_0=0$, the surface states have quartic dispersion for $n=4$ and sextic dispersion for $n=6$. The band touching with the nonlinear dispersion at $\bm{k}_\parallel=0$ is split by a finite $m_0$ into $n$ Dirac cones, the positions of which are written, with the polar coordinate $(k_x,k_y)=(k\cos\theta, k\sin\theta)$, as $k=|m_0/v_+|^{1/n}$ and $\theta=\theta_0-\theta_1+(2j-1)\pi/n \mod 2\pi$ ($j=1,\ldots,n$), where $n \theta_0 = \arg (m_0)$ and $n \theta_1 = \arg (v_+)$.
Likewise, we can calculate possible gapless states in the general form of $f(\bm{k}_{\parallel})$; see Appendix~\ref{sec:derivation}.

\subsubsection{Dirac cones in $4 \times 4$ Hamiltonians}
\label{sec:double_hami}

Next, we discuss the emergence of surface Dirac cones in the double Hamiltonian for the case (ii) in which the original $2 \times 2$ surface Hamiltonian has quadratic band touching at a high-symmetry point.
To be concrete, we examine the direct sum of two copies of the Hamiltonian~(\ref{eq:Fu_surface}), which belongs to the case (ii) when $C_{4,1}$ symmetry is preserved.
We assume the representation $C_{4,1}^-= i \sigma_y \otimes \tau_z$.
For the double Hamiltonian with the quadratic band touching, we can add perturbations
\begin{align}
M_1' \sigma_z \otimes \tau_x +M_2'   \sigma_x \otimes \tau_x
\qquad(M_1',M_2'\in\mathbb{R}) \label{eq:doubleFu_surface+}
\end{align}
to the double Hamiltonian
$H_\mathrm{Fu}(\bm{k}_\parallel)\oplus H_\mathrm{Fu}(\bm{k}_\parallel)$
without breaking $C_4$ symmetry since they commute with $C_{4,1}^-$.
The energy spectrum is given by Eq.~(\ref{eq:ene_Fusurf}) with $(M_1,M_2)$ replaced by $(M_1', M_2')$ and $(-M_1', -M_2')$, which vanishes at $\bm{k}_\parallel=\pm(k_0\cos\theta_0, k_0\sin\theta_0)$,
where $k_0=[(M_1'/v_1)^2+(M_2'/v_2)^2]^{1/4}$,
$\cos(2\theta_0) = \pm M_1'/v_1k_0^2$ and
$\sin(2\theta_0)=\pm M_2'/v_2k_0^2$. 
As a result, the energy spectrum remains gapless at the four Dirac points; see Fig.~\ref{fig:wilsonloop} (j). 

In a similar way, we can show that there are six Dirac cones in the case of $C_{6,1}^-$ and $C_{6,2}^-$.
This situation is realized as follows.
We begin with $2\times2$ Hamiltonians $H_{6,2}=v_+k_+^2\sigma_+ + \mathrm{H.c.}$ and $H_{6,4}=v_-k_-^2\sigma_+ + \mathrm{H.c.}$ for $2l=2$ and $4$ (mod 6), respectively, which have quadratic band touching at $\bm{k}_\parallel=0$.
We consider their double Hamiltonians and introduce additional symmetry-preserving perturbations $M_1' \sigma_0 \otimes \tau_z + M_2' (k_- \sigma_+ + \mathrm{H.c.}) \otimes \tau_x$ for $H_{6,2}$ and  $M_1' \sigma_0 \otimes \tau_z + M_2' (k_+ \sigma_+ + \mathrm{H.c.}) \otimes \tau_x$ for $H_{6,4}$.
The resulting energy spectrum is given by 
\begin{align}
E^2(\bm{k}_{\parallel}) =&\,
 |v_{\pm}|^2 k^4+M_1^{\prime \,2}+M_2^{\prime \,2} k^2 \notag \\
&\pm 2|v_{\pm}| k^2 \sqrt{M_1^{\prime \,2}+M_2^{\prime \,2} k^2 \cos^2 (3 \theta+\theta_0)}, \label{eq:ene_surfc6}
\end{align}
where $\theta_0 = \arg (v_{\pm})$. The perturbations change the quadratic band touching to six Dirac cones, whose positions are determined by the condition $|v_\pm|^2 k^4 = M_1^{\prime \,2}+M_2^{\prime \,2} k^2$, and $\theta = \frac{1}{3}[- \theta_0 + \pi (j-1)] \mod 2\pi$ ($j=1,\ldots,6$).

\subsection{Topology of surface states}
\label{sec:surface_anomaly}
In the remainder of the section, we will show that the gapless surface states that we discussed in Secs.~\ref{sec:surftheory_pseudo} and \ref{sec:surf_Dirac}  are anomalous in the sense that they cannot be realized in any purely 2D system, as long as the number (or representation) of bands is fixed. Instead, they can be realized on the 2D surface of 3D (fragile) topological insulators. To demonstrate this, we define crystalline-symmetry-protected surface topological numbers that characterize gapless surface states and then show that the nontrivial values of these topological numbers are inconsistent with the periodic boundary conditions imposed on purely 2D systems.
Our discussion is analogous to the one for 3D topological insulators of class AII having an odd number of Dirac cones on every surface, where each surface Dirac cone is protected by a quantized $\pi$-Berry phase~\cite{Fu2007}.
In contrast, in purely 2D systems, the total Berry phase from all the Dirac cones in the 2D Brillouin zone should be zero.
This constraint leads to the fermion doubling theorem~\cite{Nielsen1981absenceI,Nielsen1981absenceII,Nielsen1981no-go} stating that an odd number of Dirac cones are forbidden in 2D systems with time-reversal symmetry.  An odd number of Dirac cones on the surface of class AII 3D topological insulators can escape the constraints of the theorem by counting the partner Dirac cones residing on the opposite surface together.
In the following, we show similar arguments can be applied to class AI insulators of our interest.

We introduce two kinds of surface topological numbers: a $\mathbb{Z}$-valued topological number $W_1$ that is only defined for $2 \times 2$ Hamiltonians preserving $C_2 T$ symmetry, and a $\mathbb{Z}_2$-valued topological number $W_2$ that is defined for $C_2 T$-symmetric Hamiltonians of any matrix size including, for example, $4 \times 4$ Hamiltonians with $2 l \neq 0$ (mod $n$). The former topological number ensures the stability of not only Dirac cones but also gapless states with quadratic, quartic, or sextic dispersions in $2 \times 2$ Hamiltonians. The latter protects, for example, four (six) Dirac cones in $4 \times 4$ Hamiltonians with $C_4^-$ ($C_6^-$) symmetry. In addition, we prove using $W_2$ that the number of Dirac cones is constrained to a multiple of $2n$ under $n$-fold rotation symmetry in class AI 2D systems under periodic boundary conditions. This theorem is a variant of the fermion multiplication theorem in class AII~\cite{CFang19}. 

We first define the $\mathbb{Z}$-valued topological number $W_1$, which comes from a chiral symmetry $\{\Gamma,H(\bm{k})\}=0$ with $\Gamma^2=1$. The chiral symmetry is always present in $2 \times 2$ Hamiltonian matrices with $C_2T$ symmetry, because any $2 \times 2$ Hamiltonian matrix can be written as Eq.~(\ref{eq:sym_analysis}) with $g(\bm{k}_{\parallel}) =0$ due to the $C_2T$ symmetry ($C_2T=K$). Thus, we find $\Gamma = \sigma_y$.
With the polar coordinate $(k,\theta)$ measured from a gapless point, $W_1$ is defined by
\begin{equation}
 W_1 = \frac{1}{4\pi i} \oint d \theta \trace [\Gamma H^{-1}(k,\theta)\partial_{\theta}H(k,\theta)], \label{eq:W1}
\end{equation} 
where the integration path is along the circle around the gapless point with a fixed radius $k$ that is smaller than the distance to other neighboring Dirac cones. The nonlinear band touching point with $f(\bm{k}_{\parallel}) = v_+ k_{+}^n+v_- k_{-}^n$ has $W_1=-n$ for $|v_+| > |v_-|$ and $W_1=+n$ for $|v_+| < |v_-|$.  When it is split into $n$ surface Dirac cones by $C_2T$-symmetry preserving perturbations, each Dirac point has $W_1 = \pm 1$. Note that $|v_+|=|v_-|$ is a topological critical point at which Eq.~(\ref{eq:W1}) becomes ill-defined.
A nonvanishing integer value of $W_1$ implies that the gapless point in the integration loop is stable and cannot be annihilated by itself.
On the other hand, we notice, by taking the integration path in Eq.~\ref{eq:W1} to enclose the whole surface BZ, that the total sum of $W_1$ from all the gapless points should vanish in purely 2D systems.

Next, we define the $\mathbb{Z}_2$-valued topological number 
\begin{equation}
 W_2 = \exp \left( i \oint d \theta \mathcal{A}(k,\theta) \right),
 \label{eq:W2}
\end{equation}
by a loop integral of the U(1) Berry connection
\begin{equation}
\mathcal{A}(k,\theta) \equiv -i \sum_{m \in {\rm occ}} \langle u_m(k,\theta) | \partial_{\theta} | u_m (k,\theta) \rangle,
\end{equation}
where $| u_m (k,\theta) \rangle$ is an eigenvector of $H(k,\theta)$ and the summation is taken over the occupied states. Due to $C_2T$ symmetry, $W_2$ only takes $\pm 1$;  $W_2=1$ ($-1$) indicates that there are an even (odd) number of Dirac cones in the region enclosed by the integration loop.
The stability of Dirac cones for $4\times4$ Hamiltonians with $2l\ne0$ (mod $n$) discussed in Sec.~\ref{sec:double_hami} can be understood from $W_2=-1$ for an integration loop enclosing a Dirac cone.

\begin{figure}[tbp]
\centering
 \includegraphics[width=6cm]{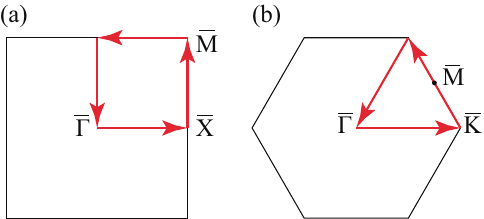}
 \caption{ 2D BZ of  (a) $C_4$-symmetric lattice and  (b) $C_6$-symmetric lattice. The area surrounded by the red arrows is a $C_n$-invariant BZ, along whose boundary the loop integral in the $\mathbb{Z}_2$-valued topological number $W_2$ in Eq.~(\ref{eq:W2}) is calculated. 
 }\label{fig:Cn_invariant_BZ}
\end{figure}

Let us derive the constraint on the number of Dirac cones in purely 2D systems by evaluating $W_2$ for the $C_n$-invariant BZ.
We define high-symmetry points in the 2D BZ: $\bar{\Gamma}=(0,0)$, $\bar{X}=(\pi,0)$, and $\bar{M}=(\pi,\pi)$ for a $C_4$ symmetric lattice and $\bar{\Gamma}=(0,0)$, $\bar{K} = 4\pi/3(1,0)$, $\bar{M}= 2\pi/\sqrt{3}(\sqrt{3}/2,1/2)$ for a $C_6$ symmetric lattice.
For the integration path of the Berry connection enclosing the $C_n$ invariant BZ shown in Fig.~\ref{fig:Cn_invariant_BZ}, 
we can evaluate $W_2$ from the eigenvalues of $C_n$ transformation\cite{CFang2012,CFang19}:
\begin{align}
 &W_2[C_4] = \prod_{i \in {\rm occ}} \xi_i(\bar{\Gamma}) \xi_i (\bar{M}) \zeta_i(\bar{X}), \\
 &W_2[C_6] = \prod_{i \in {\rm occ}} \omega_i(\bar{\Gamma}) \theta_i (\bar{K}) \zeta_i(\bar{M}),
\end{align}
where $\zeta_i$, $\theta_i$, $\xi_i$, and $\omega_i$ are the eigenvalue of $C_2$, $C_3$, $C_4$, and $C_6$, respectively, at the high-symmetry points.
For the Bloch states in the orbital pseudo-spin basis, the eigenvalues of $C_n$ at TR invariant momenta $\bar{\Gamma}$, $\bar{X}$, and $\bar{M}$ always appear in complex conjugate pairs $(e^{i\frac{2\pi}{n} m},e^{-i\frac{2\pi}{n} m})$ ($m=0,1,\cdots, n-1$);
see Eq.~(\ref{eq:doublerot}).
For the $C_6$-symmetric lattice, the $\bar{K}$ point is not a TR invariant momentum but a $C_2T$-invariant momentum; the same holds for the $C_3$ eigenvalues at the $\bar{K}$ point.  As a result, the $\mathbb{Z}_2$-valued topological numbers $W_2[C_4]$ and $W_2[C_6]$ must satisfy
\begin{align}
 W_2[C_4] = W_2[C_6] = 1. \label{eq:rot_anomaly}
\end{align}
Equation (\ref{eq:rot_anomaly}) tells us that there must be an even number of Dirac cones in $C_n$ invariant BZs, and thus the total number of Dirac cones is constrained to a multiple of $2n$ in the 2D BZs.
This is a fermion multiplication theorem for 2D class AI systems with orbital pseudo-spin basis.

We conclude from the nontrivial topological number $W_1$ and the fermion multiplication theorem that the gapless surface states discussed in Secs.~\ref{sec:surftheory_pseudo} and \ref{sec:surf_Dirac} are anomalous and cannot be realized in purely 2D systems, in which the sum of $W_1$ over all the gapless points in the 2D BZ should vanish and the number of Dirac cones in the $C_n$-invariant BZ must be an even integer (including 0).
However, such anomalous gapless states are possible on the surface of 3D systems; see Fig.~\ref{fig:Fumodelsurf} as an example. The constraints are circumvented by the presence of other gapless states residing on the opposite surface.
It should be noted that these constraints from $W_1$ and $W_2$ are only applicable to $2 \times 2$ Hamiltonian matrices and orbital pseudo-spin states, respectively. The dependence on band representations is in sharp contrast to topological insulators of class AII.

\section{Bulk Topology}
\label{sec:bulk_topology}

In this section we shift our focus to the bulk band topology.
In view of the bulk-boundary correspondence between gapless surface states and bulk topological invariants, we study the Fu model~(\ref{eq:Fumodel}) and its extensions as representative examples.
These models realize (a) two surface Dirac cones under perturbations that break $C_4$ symmetry down to $C_2$ and (b) four surface Dirac cones in the double Hamiltonian with $C_4$ symmetry, as demonstrated in Eqs.~(\ref{eq:Fu_surface+}) and (\ref{eq:doubleFu_surface+}). 
Other cases of C$_n$-symmetric fragile topological insulators are systematically studied in Sec.~\ref{sec:other_cases}.

\subsection{Topological invariants}
\subsubsection{$C_4$ and TR symmetries}

It is known that the Fu model is characterized by two $\mathbb{Z}_2$ topological invariants that are protected by TR and $C_4$ symmetries~\cite{Fu2011}: $\nu_{4} (\bar{k}_z)$ for $\bar{k}_z \in \{0,\pi\}$.
Here, $\nu_{4} (\bar{k}_z)$ is defined from the Berry phase of Bloch wave functions in the pseudo-spin basis; see Appendix~\ref{sec:C46symmetry} for the definition.
We note that $\nu_{4} (0)$ and $\nu_4(\pi)$ are weak indices protected by $C_4$ and TR symmetries~\cite{Alexandradinata2016Berry}, $\mathbb{Z}_2$-valued gauge-invariant quantities.
Depending on the values of these $C_4$ weak indices, class-AI band insulators are classified into three categories: trivial insulators when $\nu_{4}(0)=\nu_{4}(\pi)= 0$ mod 2, $C_4$ symmetry-protected weak topological insulators when $\nu_{4}(0)=\nu_{4}(\pi)= 1$ mod 2, and $C_4$ symmetry-protected strong topological insulators when $\bar{\nu}_4 \equiv \nu_{4}(\pi)-\nu_{4}(0) = 1$ mod 2.
The last class of $C_4$ symmetry-protected strong topological insulators have gapless surface states, either quadratic band touching or four Dirac cones,
on the $(001)$ surface.
We note that $C_4$ symmetry-protected strong topological insulators are fragile insulators, as we discuss in Sec.~\ref{sec:fragile_topology}.

\subsubsection{$C_2T$ symmetry}
\label{subsubsec:C_2T symmetry}

In the presence of perturbations that break $C_4$ symmetry down to $C_2$ symmetry, $\nu_{4} (\bar{k}_z)$ are no longer well-defined indices. 
In this case, we have alternative topological invariants protected by $C_2T$ symmetry: $\nu_2(\bar{k}_z)$ for $\bar{k}_z\in\{0,\pi\}$.
In the $C_2T$-invariant $k_z=\bar{k}_z$ plane,
the $C_2T$ symmetry imposes real-gauge condition on Bloch wave functions and allows to define topological invariants unique to the real vector bundle~\cite{Morimoto2014,CFang2015,Zhao2017,Ahn2018band,Bouhon2019,Ahn2019,Song2019all,Ahn2019failure,Hatcher2002,Bouhon2020,Unal2020}; see Appendix~\ref{sec:C2Tsymmetry} for more detailed discussion. Importantly, the classification of $\nu_2(\bar{k}_z)$ depends on the number of occupied bands:
$\nu_2(\bar{k}_z)\in\mathbb{Z}$ (the Euler class) for $N_{\rm occ}=2$ and
$\nu_2(\bar{k}_z)\in\mathbb{Z}_2$ (the second Stiefel-Whitney class) for $N_{\rm occ}>2$.

The 3D $C_2T$-invarianat insulators have two weak topological indices $\nu_2(0)$ and $\nu_2(\pi)$.
Similarly to the $C_4$-symmetric insulators, we have three $C_2T$-symmetric insulating phases: a trivial insulator phase when $\nu_2(0)=\nu_2(\pi)=0$, a $C_2T$ symmetry-protected weak topological phase when $\nu_{2}(0)=\nu_{2}(\pi) \neq 0$, and a $C_2T$ symmetry-protected strong topological phase when $\bar{\nu}_2 \equiv \nu_{2}(\pi)-\nu_{2}(0) \neq 0$.

In the previous studies~\cite{Shiozaki14,CFang15,Ahn2019}, it has been shown for 3D $C_2T$-invariant insulators with $N_{\rm occ}>2$ that the nontrivial strong index $\bar{\nu}_2 = 1 \mod 2$ indicates the existence of a single Dirac cone on the $C_2T$-invariant (001) surface. 
This bulk-boundary correspondence has been established for stable $C_2T$-symmetric topological insulators ($N_{\rm occ}>2$)  by the K-theory approach~\cite{Shiozaki14}.
Incidentally, the $C_2T$ symmetry-protected strong topological phase~\cite{Shiozaki14} is called 3D Stiefel-Whitney insulator phase in Ref.~\cite{Ahn2019}.
In the following analysis, we examine whether the bulk-boundary correspondence holds for $N_{\rm occ}=2$ insulators,
anticipating that the integer topological index $\bar{\nu}_2$ for $N_\mathrm{occ}=2$ correspond to the number of surface Dirac cones.

\subsection{Wilson loop characterization}

The weak indices $\nu_{4} (\bar{k}_z)$ and $\nu_{2}(\bar{k}_z)$ can be obtained from the Wilson loop method~\cite{Yu2011,Alexandradinata2014,Alexandradinata2016Berry,Alexandradinata2016,Cano2018,Ahn2018band,Bouhon2019,Bradlyn2019,Alexandradinata2020}, which is a useful approach for diagnosing the Wannier obstruction. We here summarize the Wilson loop method for the topological indices $\nu_{2,4}(\bar{k}_z)$.

Let $|u_{n}(\bm{k}) \rangle$ be the Bloch wave function of the $n$th energy band. The Wilson loop operator $\mathcal{W}_\ell$ is defined from a parallel transport of the Bloch wave function along a loop $\ell$ in the BZ. 
With the non-Abelian Berry connection
$[\mathcal{A}(\bm{k})]_{nm} \equiv
 -i \langle u_n(\bm{k}) | \partial_{\bm{k}} | u_m (\bm{k}) \rangle$,  
the Wilson loop operator is defined by
\begin{equation}
\mathcal{W}_\ell = P \exp\!\left[ i \oint_\ell \mathcal{A}(\bm{k}) \cdot d \bm{k}\right] \in U(N_{\rm occ}),
\end{equation} 
where $P$ means "path ordered'' along the loop $\ell$, and $N_{\rm occ}$ is the number of occupied bands. The eigenvalues of the Wilson loop operator,
$\{e^{i\Theta_1},e^{i\Theta_2}, \cdots, e^{i\Theta_{N_{\rm occ}}}\}$,
are gauge invariant quantities.
We can distinguish topologically nontrivial states from trivial ones using Wilson loop spectra as we explain below.

\begin{figure}[tbp]
\centering
 \includegraphics[width=8cm]{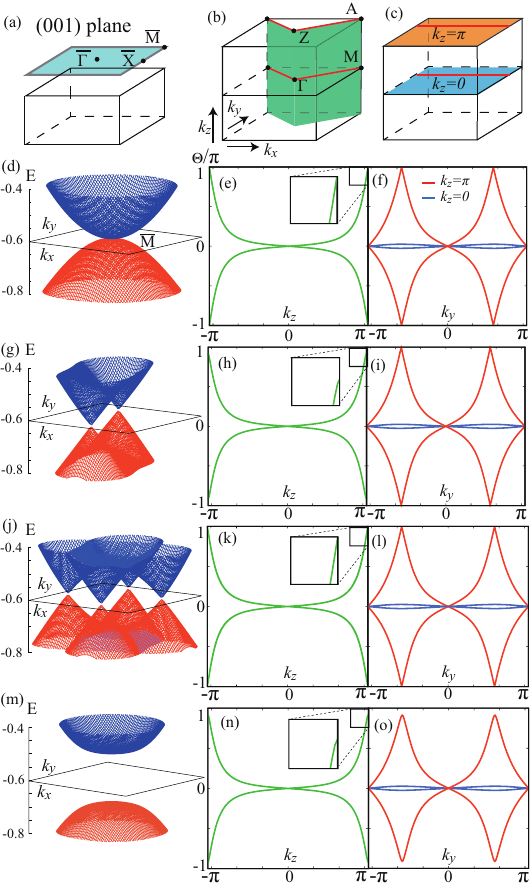}
 \caption{Surface states and Wilson loop spectra of Hamiltonians in Eqs.~(\ref{eq:Fumodel}), (\ref{eq:Fumodel+}), (\ref{eq:doubleFumodel+}), and ~(\ref{eq:doubleFumodel+gap}), where we choose the same hopping parameters as in Fig.~\ref{fig:Fumodelsurf} and the perturbation parameters as $(\mathcal{M}_1,\mathcal{M}_2,\mathcal{M}_1^{\prime},\mathcal{M}_2^{\prime},\mathcal{M}_3')=(0.1,0.08,0.1,0.08,0.1)$.
 (a) Surface BZ of the $(001)$ plane. (b) The integration path of the Wilson loop operator $\mathcal{W}_\ell$ for $C_4$ and TR-symmetric insulators.
(c) The integration path of $\mathcal{W}_\ell$ for $C_2T$-symmetric insulators. The surface energy spectra of Hamiltonians in Eqs.~(\ref{eq:Fumodel}), (\ref{eq:Fumodel+}),  (\ref{eq:doubleFumodel+}), and~(\ref{eq:doubleFumodel+gap}) are shown in (d), (g), (j), and (m), respectively.
 The surface bands are centered around $E=-0.6$. 
The Wilson loop spectra of the paths depicted in (b) are shown in green as a function of $k_z$ in (e), (h), (k), and (n) for Hamiltonians in Eqs.~\ref{eq:Fumodel}), (\ref{eq:Fumodel+}),  (\ref{eq:doubleFumodel+}), and~(\ref{eq:doubleFumodel+gap}), respectively.
The Wilson loop spectra of the paths depicted in (c) are shown in blue ($k_z=0$) and red ($k_z=\pi$) as a function of $k_y$ in (f), (i), (l), and (o) for Hamiltonians in Eqs.~\ref{eq:Fumodel}), (\ref{eq:Fumodel+}),  (\ref{eq:doubleFumodel+}), and~(\ref{eq:doubleFumodel+gap}), respectively.
 }\label{fig:wilsonloop}
\end{figure}

\subsubsection{$C_4$ and TR symmetries}

The weak indices $\nu_{4} (\bar{k}_z)$ of $C_4$-symmetric insulators are obtained from the Wilson loop spectra $\{\Theta_l(k_z) \,\, \mathrm{mod}~2\pi, l=1,\ldots,N_\mathrm{occ}\}$.
The integration path $\ell$ for the Wilson loop is chosen to be in a $k_x$-$k_y$ plane with fixed $k_z$ as the path obtained by a constant $k_z$ shift from the path $M$--$\Gamma$--$M$ on the $k_z=0$ plane.  The path $\ell$ is $A$--$Z$--$A$ at $k_z=\pi$.
The integration paths $\ell(k_z)$ are illustrated in Fig.~\ref{fig:wilsonloop} (b).
For each $k_z \in [-\pi, \pi]$, the Wilson loop spectra comprise a pair $\Theta_{2l-1}(k_z)=-\Theta_{2l}(k_z)$ and are doubly degenerate 
at $k_z=\bar{k}_z$ when $C_4$ is preserved in the pseudo-spin basis~\cite{Alexandradinata2016Berry}.
In this case, if we denote by $n_4^{(-)}(\bar{k}_z)$ the number of eigenvalues $\Theta_l/\pi = -1 (\equiv +1)$ of the Wilson loop operator $\mathcal{W}_\ell$ at $k_z=\bar{k}_z$, then the $C_4$ symmetry-protected weak indices are given by
$\nu_{4} (\bar{k}_z)=n_4^{(-)}(\bar{k}_z)/2 \mod 2$~\cite{Alexandradinata2016Berry}.
Note that $n_4^{(-)}(\bar{k}_z)$ takes an even integer value.

The Wilson loop spectra of the Fu model are shown in Fig.~\ref{fig:wilsonloop} (e), where we find $n_4^{(-)}(\pi)/2=1$ and $n_4^{(-)}(0)/2=0$.  Therefore the $C_4$ symmetry-protected strong index is $\bar{\nu}_4  =1$, which is in accordance with the existence of the surface quadratic band touching as shown in Fig.~\ref{fig:wilsonloop} (d).

\subsubsection{$C_2T$ symmetry}
Next, we discuss the relation between the Wilson loop spectra and $\nu_{2}(\bar{k}_z)$ for $C_2T$-symmetric insulators.
As we mentioned in Sec.~\ref{subsubsec:C_2T symmetry}, on the $k_z=\bar{k}_z(=0,\pi)$ planes, the Wilson loop operator is restricted to $\mathcal{W}_\ell \in SO(N_{\rm occ})$ due to the real gauge condition.
We choose a loop $\ell$ to be a line crossing the BZ along the $k_x$ direction at fixed $k_y$ on the $k_z=\bar{k}_z$ plane, as shown in Fig.~\ref{fig:wilsonloop} (c). The Wilson loop spectra $\{\Theta_l (k_y)\}$ satisfy $\{\Theta_l(k_y)\} = \{-\Theta_l(k_y)\}$ and can exhibit relative winding topologies as $k_y$ is varied from $-\pi$ to $\pi$, where $\Theta_l(k_y)$ are assumed to be smoothly connected at $\Theta_l/\pi=\pm1$.
When $N_{\rm occ}=2$, the weak index $\nu_{2}(\bar{k}_z)$ is equal to the relative winding number $n_W(\bar{k}_z)$ of the Wilson loop spectra $\{\Theta_1(k_y),\Theta_2(k_y)\}$ at $k_z=\bar{k}_z$~\cite{Ahn2019}.

We show the Wilson loop spectra of the Fu model in Fig.~\ref{fig:wilsonloop} (f). The red curves ($k_z=\pi$) have a nontrivial relative winding, as one eigenvalue ($d\Theta_1/dk_y>0$) has a winding number $2$, whereas the other one ($d\Theta_2/dk_y<0$) has a winding number $-2$.
The difference of the two winding numbers divided by two gives the relative winding number $n_W(\pi)=2$. On the other hand, $n_W(0)=0$.
Therefore the Fu model has $\bar{\nu}_2 =2$ and belongs to the $C_2T$ symmetry-protected strong topological phase with a nonzero Euler class,
as well as the $C_4$ symmetry-protected strong phase.
The Fu model can be regarded as an example of a 3D fragile $\mathbb{Z}$ topological insulator with Euler class $\bar{\nu}_2 =2$, besides being a fragile $\mathbb{Z}_2$ topological insulator with $\bar{\nu}_4=1$. 

The integer-valued Euler class is known to be well defined only for $N_{\rm occ} =2$.
When $N_{\rm occ} >2$, $\bar{\nu}_2$ is reduced to a $\mathbb{Z}_2$ index known as the second Stiefel-Whitney class.
This implies that the $C_2T$ symmetry-protected strong topological phase with $N_\mathrm{occ}=2$ can be trivialized by adding a trivial band.
This observation is consistent with the fact that the Fu model is a fragile topological insulator~\cite{Alexandradinata2020}.

\subsection{Numerical results for the extended models}

Using the Wilson loop method, we discuss the bulk band topology of the models derived from the Fu model that have multiple surface Dirac cones.

First, we consider the case where two surface Dirac cones are obtained by adding $C_4$-breaking perturbations to the Fu model as discussed in Eq.~(\ref{eq:Fu_surface+}).
The corresponding bulk Hamiltonian is given by  
\begin{align}
 \mathcal{H}_{\rm Fu}^{\prime} (\bm{k}) \equiv \mathcal{H}_{\rm Fu} (\bm{k}) + \mathcal{M}_1 \sigma_z \otimes \tau_z + \mathcal{M}_2 \sigma_x \otimes \tau_z, \label{eq:Fumodel+}
\end{align}
where the perturbation terms $\mathcal{M}_1$ and $\mathcal{M}_2$ break $\mathcal{C}_4$ down to $\mathcal{C}_2=\mathcal{C}_4^2$, where $\mathcal{C}_n$ denotes $n$-fold rotation symmetry in three spatial dimensions.
The energy band structure of the $(001)$ surface of the perturbed Hamiltonian is shown in Fig.~\ref{fig:wilsonloop} (g), which clearly exhibits two Dirac cones as expected in the surface theory.
The corresponding Wilson loop spectra are shown in Fig.~\ref{fig:wilsonloop} (h) and (i).
We find in Fig.~\ref{fig:wilsonloop} (h) that the Wilson loop eigenvalues are no longer doubly degenerate at $k_z=\pi$ and do not reach the upper and lower limits $\Theta = \pm \pi$,
which means $n_4^{(-)}(0) = n_4^{(-)}(\pi) =0$.
In contrast to $\nu_4(\bar{k}_z)$, the $C_2T$-protected index $\bar{\nu}_2$ remains to be 2, as the red (blue) curves have the relative winding number two (zero).
Thus, we obtain $\bar{\nu}_4=0$ and $\bar{\nu}_2=2$ for the model (\ref{eq:Fumodel+}).
We conclude that the two surface Dirac cones are characterized by the $C_2T$ symmetry-protected strong index $\bar{\nu}_2 =2$.

Next, we consider topology of the double Hamiltonian, two coupled copies of the Fu model, with four surface Dirac cones.
The bulk Hamiltonian with perturbations corresponding to Eq.~(\ref{eq:doubleFu_surface+}) is given by 
\begin{equation}
 \mathcal{H}_{\rm Fu} (\bm{k}) \oplus \mathcal{H}_{\rm Fu} (\bm{k})
 + \mathcal{M}_1'\sigma_z \otimes \tau_z \otimes \mu_x + \mathcal{M}_2' \sigma_x \otimes \tau_z \otimes \mu_x, \label{eq:doubleFumodel+}
\end{equation}
where the terms with $\mathcal{M}'_1$ and $\mathcal{M}'_2$ are perturbations to generate Dirac cones, and $\mu_i$ ($i=x,y,z$) are the Pauli matrices in the grading of the two $\mathcal{H}_\mathrm{Fu}$'s.
The perturbed double Hamiltonian of Eq.~(\ref{eq:doubleFumodel+}) is invariant under TR ($\mathcal{T}=\bm{1}_8 K$) and C$_4$ [$\mathcal{C}_{4}^-=\mathcal{C}_{4} \oplus (-\mathcal{C}_{4})$] symmetries.
Figure~\ref{fig:wilsonloop} (j) shows the $(001)$ surface energy band structure, which exhibits four surface Dirac cones.
The Wilson loop spectra are shown in Fig.~\ref{fig:wilsonloop} (k) and (l), in which every curve is doubly degenerate. 
Hence, the topological indices remain as $\bar{\nu}_4=2$ and $\bar{\nu}_2=4$.
At first sight this result seems to imply that the system is topologically trivial, as $\bar{\nu}_4$ and $\bar{\nu}_2$ are $\mathbb{Z}_2$ indices.
However, as we discussed in Sec.~\ref{sec:double_hami}, the four surface Dirac cones are stable as long as we keep the $C_{4,1}^-$ symmetry, and therefore the double Hamiltonian (\ref{eq:doubleFumodel+}) describes a fragile topological insulator protected by $\mathcal{C}_4^-$ and TR symmetries.
In fact, if we instead impose $\mathcal{C}_{4}^+=\mathcal{C}_{4} \oplus \mathcal{C}_{4}$ symmetry,
we can consider the double Hamiltonian with a relevant symmetry-allowed perturbation
\begin{align}
\mathcal{H}_{\rm Fu} (\bm{k}) \oplus \mathcal{H}_{\rm Fu} (\bm{k})
 +\mathcal{M}_3' \sigma_y \otimes \tau_z \otimes \mu_y, \label{eq:doubleFumodel+gap}
\end{align}
which gaps out the surface Dirac cones and makes the Wilson loop spectra unwind at the same time; see Figs.~\ref{fig:wilsonloop} (m), (n), and (o).
Thus, the topological indices become $\bar{\nu}_4=0$ and $\bar{\nu}_2=0$.

\section{Fragile topology and instability of surface states}
\label{sec:fragile_topology}

Since there is no stable topological (crystalline) insulator in class AI according to the K-theory classification, the topological crystalline insulators discussed so far should be considered as fragile topological insulators.
Therefore, adding a trivial band to occupied bands should be detrimental to their topological stability~\cite{Fu2011,Alexandradinata2014spin,Alexandradinata2020}.
However, it is not well understood to what extent added trivial bands affect surface states. To clarify this point, we extend the Hamiltonian of Eq.~(\ref{eq:Fumodel+}) to a $6 \times 6$ Hamiltonian matrix by introducing additional $s$ orbitals on A and B sites, 
\begin{align}
  \mathcal{H}_{\rm Fu}^{\prime}(\bm{k}) +\mathcal{H}_{\rm s}(\bm{k}), \label{eq:Fu+smodel}
\end{align}
where $\mathcal{H}_{\rm Fu}^{\prime}$ is embedded into the $6 \times 6$ matrix as $[\mathcal{H}_{\rm Fu}^{\prime}]_{ij} =0$ $(i,j=5,6)$. The Hamiltonian of $s$ orbitals is defined as
\begin{align}
&[\mathcal{H}_{\rm s}(\bm{k})]_{15} = i t_{sp}^A \sin(k_x), \notag \\
&[\mathcal{H}_{\rm s}(\bm{k})]_{25} = i t_{sp}^A \sin(k_y), \notag \\
&[\mathcal{H}_{\rm s}(\bm{k})]_{36} = i t_{sp}^B \sin(k_x), \notag \\
&[\mathcal{H}_{\rm s}(\bm{k})]_{46} = i t_{sp}^B \sin(k_y), \notag \\
&[\mathcal{H}_{\rm s}(\bm{k})]_{55} =   t_{s0}^A + 2 t_{s1}^A[\cos(k_x)+\cos(k_y)], \notag \\
&[\mathcal{H}_{\rm s}(\bm{k})]_{66} =   t_{s0}^B + 2 t_{s1}^B[\cos(k_x)+\cos(k_y)], \notag \\
&[\mathcal{H}_{\rm s}(\bm{k})]_{56} =   t_{s1}'+t_{s2}' e^{i k_z},
\end{align}
where $t_{sp}^a$ ($a=A,B$) is the $sp$ orbital coupling, $t_{s0}^a$ and $t_{s1}^a$ are the intrasite hopping terms, and $t_{s1}'$ and $t_{s2}'$ are the intersite hopping terms.

As discussed in Sec.~\ref{sec:bulk_topology}, the band topology of the Fu model is characterized by $\nu_4(\bar{k}_z)$ and $\nu_{2}(\bar{k}_z)$.
Since these topological indices are fragile against the addition of trivial bands such as $s$ orbitals, we expect that the $sp$ orbital coupling should trivialize these topological indices, whereas its effect on the surface states can depend on whether the influence of additional $s$-orbital bands reaches the surface.
To demonstrate this, we consider two parameter regimes, the one where surface states of $s$ orbitals are present, and the other where they are absent. Here, we introduce the surface states of $s$ orbitals in the $(001)$ plane by implementing a 2D array of the 1D Su-Schrieffer-Heeger model~\cite{Su1979} of $s$ orbitals along the $z$ direction, as given by $[\mathcal{H}_s(\bm{k})]_{ij}$ ($i,j=5,6$).
In this setup, when $|t_{s2}'/t_{s1}'| > 1$, there exist surface charges associated with a nontrivial Zak phase in the $z$ direction. Including the $sp$ orbital coupling and the hopping in the $xy$ plane lifts the degeneracy of the $s$-orbital surface states, but they remain in the band gap as long as these terms in the Hamiltonian are sufficiently small.  

\begin{figure}[tbp]
\centering
 \includegraphics[width=8cm]{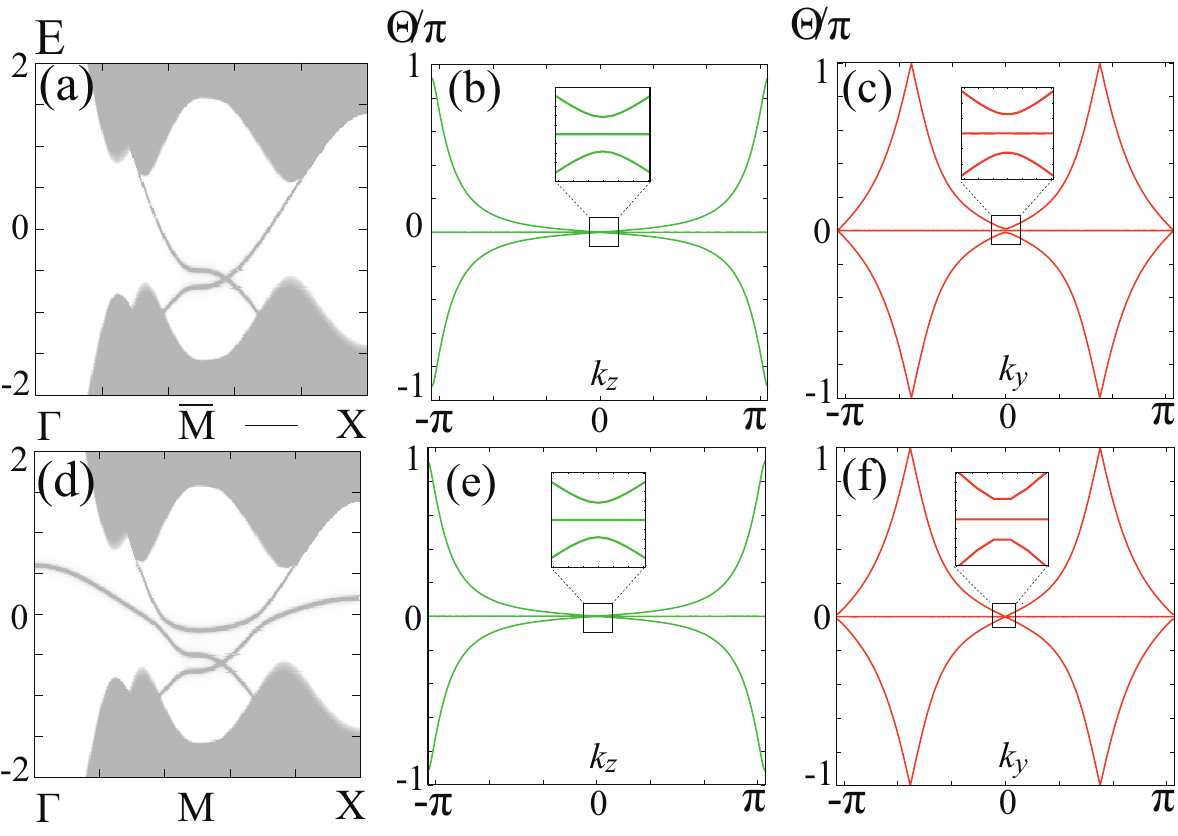}
 \caption{
(a) and (d) Energy spectra of $\mathcal{H}_\mathrm{Fu}^{\prime}+\mathcal{H}_s$.
(b) and (e) Wilson loop spectra for the path in Fig.~\ref{fig:wilsonloop} (b).
(c) and (f) Wilson loop spectra for the path in Fig.~\ref{fig:wilsonloop} (c).
 The perturbed Fu model $\mathcal{H}_\mathrm{Fu}^{\prime}$ with the perturbation terms $(\mathcal{M}_1,\mathcal{M}_2)=(0.1,0)$ in Eq.~(\ref{eq:Fumodel+}) has two surface Dirac cones located on the line connecting $\bar{M}$ and $\bar{X}$, split from the quadratic band touching at the $\bar{M}$ point in the Fu model (Fig.~\ref{fig:Fumodelsurf}). The parameters in $\mathcal{H}_s$ are chosen as 
$(t^A_{sp}, t^B_{sp}, t_{s0}^A, t_{s1}^A, t_{s0}^B,t_{s1}^B) = (0.1,-0.1,0.2,0.1,-0.2,-0.1)$.
In (a), (b), and (c), the parameters of the hopping along the $z$ axis are $(t_{s1}',t_{s2}')=(4,1)$, so that
$\mathcal{H}_s$ has no $s$-orbital surface state.
In (d), (e), and (f), the hopping parameters are $(t_{s1}',t_{s2}')=(1,4)$, and $\mathcal{H}_s$ has $s$-orbital surface states.
The surface energy spectrum is gapped by the hybridization with an $s$-orbital surface band in (d).
The unwinding of Wilson loop spectra in (b), (c), (e), and (f) indicate that $\nu_4$ and $\nu_2$ are trivial. 
Note that the addition of $s$ orbitals lifts the degeneracy of the Wilson bands at $k_y=0$ in (b) and (c), implying that the $\mathbb{Z}_2$ invariant $\bar{\nu}_2$ becomes ill-defined even when $\mathcal{M}_1=\mathcal{M}_2=0$.
 }\label{fig:Fumodel_fragile}
\end{figure}

Figures~\ref{fig:Fumodel_fragile} (a)-(f) show the surface density of states and Wilson loop spectra of Eq.~(\ref{eq:Fu+smodel}).
Specifically, Figs.~\ref{fig:Fumodel_fragile} (a)-(c) represent the case without $s$-orbital surface states ($|t_{s2}'/t_{s1}'| < 1$), and (d)-(f) the case with $s$-orbital surface states ($|t_{s2}'/t_{s1}'| > 1$).
In the absence of the $s$-orbital surface states, Fig.~\ref{fig:Fumodel_fragile} (a) shows that the surface Dirac cones are robust and remain to be gapless, while Fig.~\ref{fig:Fumodel_fragile} (b) and (c) indicate that the Wilson loop spectra are unwound by the the $sp$ orbital hybridization, implying that $\nu_4 $ and $\nu_2 $ are trivialized.
On the other hand, in Fig.~\ref{fig:Fumodel_fragile} (d), a surface-localized quadratic band hybridizes with an $s$-orbital surface band, and as a result a gap opens.
The unwinding of the Wilson loop spectra is also seen in Figs.~\ref{fig:Fumodel_fragile} (e) and (f), implying $\bar{\nu}_4=\bar{\nu}_2=0$.

We conclude that surface states of a fragile topological insulator are not always gapped out even when the bulk topology is trivialized by the addition of trivial bands.
Gapping surface states requires hybridization with another surface band localized at the surface.
This mechanism is also briefly discussed in Ref.~\onlinecite{turner2021}.

\section{More lattice models}
\label{sec:other_cases}

In the previous sections we have discussed gapless surface states and related bulk topological invariants for the Fu model and its variants.
In this section we introduce another series of toy models defined on tetragonal and hexagonal lattices, for which the surface Hamiltonian can be easily derived in the continuum limit $\bm{k}\to0$, and discuss their surface states and topological invariants.

The bulk Hamiltonian of the toy model is written in the continuum limit (i.e., near the $\Gamma$ point) as
\begin{align}
 \mathcal{H}_{l}(\bm{k}) =\, & M_{z} \bm{1}_2 \otimes \tau_z+\{f_+(\bm{k}_{\parallel}) \sigma_+ + [f_+(\bm{k}_{\parallel})]^{\ast} \sigma_-\} \otimes \tau_x \notag \\
                              & + t_z k_z \sigma_y \otimes \tau_x , \label{eq:bulk_hami}
\end{align}
where $\sigma_i$ and $\tau_i$ are the Pauli matrices in orbital and sublattice spaces, $f_+(\bm{k}_{\parallel})$ is related by $f_+=if$ to the symmetry-allowed function $f(\bm{k}_{\parallel})$ discussed in Sec.~\ref{sec:surftheory_pseudo}, $\sigma_{\pm} \equiv (\sigma_z \pm i \sigma_x)/2$, and $M_z$ and $t_z$ are real numbers.
The Hamiltonian $\mathcal{H}_l$ enjoys TR symmetry and $C_n$ rotation symmetry with orbital angular momentaum $l$ such that
\begin{align}
 &\mathcal{T} \mathcal{H}_{l}(\bm{k}) \mathcal{T}^{-1} =  \mathcal{H}_{l}(-\bm{k}), \quad \mathcal{T} =\bm{1}_4 K, \label{eq:tbTR_sym}\\
 &\mathcal{C}_{n,l}  \mathcal{H}_{l}(\bm{k}) \mathcal{C}_{n,l}^{-1} = \mathcal{H}_{l}(R_n \bm{k}), \quad
   \mathcal{C}_{n,l}  =e^{i \frac{2\pi l}{n} \sigma_y} \otimes \bm{1}_2, \label{eq:tbrot_sym}
\end{align}
where $\mathcal{T}^2=(\mathcal{C}_{n,l})^n = \bm{1}_{4 }$ and $[\mathcal{T},\mathcal{C}_{n,l}]=0$.
One can derive the surface theory $H(\bm{k}_\parallel)$ discussed in Sec.~\ref{sec:surftheory_pseudo} from $\mathcal{H}_l(\bm{k})$ by introducing a kink in the Dirac mass $M$;
see Appendix~\ref{app:surfH}.
For the double Hamiltonian $\mathcal{H}_{l}(\bm{k}) \oplus \mathcal{H}_{l}(\bm{k})$
we impose the TR symmetry $\mathcal{T}=\bm{1}_8 K$ and the extended $C_n$ symmetry
\begin{align}
 \mathcal{C}_{n,l}^{\pm}  =\mathcal{C}_{n,l} \oplus (\pm \mathcal{C}_{n,l}). \label{eq:tbdoublerot_sym}
\end{align}

In order to study the bulk-boundary correspondence between the existence of surface Dirac cones and the bulk topological invariants,
we use tight-binding models $\mathcal{H}_l^\mathrm{lattice}$ that are reduced to $\mathcal{H}_l$ with $f(\bm{k}_\parallel)-m_0\delta_{2l,n} \propto k_+^{2l}$ at the $\Gamma$ point $\bm{k}=0$ ($m_0\in\mathbb{C}$ defined in the caption of Table \ref{tab:kp_theory}), 
 except for $\mathcal{H}_{l=2}^\mathrm{hexa}$, in which $f(\bm{k}_\parallel) \propto k_-^2$.
The tight-binding models are defined on a tetragonal lattice for C$_4$ symmetry and a hexagonal lattice for C$_6$ symmetry,
and they are presented in Appendix~\ref{app:tbmodel}, including the perturbations that change the gapless surface band structure from the quadratic band touching at $\bm{k}_\parallel=0$ to multiple Dirac cones.

The bulk band topology is characterized by $\bar{\nu}_{2}$ when $n=2$, $(\bar{\nu}_{2},\bar{\nu}_4)$ when $n=4$, and $(\bar{\nu}_{2},\bar{\nu}_{6})$ when $n=6$.
Here a $C_6$ symmetry-protected weak index $\nu_6(\bar{k}_z) \in \{0,1\}$ is introduced to characterize systems with C$_6$ symmetry; see Appendix~\ref{sec:C46symmetry} for its definition.  The strong index is defined by $\bar{\nu}_6 \equiv \nu_6(\pi)-\nu_6(0)$ and is determined from the Wilson loop calculation in a similar way to $\bar{\nu}_4$~\cite{Alexandradinata2016Berry}. 
We note that, when the number of occupied bands is two, the exponent $2l$ in the dispersion $f(k_\parallel)\propto k_+^{2l}$ equals the strong index $\bar{\nu}_2$, the Euler class.

We have calculated the surface spectra and the bulk topological invariants for the tight-binding models.
The results are summarized in Table~\ref{tab:tb_model} and discussed below.

\begin{table}[tb]
\caption{
Surface states and topological invariants for the tight-binding models whose continuum limit has the form in Eq.~(\ref{eq:bulk_hami}).
The tight-binding models are presented in Appendix~\ref{app:tbmodel}.
The third and fourth columns show the size of bulk Hamiltonians [i.e., a $4\times 4$ Hamiltonian (minimal) or a direct sum of two $4\times 4$ Hamiltonians (double)], and their TR and C$_n$ symmetry defined in Eq.~(\ref{eq:tbTR_sym}) and (\ref{eq:tbrot_sym}). The fifth and sixth columns describe the gapless structure of surface states and the relevant topological invariants, which we determine from the numerical calculation using tight-binding models.
The topological invariants are defined by $\bar{\nu}_n =\nu_n (k_z=\pi) - \nu_n (k_z=0)$. Note that there is a sign ambiguity in $\bar{\nu}_{n}$, and that $\bar{\nu}_4=2$ and $\bar{\nu}_6=2$ in the double Hamiltonian remains intact under $\mathcal{C}_{4,1}^-$ and $\mathcal{C}_{6,1}^-$ which forbid any gap-opening mass term in the surface Hamiltonian.
}
\label{tab:tb_model}
\begin{tabular}{cccccc}
\hline\hline
$n$ & $2l$ & $H_\mathrm{bulk}$ & Symmetry & Surface & Top.~Invariants \\
\hline 
2 & 2 & Minimal & $\{\mathcal{T},\mathcal{C}_{2,1}\}$ & 2 Dirac cones & $\bar{\nu}_2 =2$\\
2 & 2 & Double & $\{\mathcal{T},\mathcal{C}_{2,1}^{\pm}\}$ & Gapped & $\bar{\nu}_2 =0$\\
4 & 2 & Minimal & $\{\mathcal{T},\mathcal{C}_{4,1}\}$ & Quadratic & $(\bar{\nu}_2,\bar{\nu}_4) =(2,1)$\\
4 & 2 &  Double  & $\{\mathcal{T},\mathcal{C}_{4,1}^+\}$ & Gapped& $(\bar{\nu}_2,\bar{\nu}_4) =(0,0)$\\
4 & 2 & Double  & $\{\mathcal{T},\mathcal{C}_{4,1}^-\}$& 4 Dirac cones& $(\bar{\nu}_2,\bar{\nu}_4) =(0,2)$\\
4 & 4 & Minimal & $\{\mathcal{T},\mathcal{C}_{4,2}\}$ & 4 Dirac cones& $(\bar{\nu}_{2},\bar{\nu}_{4}) =(4,0)$\\
4 & 4 & Double & $\{\mathcal{T},\mathcal{C}_{4,2}^{\pm}\}$ & Gapped & $(\bar{\nu}_{2},\bar{\nu}_{4}) =(0,0)$\\
6 & 2 & Minimal & $\{\mathcal{T},\mathcal{C}_{6,1}\}$& Quadratic & $(\bar{\nu}_{2},\bar{\nu}_{6}) =(2,1)$\\
6 & 2 &  Double & $\{\mathcal{T},\mathcal{C}_{6,1}^+\}$ &  Gapped&$(\bar{\nu}_2,\bar{\nu}_6) =(0,0)$\\
6 & 2 &  Double &  $\{\mathcal{T},\mathcal{C}_{6,1}^-\}$ &6 Dirac cones& $(\bar{\nu}_2,\bar{\nu}_6) =(0,2)$\\
6 & 4 & Minimal & $\{\mathcal{T},\mathcal{C}_{6,2}\}$& Quadratic & $(\bar{\nu}_{2},\bar{\nu}_{6}) =(2,1)$\\
6 & 4 &  Double & $\{\mathcal{T},\mathcal{C}_{6,2}^+\}$ &  Gapped&$(\bar{\nu}_2,\bar{\nu}_6) =(0,0)$\\
6 & 4 &  Double &  $\{\mathcal{T},\mathcal{C}_{6,2}^-\}$ &6 Dirac cones& $(\bar{\nu}_2,\bar{\nu}_6) =(0,2)$\\
6 & 6 & Minimal &  $\{\mathcal{T},\mathcal{C}_{6,3}\}$  & 6 Dirac cones & $(\bar{\nu}_{2},\bar{\nu}_{6}) =(6,0)$\\
6 & 6 & Double &  $\{\mathcal{T},\mathcal{C}_{6,3}^{\pm}\}$  &Gapped & $(\bar{\nu}_{2},\bar{\nu}_{6}) =(0,0)$\\
\hline\hline
\end{tabular} 
\end{table}

When $2l =0 \mod n$, the surface Dirac cones are characterized by the $C_2T$ symmetry-protected strong index $\bar{\nu}_2$ only, i.e., the Euler class, which determines the number of surface Dirac cones.
The invariants $\bar{\nu}_4$ and $\bar{\nu}_{6}$ are zero because the rotation symmetry has a trivial representation, $\mathcal{C}_{n,n/2} = -\bm{1}_4$.
In this case, our lattice models realize 3D fragile $\mathbb{Z}$ topological insulators.
For their double Hamiltonian, $\bar{\nu}_2$ becomes trivial, and the surface Dirac cones are gapped out.

When $2l \neq  0 \mod n$ ($n=4,6$), surface bands have a quadratic band touching, and the $C_n$ symmetry-protected strong index becomes nontrivial, $\bar{\nu}_n=1$, while the $C_2T$ symmetry-protected strong index $\bar{\nu}_2=2$. 
As for the double Hamiltonian with symmetry-preserving perturbations, the quadratic band touching is gapped out for the $C_{4,1}^+$ and $C_{6,l=1,2}^+$ symmetries, whereas it remains stable under $C_{4,1}^-$ and $C_{6,l=1,2}^-$.
The bulk topological invariants are given by $(\bar{\nu}_{2},\bar{\nu}_4) = (0,0)$ and $(\bar{\nu}_{2},\bar{\nu}_6) = (0,0)$ for $\mathcal{C}_{4,1}^+$ and $\mathcal{C}_{6,l=1,2}^+$ symmetries, and $(\bar{\nu}_{2},\bar{\nu}_4) = (0,2)$ and $(\bar{\nu}_{2},\bar{\nu}_6) = (0,2)$ for $\mathcal{C}_{4,1}^-$ and $\mathcal{C}_{6,l=1,2}^-$ symmetries. 
Since $\bar{\nu}_4$ and $\bar{\nu}_6$ are $\mathbb{Z}_2$ invariants, these topological invariants of double Hamiltonians are trivial.
Nevertheless, no gap-opening perturbation to the surface state is allowed by $C_{4,1}^-$ and $C_{6,l=1,2}^-$ symmetries, which implies that a nontrivial $C_{4}^-$ or $C_{6}^-$ symmetry-protected topology exists even for the double Hamiltonians.

\section{Concluding remarks}
\label{sec:conclude}
In this paper, we have discussed $T$ and $C_n$ symmetry-protected fragile topological phases in class AI. By developing the surface theory in the pseudo-spin basis with orbital angular momentum $l$, we found a series of fragile topological phases with $n$ surface Dirac cones, which emerge in the two different mechanisms.
As a representative model we have used the Fu model, which has $C_{4,1}$ rotation symmetry in our notation.

One mechanism works when the orbital angular momentum $l$ satisfy $2l=0 \mod n$.  In this case, $n$ surface Dirac cones appear in the energy spectrum of  $2 \times 2$ surface Hamiltonian. For example, we have demonstrated that the perturbed Fu model with C$_{2,1}$ symmetry has two surface Dirac cones, which are dictated by the $C_2T$ symmetry-protected strong (Euler) index $\bar{\nu}_2=2$. We have shown that this type of fragile topological insulators are 3D fragile $\mathbb{Z}$ topological insulators with an even integer $\bar{\nu}_2$ specifying the number of surface Dirac cones.
We expect the photonic crystals and metamaterials~\cite{Yannopapas2011,Lu2014topological,Slobozhanyuk2017three,Ochiai2017,Ozawa2019,Yang2019realization,Kim2020recent,He2020acoustic} to be a potential platform for realizing these fragile topological phases with multiple surface Dirac cones.  

The second mechanism is effective when $2l\ne0 \mod n$.
In this case, the surface bands have a quadratic band touching at a high-symmetry point in the $2\times2$ surface Hamiltonian, while the doubled surface Hamiltonian has four (six) surface Dirac cones protected by $C_4^-$ ($C_6^-$) symmetry, which suggests $\mathbb{Z}_2 \times \mathbb{Z}_2$ topological structure. For example, we have found four surface Dirac cones in the double Fu model with symmetry-preserving perturbations.
It was found, however, that the bulk $\mathbb{Z}_2$ topological invariants $\bar{\nu}_4$ and $\bar{\nu}_2$ fail to identify this topological phase with four surface Dirac cones, because $\bar{\nu}_4$ and $\bar{\nu}_2$ become even, thus trivial, in the double bulk Hamiltonian.
It seems that another topological invariant that can distinguish $(2,0)$ and $(1,1) \in \mathbb{Z}_2 \times \mathbb{Z}_2$ is needed, but is still lacking.
Identifying such a topological invariant is left for future work.

Furthermore, for the perturbed Fu model coupled with an $s$-orbital band, we have shown that the surface Dirac cones have some robustness against the addition of a trivial band that resolves the Wannier obstruction in the perturbed Fu model.
We have demonstrated that the hybridization with a surface-localized trivial band is necessary for gapping out the surface states.
We expect similar robustness for other surface Dirac cones since they are also protected by a representation-dependent topology.

From the results shown in Sec. III and Sec.~IV, we can conclude that the bulk-boundary correspondence holds if the prerequisite conditions on symmetry (time-reversal and rotation) and band representations (a fixed number of bands in the pseudo-spin basis) are respected.
When the conditions are violated (for example, by the addtion of $s$-orbital bands), we cannot expect the bulk-boundary correspondence to hold anymore.

Finally, we comment on the analogy and difference between classes AI and AII.
Recently multiple surface Dirac cones have also been studied in class AII topological crystalline insulators having surface rotation anomaly~\cite{CFang19}, which are protected by rotation symmetry of the C$_{n}^-$ type ($n=2,4,6$).
Thus one may wonder if the second mechanism we have discussed above can be thought of as an analog of the surface rotation anomaly.
However, there is an important difference.  The class AII topological crystalline insulators with rotation symmetry are shown\cite{CFang19} to have hinge states on the side surfaces that are parallel to the rotation axis. Since the rotation symmetry is not a good symmetry on the side surface, Dirac fermions on the side surface acquire a Dirac mass $M(\bm{r})$, which changes its sign under the rotation as $M(R_n\bm{r}) = -M(\bm{r})$. As a result, the side surface always has domain walls, along which gapless edge (hinge) states must be present.  The existence of the gapless hinge states is related to the topological classification $\mathbb{Z}_2$ of class AII insulators in two dimensions.  By contrast, 2D class AI insulators have only a trivial insulator phase as a stable phase, and therefore 3D class AI insulators cannot have any hinge states unless other additional symmetry is assumed.

\section*{acknowledgements}
S.K. thanks Ai Yamakage for the numerical calculation of the Wilson loop. 
This work was supported by JSPS KAKENHI (Grant Nos.\ 19K03680,  19K14612, JP19H01824) and JST CREST (Grant Nos.\ JPMJCR16F2, JPMJCR19T2).

\appendix

\section{Momentum dependent mass term}
\label{sec:massterm}

For the surface theory in class AI, momentum dependent mass terms can gap out Dirac points located away from high-symmetry points. Here we explain this gap-opening mechanism for a $2 \times 2$ Hamiltonian and a $4 \times 4$ Hamiltonian.

We take the $2\times2$ Hamiltonian defined by Eq.~(\ref{eq:sym_analysis}) with the mometum dependent mass term $g(\bm{k}_{\parallel})\sigma_y$,
where $f(\bm{k}_\parallel)$ and $g(\bm{k}_\parallel)$ are complex and real functions of $\bm{k}_\parallel$. The energy spectrum is given by
\begin{align}
E(\bm{k}_{\parallel}) = \pm \sqrt{f_\mathrm{r}(\bm{k}_{\parallel})^2 + f_\mathrm{i}(\bm{k}_\parallel)^2 + g(\bm{k}_{\parallel})^2}, \label{eq:app_energy}
\end{align}
where we have introduced the real and imaginary parts of the complex function,
$f(\bm{k}_\parallel)=f_\mathrm{r}(\bm{k}_\parallel) + i f_\mathrm{i}(\bm{k}_\parallel)$.
The two surface bands with positive and negative energies can touch at the momenta where the three conditions
$f_\mathrm{r}(\bm{k}_{\parallel}) = f_\mathrm{i}(\bm{k}_{\parallel}) = g(\bm{k}_{\parallel}) =0$ are satisfied simultaneously. Since we only have two variables
$\bm{k}_\parallel=(k_x,k_y)$, the conditions are not satisfied in general, which means that the surface states are gapped. However, the presence of both TR and $C_n$ symmetries demands $g=0$, so that there are solutions for $E(\bm{k}_{\parallel})=0$, which correspond to either quadratic band touching at a high-symmetry point or $n$ Dirac points at generic momenta.

Similarly, we can show the gap-opening for $4\times4$ Hamiltonian as follows. We consider the double Hamiltonian with a mometum dependent mass term $g'(\bm{k}_{\parallel})$, 
\begin{align}
 H(\bm{k}_{\parallel}) \oplus H(\bm{k}_{\parallel}) + g'(\bm{k}_{\parallel}) \sigma_y \otimes \tau_y,
\end{align}
with $H(\bm{k}_{\parallel}) = f(\bm{k}_{\parallel}) \sigma_+ + f^{\ast}(\bm{k}_{\parallel}) \sigma_-$.
Since the gamma matrices, $\sigma_z\otimes\tau_0$, $\sigma_x\otimes\tau_0$, and $\sigma_y\otimes\tau_y$, mutually anti-commute, the energy spectrum is given by Eq.~(\ref{eq:app_energy}) with $g$ replaced by $g'$. The following discussion goes in parallel to the case of the $2 \times 2$ Hamiltonian.
Under the representation $C_{n,l}^+=C_{n,l}\otimes \tau_0$, in which the momentum-independent mass term $M \sigma_y\otimes\tau_y$ is allowed, the spectrum is fully gapped in both cases (i) and (ii) discussed in Sec.~\ref{sec:surftheory_pseudo}.
On the other hand, under the representation $C_{n,l}^-=C_{n,l}\otimes\tau_z$, in which the allowed momentum-dependent mass term $g'(\bm{k}_\parallel)\sigma_y\otimes\tau_y$ vanishes at $\bm{k}_\parallel=0$, surface states are fully gapped in case (i) and remain gapless in case (ii).

\section{Multiple Dirac points in $2 \times 2$ surface Hamiltonian}
\label{sec:derivation}

According to the surface theory discussed in Sec.~\ref{sec:surftheory_pseudo}, $f(\bm{k}_{\parallel})$ includes multiple terms in the leading order.
We here show the emergence of $n$ Dirac points for the case (i) $2l=0$ (mod $n$) discussed in Sec.~\ref{sec:surftheory_pseudo}, assuming the general form of $f(\bm{k}_{\parallel}) = M(\bm{k}_{\parallel}) + v_+ k_+^n + v_- k_-^n$,
where $M(\bm{k}_{\parallel})=m_0+v_0|\bm{k}_{\parallel}|^n$ and $m_0,v_0,v_{\pm} \in \mathbb{C}$.
For the purpose of convenience, we use the polar coordinate $\bm{k}_\parallel=(k_x,k_y) =(k\cos\theta,k\sin\theta)$
to write $f(\bm{k}_{\parallel}) = m_0 + v_0 k^n + v_+ k^n e^{i n \theta} + v_- k^n e^{-i n \theta}$.

\begin{figure*}[tbp]
\centering
 \includegraphics[width=14cm]{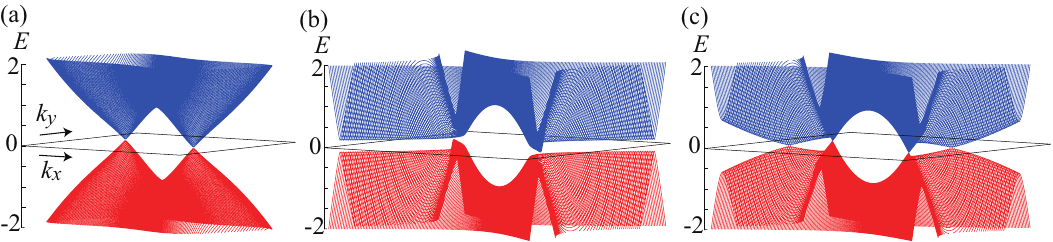}
 \caption{
Energy spectra of Eq.~(\ref{eq:Hami_app_2dim}) with $n=2$.
We set $v_1=v_2=2$ and $\phi_1=0$ for all figures.
Other parameters are chosen differently for each figure:
(a) $(m_0^{\rm r},v_0^{\rm r},m_0^{\rm i},v_0^{\rm i}, \phi_2)=(0.4,0.8,0.8,0.4,\pi/5)$, (b) $(0.8,0.3,0.6,0.3,\pi/2)$, and (c) $(0.4,0.8,0.8,0.4,\pi/2)$.
These figures exhibit (a) two Dirac cones, (b) gapped spectrum, and (c) four Dirac cones.
 }\label{fig:surfaceband}
\end{figure*}

To being with, we divide $f(\bm{k}_{\parallel})$ into the real and imaginary parts, $f(\bm{k}_{\parallel})=f_{\rm r}(\bm{k}_{\parallel})+if_{\rm i}(\bm{k}_{\parallel})$, and rewrite the surface Hamiltonian as 
\begin{align}
 H(\bm{k}_{\parallel})=f_{\rm r}(\bm{k}_{\parallel}) \sigma_z -f_{\rm i}(\bm{k}_{\parallel}) \sigma_x, \label{eq:Hami_app_2dim}
\end{align}
with
\begin{subequations}
\begin{align}
&f_{\rm r}(\bm{k}_{\parallel}) = M_1(\bm{k}_{\parallel}) +v_1 k^n \cos (n \theta + \phi_1), \\
&f_{\rm i}(\bm{k}_{\parallel}) = M_2(\bm{k}_{\parallel}) +v_2 k^n \sin (n \theta + \phi_2),
\end{align}
\end{subequations}
where $M_1(\bm{k}_{\parallel}) \equiv m_0^{\rm r}+v_0^{\rm r} k^n $ and $M_2(\bm{k}_{\parallel}) \equiv m_0^{\rm i}+v_0^{\rm i} k^n $. The superscripts ``r'' and ``i'' represent the real and imaginary parts of the parameters, e.g., $m_0 = m_0^{\rm r} + i m_0^{\rm i}$. The parameters $v_1$, $v_2$, $\phi_1$, and $\phi_2$ are defined by
\begin{subequations}
\begin{align}
 &v_1 = \sqrt{(v_+^{\rm r}+v_-^{\rm r})^2+(v_+^{\rm i}-v_-^{\rm i})^2}, \\
 &v_2 = \sqrt{(v_+^{\rm r}-v_-^{\rm r})^2+(v_+^{\rm i}+v_-^{\rm i})^2}, \\
 &\tan (\phi_1) = \frac{v_+^i-v_-^i}{v_+^{\rm r}+v_-^{\rm r}}, \\
 &\tan (\phi_2) = \frac{v_+^i+v_-^i}{v_+^{\rm r}-v_-^{\rm r}},
\end{align}
\end{subequations}
where $|\phi_{1,2}|<\pi/2$.
Since the energy spectrum of Eq.~(\ref{eq:Hami_app_2dim}) is given by $E(\bm{k}_{\parallel}) = \pm \sqrt{f_{\rm r}^2(\bm{k}_{\parallel})+f_{\rm i}^2(\bm{k}_{\parallel})}$, the gapless spectra appear at momenta that satisfy $f_{\rm r}(\bm{k}_{\parallel}) = f_{\rm i}(\bm{k}_{\parallel})=0$. These conditions lead to
\begin{subequations} 
\label{cos(n theta + phi1)} 
\begin{align}
 &\cos(n \theta + \phi_1) = - \frac{M_1(\bm{k}_{\parallel})}{v_1k^n}, \\
 & \sin(n \theta + \phi_2) = - \frac{M_2(\bm{k}_{\parallel})}{v_2k^n},
\end{align}
\end{subequations}
when $k \neq 0$.
We can transform Eq.~(\ref{cos(n theta + phi1)}) to the following equation that determines $k$:
\begin{align}
 k^{2n} \cos^2 (\phi_1-\phi_2) 
                    = & \left(\frac{M_1(\bm{k}_{\parallel})}{v_1} \right)^2 + \left(\frac{M_2(\bm{k}_{\parallel})}{v_2} \right)^2 \notag \\
                      & + \frac{2 M_1(\bm{k}_{\parallel}) M_2(\bm{k}_{\parallel})}{v_1v_2} \sin (\phi_1 - \phi_2). \label{eq:caseI}
\end{align}
We analyze these equations for the following two cases separately.
\begin{itemize}
\item[(I)] When $\phi_1 \neq \phi_2 - (-1)^l \frac{\pi}{2}$ with $l = 0$ or $1$, $k$ is determined from Eq.\ (\ref{eq:caseI}).
Then we substitute the solution to the right-hand side of Eq.\ (\ref{cos(n theta + phi1)}) to obtain $n$ solutions of $\theta$ modulo $2\pi$.
These solutions give $n$ Dirac points; see Fig.~\ref{fig:surfaceband} (a). In particular, when $\phi_1 =\phi_2$, Eq.~(\ref{eq:caseI}) is reduced to
\begin{align}
 k^{2n} = \left(\frac{M_1(\bm{k}_{\parallel})}{v_1} \right)^2 + \left(\frac{M_2(\bm{k}_{\parallel})}{v_2} \right)^2,
\end{align}
 as discussed in Sec.~\ref{sec:single_hami}.
\item[(II)] When $\phi_1 = \phi_2 - (-1)^l \frac{\pi}{2}$ with $l = 0$ or $1$, Eq.\ (\ref{cos(n theta + phi1)}) as well as Eq.~(\ref{eq:caseI})
leads to the condition
\begin{align}
 \frac{M_1(\bm{k}_{\parallel})}{v_1} =(-1)^l \frac{M_2(\bm{k}_{\parallel})}{v_2},
\end{align}
which is written as
\begin{equation}
k^n=-\frac{m_0^{\rm r} v_2 - (-1)^l m_0^{\rm i} v_1}{v_0^{\rm r}v_2 - (-1)^l v_0^{\rm i} v_1}.
\end{equation}
If the right-hand side is negative, there is no solution; the surface states are gapped as shown in Fig.~\ref{fig:surfaceband} (b).
If the right-hand side of the above equation is positive, then $k(>0)$ is determined, and
substituting it into Eq.\ (\ref{cos(n theta + phi1)}) yields $2n$ solutions of $\theta$.
We find $2n$ Dirac points; see Fig.~\ref{fig:surfaceband} (c). 
\end{itemize}

\section{Double Hamiltonian with different $l$'s}
\label{sec:doublehami}

In this appendix, we discuss general form of double Hamiltonian and its stability for surface states.
For systems with TR and C$_n$ symmetries, $4 \times 4$ surface Hamiltonian in the pseudo spin basis is constructed by combining two $2 \times 2$ surface Hamiltonian with orbital pseudo spin $l$ and $l'$, 
\begin{align}
H_{l,l'} \equiv H_{l} \oplus H_{l'},
\end{align}
where TR and C$_n$ symmetries are represented as
\begin{align}
&T \equiv \bm{1}_4 K, \\
&C_{n,(l,l')}^{\pm} \equiv \begin{pmatrix} e^{i \frac{2\pi l }{n} \sigma_y} & 0 \\ 0 & \pm e^{ i \frac{2\pi l'}{n}  \sigma_y} \end{pmatrix}.
\end{align}
The double Hamiltonian $H_{l,l'}$ with $l = l'$ is discussed in Sec.~\ref{sec:surftheory_pseudo}.
Here we focus on the case with $l \neq l'$, i.e., $H_{0,1}$ for $n=4$; $H_{0,1}$, $H_{0,2}$, and $H_{1,2}$ for $n=6$. 

We first notice that $H_{0,1}$ and $H_{0,2}$ are adiabatically connected to a trivial gapped surface Hamiltonian, since the basis of $H_0$ is an $s$-orbital-like. That is, an $sp$-or $sd$-orbital mixing term can open a gap as discussed in Sec.~\ref{sec:fragile_topology}.

The stability of $H_{1,2}$ depends on whether the symmetry of interest is $C_{6,(1,2)}^{+}$ or $C_{6,(1,2)}^-$, because
$H_{1,2}$ with $C_{6,(1,2)}^{\pm}$ can be mapped to $H_{1,1}$ with $C_{6,1}^{\mp}$.
To see this, we consider a concrete example of $H_{1,2}$,
\begin{align}
H_{1,2}(\bm{k}_{\parallel}) = v (k_x^2 -k_y^2) \sigma_z \otimes \tau_0 - 2v k_x k_y\sigma_x \otimes \tau_z,
\end{align}
which satisfies $C_{6,(1,2)}^{\pm} H_{1,2}(\bm{k}_{\parallel})(C_{6,(1,2)}^{\pm})^{\dagger} =H_{1,2}(R_6\bm{k}_{\parallel})$ with
\begin{align}
 C_{6,(1,2)}^{\pm} = \begin{pmatrix} e^{i \frac{2\pi }{6} \sigma_y} & 0 \\ 0 & \pm e^{ i \frac{4\pi}{6}  \sigma_y} \end{pmatrix}.
\end{align}
Performing the unitary transformation $U = \bm{1}_2 \oplus \sigma_z$ on $H_{1,2}$ and $C_{6,(1,2)}^{\pm}$, we obtain
\begin{align*}
U H_{1,2}(\bm{k}_{\parallel}) U^{\dagger}& = v (k_x^2 -k_y^2) \sigma_z \otimes \tau_0 - 2v k_x k_y \sigma_x \otimes \tau_0, \\
U C_{6,(1,2)}^{\pm}  U^{\dagger}&= \begin{pmatrix} e^{i \frac{2\pi }{6} \sigma_y} & 0 \\ 0 & \mp e^{ i \frac{2\pi}{6}  \sigma_y} \end{pmatrix}
=C^{\mp}_{6,1}.
\end{align*}
Following the discussion in Sec.~\ref{sec:surftheory_pseudo}, we can conclude that
$H_{1,2}$ with $C_{6,(1,2)}^{+}$ ($C_{6,(1,2)}^-$) is stable (unstable) against symmetry-preserving perturbations.

\section{$C_n$ symmetry-protected topological invariants}

\subsection{$C_n$ $(n=4,6)$ and TR symmetries}
\label{sec:C46symmetry}
\begin{figure}[tbp]
\centering
 \includegraphics[width=6cm]{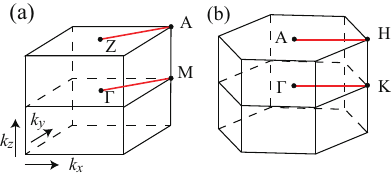}
 \caption{ 3D BZ of  (a) a tetragonal lattice and  (b)  a hexagonal lattice. The red lines indicate the integration paths in the definition of $\mathbb{Z}_2$ topological invariants~(\ref{eq:c46_inv_weak}). 
 }\label{fig:integralpath}
\end{figure}

We review the $\mathbb{Z}_2$ invariants protected by C$_n$ $(n=4,6)$ and TR symmetries, which are defined only when the representations of the $C_n$ operator forms a 2D representation under TR symmetry.
Now, let $| u_m(\bm{k}) \rangle $ ($m=1,\cdots, 2N$) be the occupied states of the bulk Hamiltonian and $C_n$ and $T$ the symmetry operators acting on this basis, where $m$ labels energy bands, a pair of $\{| u_{2m-1}(\bm{k}) \rangle, | u_{2m}(\bm{k}) \rangle \}$ describes the basis of the 2D representation, and we choose the rotation axis to be along the $z$ direction.
The $\mathbb{Z}_2$ invariant $\nu_{n,\bm{k}_1\bm{k}_2}$ is defined by~\cite{Fu2011,Alexandradinata2016Berry}
\begin{align}
 &(-1)^{\nu_{n, \bm{k}_1 \bm{k}_2}} = \exp \left( i \int_{\bm{k}_1}^{\bm{k}_2} d \bm{k} \cdot \mathcal{A}(\bm{k}) \right) \frac{\pf[V_4(\bm{k}_2)]}{\pf[V_4(\bm{k}_1)]}, \label{eq:c46_inv_weak}
\end{align}
where $\bm{k}_{i=1,2}$ are $C_nT$ invariant points, $\mathcal{A}(\bm{k})$ is the U(1) Berry connection defined by
\begin{equation}
\mathcal{A}(\bm{k}) \equiv -i \sum_{m \in {\rm occ}} \langle u_m(\bm{k}) | \partial_{\bm{k}} | u_m (\bm{k}) \rangle,
\end{equation}
and $V_n(\bm{k}_i)$ is a skew-symmetric part of the matrix
\begin{equation}
[w_n(\bm{k}_i)]_{mm'} \equiv \langle u_m(\bm{k}_i) | C_nT | u_{m'} (\bm{k}_i) \rangle.
\end{equation}
Specifically, we have $V_4 (\bm{k}_i) = w_4(\bm{k}_i)$ for C$_4$ symmetry and $V_6 (\bm{k}_i) = [w_6(\bm{k}_i)-w_6^T(\bm{k}_i)]/2$ for C$_6$ symmetry. Note that $w_4$ is by itself a skew-symmetric matrix because of $(C_4T)^2=-1$. Equation~(\ref{eq:c46_inv_weak}) is invariant under the gauge transformation, $|u_m (\bm{k})\rangle \to |u_{m'} (\bm{k}) \rangle [U_{\bm{k}}]_{m'm}$ with $U(\bm{k}) \in U(2N)$.
The integration path is confined in the $k_z=\bar{k}_z$ plane ($\bar{k}_z=0$ or $\pi$), with the end points being $C_nT$ invariant momenta; see Fig.~\ref{fig:integralpath}.
For the $k_z=0$ plane, we define $(\bm{k}_1, \bm{k}_2) = (\Gamma, M)$ for C$_4$ symmetry and $(\Gamma,K)$ for C$_6$ symmetry.
For the $k_z=\pi$ plane, we define $(\bm{k}_1, \bm{k}_2) = (Z, A)$ for C$_4$ symmetry and $(A, H)$ for C$_6$ symmetry.
Then we can define the $C_n$ symmetry-protected weak indices as
\begin{equation}
\nu_n(\bar{k}_z) = \nu_{n, \bm{k}_1 \bm{k}_2}
\end{equation}
for $\bar{k}_z=0,\pi$, and the $C_n$ symmetry-protected strong index as 
\begin{equation}
\bar{\nu}_{n}\equiv \nu_n(\pi) - \nu_n(0) \mod 2.
\end{equation}
When $\bar{\nu}_{n}=1$, gapless surface states exist on the $C_nT$-invariant $(001)$ surface from the bulk-boundary correspondence. The relation between $\bar{\nu}_{n}$ and $\nu_n(\bar{k}_z)$ is in analogy with the strong and weak $\mathbb{Z}_2$ indices in class AII topological insulators~\cite{Fu2007}. 

\subsection{$C_2T$ symmetry}
\label{sec:C2Tsymmetry}

We here discuss topological invariants related to the $C_2T$ symmetry. We consider a 3D system with $C_2T$ symmetry, whose twofold rotation axis is the $z$ axis.
Then, the $C_2T$ operation acts as $(k_x,k_y,k_z) \to (k_x,k_y,-k_z)$, and the $k_z=0,\pi$ planes are $C_2T$-invariant planes. Recalling that $C_2T$ is an anti-unitary operator satisfying $(C_2T)^2=1$, the $C_2T$ symmetry imposes the real gauge condition on the wave functions in the $C_2T$-invariant planes; namely, letting $|u_m(\bm{k}) \rangle$ be an eigenstate of the 3D bulk Hamiltonian and taking the unitary part of the $C_2T$ operator to be the identity matrix, we find $ |u_m(k_x,k_y,\bar{k}_z) \rangle^{\ast}= |u_m(k_x,k_y,\bar{k}_z) \rangle$. Therefore, the gauge transformation is restricted to the orthogonal group, and the corresponding classifying space of the gapped Hamiltonian is equivalent to the real Grassmannian $G_{N,N_{\rm occ}} = O(N)/[O(N_{\rm occ}) \times O(N-N_{\rm occ})]$, where $N$ and  $N_{\rm occ}$ are the number of total energy bands and that of occupied bands, respectively. The topology of the gapped Hamiltonian on the $C_2T$-invariant planes is characterized by the second homotopy group~\footnote{Exactly speaking, when the first homotopy group is nontrivial, it also affects the band topology through the Whitney sum formula~\cite{Ahn2018band}. However, in the pseudo-spin basis, the first homotopy group becomes trivial.}. In the limit $N \to \infty$, the second homotopy group of the classifying space is~\cite{Hatcher2002,Ahn2019,Bouhon2020}
\begin{align}
 \pi_2 \left[ G_{N,N_{\rm occ}} \right]_{N \to \infty}  \simeq \begin{cases} 0 & N_{\rm occ}=1, \\
                                  \mathbb{Z} & N_{\rm occ}=2, \\
                                  \mathbb{Z}_2 & N_{\rm occ} \ge 3,   
               \end{cases} \label{eq:homotopy_O}
\end{align}
where $\mathbb{Z}$ and $\mathbb{Z}_2$, respectively, indicate the Euler class and the second Stiefel-Whitney class, both of which we label by $\nu_{2}(\bar{k}_z)$. In 3D systems, we have two weak topological invariants $\{\nu_{2}(0),\nu_{2}(\pi)\}$ and a strong topological invariant,
\begin{align}
 \bar{\nu}_2 \equiv \nu_{2} (\pi) - \nu_{2} (0), \label{eq:3dinv}
\end{align}
which can distinguish a strong 3D topological phase from a layer structure of 2D topological phases, i.e., a weak topological phase. 
By definition, $\bar{\nu}_2$ in Eq.~(\ref{eq:3dinv}) takes an integer value when $N_{\rm occ}=2$ but is reduced to a binary index when $N_{\rm occ} \ge 3$. Thus, a $C_2T$-symmetry protected strong insulator with $N_{\rm occ}=2$ and $\bar{\nu}_2\in2\mathbb{Z}$ is fragile against addition of a trivial occupied band. In particular, for energy bands in the pseudo-spin basis, we find that $\nu_{2}(\bar{k}_z)=0 \mod 2$ is always satisfied since the weak topological indices are related to the product of $C_2$ eigenvalues~\cite{Ahn2018band,Ahn2019}, which is trivial ($-1\times-1=1$) in the pseudo-spin basis.

\section{Derivation of surface Hamiltonian by domain-wall projection}
\label{app:surfH}

We derive the surface Hamiltonian for the bulk Hamiltonian in Eq.~(\ref{eq:bulk_hami}) using the method of boundary projection for domain-wall states~\cite{Jackiw-Rebbi,Khalaf18,Khalaf2018high,Geier2018,Trifunovic2019}.
We start with the bulk Hamiltonian,
\begin{align}
\mathcal{H}(\bm{k}) = M \bm{1}_2 \otimes \tau_z + \bm{f}(\bm{k}) \cdot \bm{\sigma} \otimes \tau_x, \label{eq:bulkH}
\end{align}
where $M$ is a mass and $\bm{f}(\bm{k}) = (f_x(\bm{k}), f_y(\bm{k}), f_z(\bm{k})) \in \mathbb{R}^3$. $\sigma_i$ and $\tau_i$ ($i=x,y,z$) are the Pauli matrices in the orbital and sublattice degrees of freedom. Note that Eq.~(\ref{eq:bulkH}) is similar to models of 3D topological insulators, but here $\sigma_i$ are not the spin Pauli matrices.
We write $\bm{f}(\bm{k})$ in the form
\begin{align}
&f_x(\bm{k}) = f_1(\bm{k}_{\parallel}), \\
&f_y(\bm{k}) = t_z \, k_z, \\
&f_z(\bm{k}) = f_2(\bm{k}_{\parallel}), 
\end{align}
and assume that $\mathcal{H}(\bm{k})$ in Eq.~(\ref{eq:bulkH}) is invariant under TR ($\mathcal{T}=\bm{1}_4K$) and C$_n$ [$\mathcal{C}_{n,l}  =\exp\!\left( i \frac{2\pi l}{n} \sigma_y\right) \otimes \bm{1}_2$] symmetries, with the rotation axis fixed to the $z$ axis.

To find surface states that are localized at the $(001)$ surface, we suppose that the mass is a function of $z$ and forms a domain wall at $z=0$, where the mass changes its sign as $M(z \to \pm\infty) \to \pm M_0$.
The surface states are obtained as those localized at the domain wall~\cite{Jackiw-Rebbi}.
We replace $k_z$ with $-i \partial_z$ to write Eq.~(\ref{eq:bulkH}) as
\begin{align}
 \mathcal{H}(\bm{k}_{\parallel},z)=\,&M(z) \bm{1}_2 \otimes \tau_z + [f_1(\bm{k}_{\parallel}) \sigma_x + f_2(\bm{k}_{\parallel}) \sigma_z] \otimes \tau_x \notag \\
                                              &- i t_z \partial_z \,\sigma_y \otimes \tau_x. 
\end{align}
Thus, the problem is reduced into the eigenvalue problem
\begin{equation}
\mathcal{H}(\bm{k}_{\parallel},z) \Psi(\bm{k}_{\parallel},z)
= E_{\bm{k}_{\parallel}}\Psi(\bm{k}_{\parallel},z) .
\label{domain wall}
\end{equation}
When $M_0/t_z>0$, we assume the domain-wall states to have wave functions of the form
\begin{align}
 \Psi(\bm{k}_{\parallel},z) = N \exp \!\left[ -\int_0^z M(z')/t_z \, dz \right]\! \psi(\bm{k}_{\parallel}), \label{eq:wave_temp}
\end{align}
where $N$ is a normalization constant. Substituting Eq.~(\ref{eq:wave_temp}) into the eigenvalue equation (\ref{domain wall}), we obtain the simultaneous equations
which $\psi(\bm{k}_\parallel)$ must satisfy:
\begin{align}
 &(\bm{1}_{4} - \sigma_y \otimes \tau_y) \psi(\bm{k}_{\parallel}) =0, \label{eq:cond1_app}\\
 &[f_1(\bm{k}_{\parallel}) \,\sigma_x + f_2(\bm{k}_{\parallel}) \,\sigma_z]\otimes \tau_x \psi(\bm{k}_{\parallel}) =E_{\bm{k}_{\parallel}} \psi(\bm{k}_{\parallel}). \label{eq:cond2_app} 
\end{align}
Equation~(\ref{eq:cond1_app}) implies that the domain-wall states $\psi(\bm{k}_\parallel)$ obey the condition $P_+\psi(\bm{k}_{\parallel}) =\psi(\bm{k}_{\parallel}) $,
where the projection operator $P_+=(\bm{1}_{4} + \sigma_y \otimes \tau_y) /2$.
We diagonalize $P_+$ by the unitary transformation
$UP_+U^\dagger=(\bm{1}_4+\bm{1}_2\otimes\tau_z)/2$ with
$U=\exp(-i\frac{\pi}{4}\sigma_y\otimes\tau_x)$.
In this basis, the domain-wall states are eigenstates of $\bm{1}_2\otimes\tau_z$ with
the eigenvalue $+1$.
Since the unitary operator $U$ commutes with $\mathcal{T}$ and
$\mathcal{C}_{n,l}$, the TR and $C_n$ operators for the surface states are given by
$T=\bm{1}_2 K$ and $C_{n,l} =\exp( i \frac{2\pi l}{n} \sigma_y)$.
Performing the unitary transformation of the operators on the left-hand side of Eq.\ (\ref{eq:cond2_app}),
\begin{subequations}
\begin{align}
&
U\sigma_x\otimes\tau_x U^\dagger=-\sigma_z\otimes\bm{1}_2,
\\
&
U\sigma_z\otimes\tau_x U^\dagger=\sigma_x\otimes\bm{1}_2,
\end{align}
\end{subequations}\noindent
and projecting Eq.\ (\ref{eq:cond2_app}) onto the subspace $\bm{1}_2\otimes\tau_z=1$,
we obtain the surface Hamiltonian 
\begin{align}
 H(\bm{k}) &= f_2(\bm{k}_{\parallel}) \sigma_x  -f_1(\bm{k}_{\parallel}) \sigma_z \nonumber\\
 &= -[f_1(\bm{k}_\parallel)+if_2(\bm{k}_\parallel)]\sigma_+ - [f_1(\bm{k}_\parallel)-if_2(\bm{k}_\parallel)]\sigma_-, \label{eq:surf_Hami_app}
\end{align}
which can be compared with the surface Hamiltonian in Eq.\ (\ref{eq:sym_analysis}). 
%

\section{Tight-binding models}
\label{app:tbmodel}

\begin{figure*}[tbp]
\centering
 \includegraphics[width=16cm]{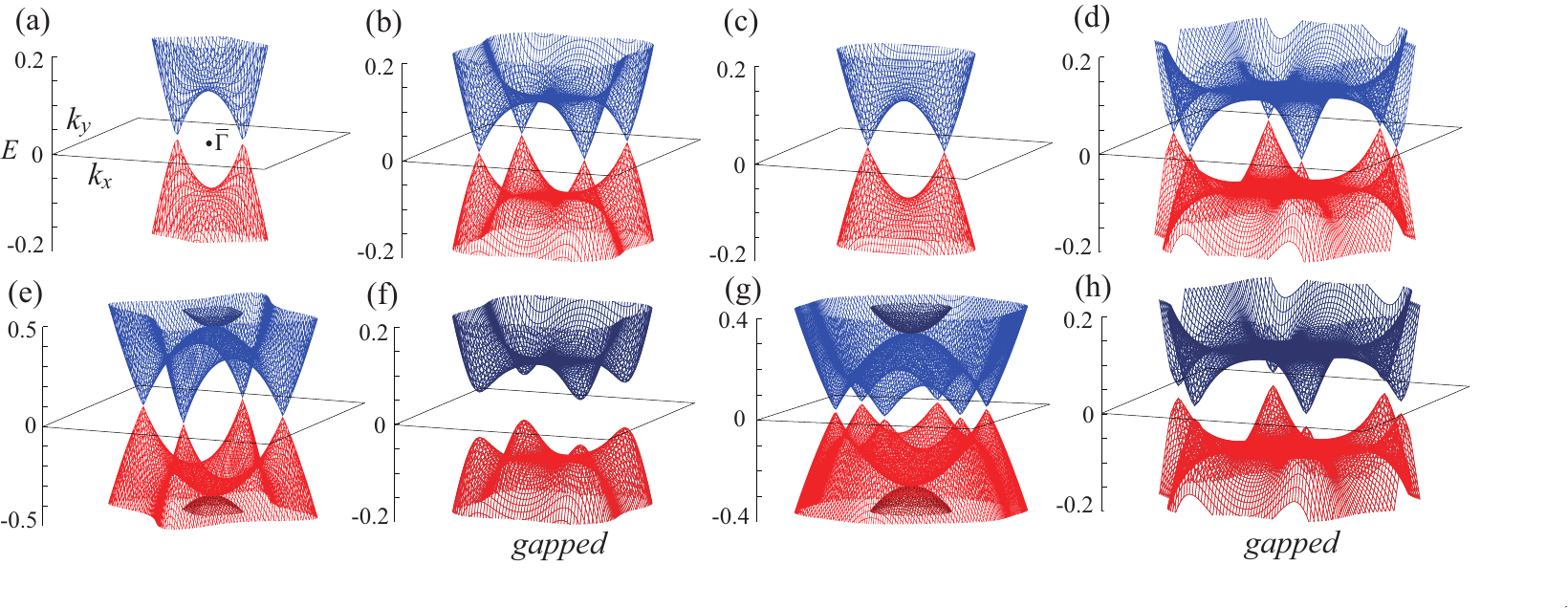}
 \caption{Surface energy spectra of the tight-binding Hamiltonians, where we choose the same parameters $(M_{3},m_{3},t,t_z)=(5,-2,1,1)$. The surface energy spectra of Hamiltonians in Eqs.~(\ref{eq:mc41}), (\ref{eq:mc42}), (\ref{eq:mc61}), (\ref{eq:mc63}) with $(\mathcal{M}_1,\mathcal{M}_2)=(0.01,0)$ are shown in (a), (b), (c), and (d). Note that the Hamiltonians in Eq.~(\ref{eq:mc62}) and Eq.~(\ref{eq:mc61}) have the same the surface energy spectra. The surface energy spectra of the double Hamiltonians in Eqs.~(\ref{eq:pc41}), (\ref{eq:pc42}), (\ref{eq:pc61}), and (\ref{eq:pc63}) are shown in (e), (f), (g), and (h), where the perturbations are chosen as $(\mathcal{M}_1,\mathcal{M}_2,\mathcal{M}_1',\mathcal{M}_2',\mathcal{M}_3',\mathcal{M}_4')=(0,0,0,0.03,0,03,0)$, $(0.01,0,0,0.1,0.1,0)$, $(0,0,0,0.03,0.03,0)$, and $(0.01,0,0,0.1,0,0)$, respectively. }\label{fig:multi_Dirac}
\end{figure*}

In this appendix, we present tight-binding models on a tetragonal lattice for $C_4$ and a hexagonal lattice for $C_6$ symmetry that are reduced to the toy model in Eq.~(\ref{eq:bulk_hami}) in the continuum limit.
We include perturbations that change the surface band touching at $\bm{k}_\parallel=0$ into multiple Dirac cones.

For the tetragonal lattice, the tight-binding models corresponding to Eq.~(\ref{eq:bulk_hami}) with $l=1$ and $2$ are given by
\begin{align}
 \mathcal{H}_{l=1}^{\rm tetra} (\bm{k})=&
\Big(M_{3}+m_{3} \sum_{i=x,y,z} \cos (k_i)\Big)\Gamma_{03} +t_z \sin (k_z) \Gamma_{21}
 \notag \\
 & +2t [\cos (k_x)-\cos (k_y)] \Gamma_{31} \notag \\
 & + 2t \sin (k_x) \sin (k_y) \Gamma_{11}
 +\mathcal{M}_1 \Gamma_{31} + \mathcal{M}_2 \Gamma_{11} \label{eq:mc41}
\end{align} 
and
\begin{align}
 \mathcal{H}_{l=2}^{\rm tetra} (\bm{k})=&
 \Big(M_{3}+m_{3} \sum_{i=x,y,z} \cos (k_i)\Big)\Gamma_{03} +t_z \sin (k_z) \Gamma_{21} \notag \\
                   &+2t \{8[\cos (k_x)+\cos (k_y)]-12 \cos(k_x) \cos(k_y) \notag \\
                   &\qquad+\cos(2k_x) + \cos(2 k_y) -6\} \Gamma_{31}  \notag \\
                   & +8t \sin (k_x) \sin (k_y)[\cos (k_x)-\cos (k_y)]\Gamma_{11} \notag \\
                   & +\mathcal{M}_1 \Gamma_{31} + \mathcal{M}_2 \Gamma_{11}, \label{eq:mc42}
\end{align}    
where we have used the notation $\Gamma_{ij} \equiv \sigma_i \otimes \tau_j$ ($i,j=0,1,2,3$) and $\sigma_0=\tau_0=\bm{1}_2$.
The perturbations $\mathcal{M}_1\Gamma_{31} + \mathcal{M}_2 \Gamma_{11}$ break the $\mathcal{C}_{4,1}$ symmetry down to $\mathcal{C}_{2,1}$ but keep the $\mathcal{C}_{4,2}$ symmetry. 
As a result these perturbations lead to two (four) surface Dirac cones for $\mathcal{H}_{l=1}^{\rm tetra}$ ($\mathcal{H}_{l=2}^{\rm tetra}$), as we discussed in Sec.~\ref{sec:surftheory_pseudo}.

The tight-binding models for the hexagonal lattice are given by
\begin{align}
\mathcal{H}_{l=1}^{\rm hexa} (\bm{k})=&
 \Bigg\{M_{3}+m_{3} \Bigg[\cos(k_x)+2 \cos\left(\frac{k_x}{2}\right) \cos\left(\frac{\sqrt{3} k_y}{2}\right) \notag \\
   &\quad + \cos(k_z)\Bigg]\Bigg\}\Gamma_{03} +t_z \sin (k_z) \Gamma_{21}  \notag \\
   &+t\left[\cos(k_x) - \cos\left(\frac{k_x}{2}\right) \cos \left(\frac{\sqrt{3} k_y}{2}\right) \right] \Gamma_{31} \notag \\
   &+ t\sqrt{3} \sin \left(\frac{k_x}{2}\right) \sin \left(\frac{\sqrt{3} k_y}{2}\right)  \Gamma_{11} \notag \\
                   & +\mathcal{M}_1 \Gamma_{31} + \mathcal{M}_2 \Gamma_{11}, \label{eq:mc61}
\end{align}
\begin{align}
\mathcal{H}_{l=2}^{\rm hexa} (\bm{k})=&
 \Bigg\{M_3+m_3 \Bigg[\cos(k_x)+2 \cos\left(\frac{k_x}{2}\right) \cos\left(\frac{\sqrt{3} k_y}{2}\right) \notag \\
  &\quad + \cos(k_z)\Bigg]\Bigg\}\Gamma_{03} +t_z \sin (k_z) \Gamma_{21}  \notag \\
  &+t\left[\cos(k_x) - \cos\left(\frac{k_x}{2}\right) \cos \left(\frac{\sqrt{3} k_y}{2}\right) \right] \Gamma_{31} \notag \\
                   & - t\sqrt{3} \sin \left(\frac{k_x}{2}\right) \sin \left(\frac{\sqrt{3} k_y}{2}\right)  \Gamma_{11} \notag \\
                   & +\mathcal{M}_1 \Gamma_{31} + \mathcal{M}_2 \Gamma_{11}, \label{eq:mc62}
\end{align}
\begin{align}
\mathcal{H}_{l=3}^{\rm hexa} (\bm{k})=&
 \Bigg\{M_3+m_3 \Bigg[\cos(k_x)+2 \cos\left(\frac{k_x}{2}\right) \cos\left(\frac{\sqrt{3} k_y}{2}\right) \notag \\
  &\quad + \cos(k_z)\Bigg]\Bigg\}\Gamma_{03} +t_z \sin (k_z) \Gamma_{21}  \notag \\
  &+\frac{15}{16}t\big\{\big[\cos(4k_x) +2 \cos(2k_x) \cos(2\sqrt{3} k_y)\big] \notag \\
  &\qquad -\big[\cos(4k_y) +2 \cos(2k_y) \cos(2\sqrt{3} k_x)\big]\big\} \Gamma_{31} \notag \\
  &+ 2t \big[\sin (2k_x)-2\sin(k_x)\cos(\sqrt{3}k_y)\big] \notag \\
  &\qquad\times \big[\sin (2k_y)-2\sin(k_y)\cos(\sqrt{3}k_x)\big] \Gamma_{11} \notag \\
  & +\mathcal{M}_1 \Gamma_{31} + \mathcal{M}_2 \Gamma_{11}. \label{eq:mc63}
\end{align}
The perturbations $\mathcal{M}_1 \Gamma_{31} + \mathcal{M}_2 \Gamma_{11}$ are not invariant under $\mathcal{C}_{6,1}$ and $\mathcal{C}_{6,2}$ but invariant under $\mathcal{C}_{6,3}$, thereby changing the nonlinear band touching at $\bm{k}_\parallel=0$ in the surface spectra into two Dirac cones for $\mathcal{H}_{l=1}^{\rm hexa}$ and $\mathcal{H}_{l=2}^{\rm hexa}$, and into six Dirac cones for $\mathcal{H}_{l=3}^{\rm hexa}$.
We have computed the surface spectra of $\mathcal{H}_{l=1,2,3}^{\rm hexa}$ 
and confirmed the existence of multiple surface Dirac cones around the $\bar{\Gamma}$ point; see Figs.~\ref{fig:multi_Dirac} (a), (b), (c), and (d).

Likewise, we obtain the double Hamiltonian by stacking $\mathcal{H}_{l=1,2}^{\rm tetra}$ or $\mathcal{H}_{l=1,2,3}^{\rm hexa}$.
Under the $\mathcal{C}_n^+$ symmetry [Eq.~(\ref{eq:tbdoublerot_sym})], we have a relevant symmetry-allowed perturbation,
\begin{equation}
 \mathcal{M}_1' \Gamma_{022},
\end{equation}
which opens a gap in the surface spectra.
Under the $\mathcal{C}_n^-$ symmetry [Eq.~(\ref{eq:tbdoublerot_sym})], we have the following symmetry-allowed perturbations:
\begin{align}
 \mathcal{M}_2' \Gamma_{003} + \mathcal{M}_3' \Gamma_{331} + \mathcal{M}_4' \Gamma_{131} \label{eq:pc41}
\end{align} 
for $\mathcal{H}_{l=1}^{\rm tetra} \otimes \bm{1}_2$,
\begin{align}
 \left\{\mathcal{M}_2' [\cos(k_x)-\cos(k_y)] + \mathcal{M}_3' \sin (k_x) \sin(k_y)\right\} \Gamma_{022} \label{eq:pc42}
\end{align} 
for $\mathcal{H}_{l=2}^{\rm tetra} \otimes \bm{1}_2$,
\begin{align}
  &\mathcal{M}_2' \Gamma_{003}
  + \mathcal{M}_3' \left\{\!\left[ \sin (k_x) + \sin \left( \frac{k_x}{2} \right) \cos \left( \frac{\sqrt{3}k_y}{2} \right) \right] \Gamma_{331} \right. \notag \\
  &\left. \qquad\qquad\qquad\quad \mp \sqrt{3} \cos \left( \frac{k_x}{2} \right) \sin \left( \frac{\sqrt{3}k_y}{2} \right) \Gamma_{131} \right\}
\label{eq:pc61}
\end{align}
for $\mathcal{H}_{l=1}^{\rm hexa} \otimes \bm{1}_2$ (minus sign) and $\mathcal{H}_{l=2}^{\rm hexa} \otimes \bm{1}_2$ (plus sign), and
\begin{align}
 &\Bigg\{\mathcal{M}_2' \left[\sin(k_x)-2\sin\left(\frac{k_x}{2}\right)\cos\left(\frac{\sqrt{3}k_y}{2}\right)\right] \notag \\
 &+ \mathcal{M}_3' \left[\sin(k_y)-2\sin\left(\frac{k_y}{2}\right)\cos\left(\frac{\sqrt{3}k_x}{2}\right)\right]\Bigg\} \Gamma_{022} \label{eq:pc63}
\end{align} 
for $\mathcal{H}_{l=3}^{\rm hexa} \otimes \bm{1}_2$.
Here $\Gamma_{ijk} \equiv \sigma_i \otimes \tau_j \otimes \mu_k $ ($i,j,k=0,1,2,3$), and $\mu_i$'s are Pauli matrices in the grading of the double Hamiltonian.
The perturbations $\mathcal{M}_3'$ and $\mathcal{M}_4'$ in Eqs.~(\ref{eq:pc41}) and (\ref{eq:pc61}) generate four (six) Dirac cones for $\mathcal{H}_{l=1}^{\rm tetra} \otimes \bm{1}_2$ ($\mathcal{H}_{l=1,2}^{\rm hexa} \otimes \bm{1}_2$), whereas the perturbations $\mathcal{M}_2'$ and $\mathcal{M}_3'$ in Eqs.~(\ref{eq:pc42}) and (\ref{eq:pc63}) open a gap for $\mathcal{H}_{l=2}^{\rm tetra} \otimes \bm{1}_2$ and $\mathcal{H}_{l=3}^{\rm hexa} \otimes \bm{1}_2$.
We note that the $\mathcal{M}_3'$ perturbation in Eq.\ (\ref{eq:pc61}) vanishes at $\bm{k}_\parallel=0$.
The $\mathcal{M}_2'$ perturbation plays an important of generating a line node in the surface spectra by giving positive and negative energy shifts to two $\mathcal{H}_l$.
The line node is changed by the $\mathcal{M}_3'$ term into six Dirac cones.
We have simulated the surface spectra for the tight-binding Hamiltonians $\mathcal{H}_l^\mathrm{tetra}$ and $\mathcal{H}_l^\mathrm{hexa}$ to confirm the existence of multiple Dirac cones in Figs.~\ref{fig:multi_Dirac} (e) and (g) and a gap-opening in Figs.~\ref{fig:multi_Dirac} (f) and (h).

\bibliography{fragile_phase}

\begin{thebibliography}{88}%
\makeatletter
\providecommand \@ifxundefined [1]{%
 \@ifx{#1\undefined}
}%
\providecommand \@ifnum [1]{%
 \ifnum #1\expandafter \@firstoftwo
 \else \expandafter \@secondoftwo
 \fi
}%
\providecommand \@ifx [1]{%
 \ifx #1\expandafter \@firstoftwo
 \else \expandafter \@secondoftwo
 \fi
}%
\providecommand \natexlab [1]{#1}%
\providecommand \enquote  [1]{``#1''}%
\providecommand \bibnamefont  [1]{#1}%
\providecommand \bibfnamefont [1]{#1}%
\providecommand \citenamefont [1]{#1}%
\providecommand \href@noop [0]{\@secondoftwo}%
\providecommand \href [0]{\begingroup \@sanitize@url \@href}%
\providecommand \@href[1]{\@@startlink{#1}\@@href}%
\providecommand \@@href[1]{\endgroup#1\@@endlink}%
\providecommand \@sanitize@url [0]{\catcode `\\12\catcode `\$12\catcode
  `\&12\catcode `\#12\catcode `\^12\catcode `\_12\catcode `\%12\relax}%
\providecommand \@@startlink[1]{}%
\providecommand \@@endlink[0]{}%
\providecommand \url  [0]{\begingroup\@sanitize@url \@url }%
\providecommand \@url [1]{\endgroup\@href {#1}{\urlprefix }}%
\providecommand \urlprefix  [0]{URL }%
\providecommand \Eprint [0]{\href }%
\providecommand \doibase [0]{https://doi.org/}%
\providecommand \selectlanguage [0]{\@gobble}%
\providecommand \bibinfo  [0]{\@secondoftwo}%
\providecommand \bibfield  [0]{\@secondoftwo}%
\providecommand \translation [1]{[#1]}%
\providecommand \BibitemOpen [0]{}%
\providecommand \bibitemStop [0]{}%
\providecommand \bibitemNoStop [0]{.\EOS\space}%
\providecommand \EOS [0]{\spacefactor3000\relax}%
\providecommand \BibitemShut  [1]{\csname bibitem#1\endcsname}%
\let\auto@bib@innerbib\@empty
\bibitem [{\citenamefont {Schnyder}\ \emph {et~al.}(2008)\citenamefont
  {Schnyder}, \citenamefont {Ryu}, \citenamefont {Furusaki},\ and\
  \citenamefont {Ludwig}}]{Schnyder08}%
  \BibitemOpen
  \bibfield  {author} {\bibinfo {author} {\bibfnamefont {A.~P.}\ \bibnamefont
  {Schnyder}}, \bibinfo {author} {\bibfnamefont {S.}~\bibnamefont {Ryu}},
  \bibinfo {author} {\bibfnamefont {A.}~\bibnamefont {Furusaki}},\ and\
  \bibinfo {author} {\bibfnamefont {A.~W.~W.}\ \bibnamefont {Ludwig}},\
  }\bibfield  {title} {\bibinfo {title} {Classification of topological
  insulators and superconductors in three spatial dimensions},\ }\href
  {https://doi.org/10.1103/PhysRevB.78.195125} {\bibfield  {journal} {\bibinfo
  {journal} {Phys. Rev. B}\ }\textbf {\bibinfo {volume} {78}},\ \bibinfo
  {pages} {195125} (\bibinfo {year} {2008})}\BibitemShut {NoStop}%
\bibitem [{\citenamefont {Kitaev}(2009)}]{Kitaev09}%
  \BibitemOpen
  \bibfield  {author} {\bibinfo {author} {\bibfnamefont {A.}~\bibnamefont
  {Kitaev}},\ }\bibfield  {title} {\bibinfo {title} {Periodic table for
  topological insulators and superconductors},\ }\href
  {https://doi.org/10.1063/1.3149495} {\bibfield  {journal} {\bibinfo
  {journal} {AIP Conference Proceedings}\ }\textbf {\bibinfo {volume} {1134}},\
  \bibinfo {pages} {22} (\bibinfo {year} {2009})}\BibitemShut {NoStop}%
\bibitem [{\citenamefont {Schnyder}\ \emph {et~al.}(2009)\citenamefont
  {Schnyder}, \citenamefont {Ryu}, \citenamefont {Furusaki},\ and\
  \citenamefont {Ludwig}}]{Schnyder09}%
  \BibitemOpen
  \bibfield  {author} {\bibinfo {author} {\bibfnamefont {A.~P.}\ \bibnamefont
  {Schnyder}}, \bibinfo {author} {\bibfnamefont {S.}~\bibnamefont {Ryu}},
  \bibinfo {author} {\bibfnamefont {A.}~\bibnamefont {Furusaki}},\ and\
  \bibinfo {author} {\bibfnamefont {A.~W.~W.}\ \bibnamefont {Ludwig}},\
  }\bibfield  {title} {\bibinfo {title} {Classification of topological
  insulators and superconductors},\ }\href {https://doi.org/10.1063/1.3149481}
  {\bibfield  {journal} {\bibinfo  {journal} {AIP Conference Proceedings}\
  }\textbf {\bibinfo {volume} {1134}},\ \bibinfo {pages} {10} (\bibinfo {year}
  {2009})}\BibitemShut {NoStop}%
\bibitem [{\citenamefont {Ryu}\ \emph {et~al.}(2010)\citenamefont {Ryu},
  \citenamefont {Schnyder}, \citenamefont {Furusaki},\ and\ \citenamefont
  {Ludwig}}]{Ryu10}%
  \BibitemOpen
  \bibfield  {author} {\bibinfo {author} {\bibfnamefont {S.}~\bibnamefont
  {Ryu}}, \bibinfo {author} {\bibfnamefont {A.~P.}\ \bibnamefont {Schnyder}},
  \bibinfo {author} {\bibfnamefont {A.}~\bibnamefont {Furusaki}},\ and\
  \bibinfo {author} {\bibfnamefont {A.~W.~W.}\ \bibnamefont {Ludwig}},\
  }\bibfield  {title} {\bibinfo {title} {Topological insulators and
  superconductors: tenfold way and dimensional hierarchy},\ }\href
  {http://stacks.iop.org/1367-2630/12/i=6/a=065010} {\bibfield  {journal}
  {\bibinfo  {journal} {New Journal of Physics}\ }\textbf {\bibinfo {volume}
  {12}},\ \bibinfo {pages} {065010} (\bibinfo {year} {2010})}\BibitemShut
  {NoStop}%
\bibitem [{\citenamefont {Morimoto}\ and\ \citenamefont
  {Furusaki}(2013)}]{Morimoto13}%
  \BibitemOpen
  \bibfield  {author} {\bibinfo {author} {\bibfnamefont {T.}~\bibnamefont
  {Morimoto}}\ and\ \bibinfo {author} {\bibfnamefont {A.}~\bibnamefont
  {Furusaki}},\ }\bibfield  {title} {\bibinfo {title} {Topological
  classification with additional symmetries from {Clifford} algebras},\ }\href
  {https://doi.org/10.1103/PhysRevB.88.125129} {\bibfield  {journal} {\bibinfo
  {journal} {Phys. Rev. B}\ }\textbf {\bibinfo {volume} {88}},\ \bibinfo
  {pages} {125129} (\bibinfo {year} {2013})}\BibitemShut {NoStop}%
\bibitem [{\citenamefont {Chiu}\ \emph {et~al.}(2013)\citenamefont {Chiu},
  \citenamefont {Yao},\ and\ \citenamefont {Ryu}}]{Chiu13}%
  \BibitemOpen
  \bibfield  {author} {\bibinfo {author} {\bibfnamefont {C.-K.}\ \bibnamefont
  {Chiu}}, \bibinfo {author} {\bibfnamefont {H.}~\bibnamefont {Yao}},\ and\
  \bibinfo {author} {\bibfnamefont {S.}~\bibnamefont {Ryu}},\ }\bibfield
  {title} {\bibinfo {title} {Classification of topological insulators and
  superconductors in the presence of reflection symmetry},\ }\href
  {https://doi.org/10.1103/PhysRevB.88.075142} {\bibfield  {journal} {\bibinfo
  {journal} {Phys. Rev. B}\ }\textbf {\bibinfo {volume} {88}},\ \bibinfo
  {pages} {075142} (\bibinfo {year} {2013})}\BibitemShut {NoStop}%
\bibitem [{\citenamefont {Freed}\ and\ \citenamefont
  {Moore}(2013)}]{Freed2013}%
  \BibitemOpen
  \bibfield  {author} {\bibinfo {author} {\bibfnamefont {D.~S.}\ \bibnamefont
  {Freed}}\ and\ \bibinfo {author} {\bibfnamefont {G.~W.}\ \bibnamefont
  {Moore}},\ }\bibfield  {title} {\bibinfo {title} {Twisted equivariant
  matter},\ }\href {https://doi.org/10.1007/s00023-013-0236-x} {\bibfield
  {journal} {\bibinfo  {journal} {Annales Henri Poincaré}\ }\textbf {\bibinfo
  {volume} {14}},\ \bibinfo {pages} {1927} (\bibinfo {year}
  {2013})}\BibitemShut {NoStop}%
\bibitem [{\citenamefont {Shiozaki}\ and\ \citenamefont
  {Sato}(2014)}]{Shiozaki14}%
  \BibitemOpen
  \bibfield  {author} {\bibinfo {author} {\bibfnamefont {K.}~\bibnamefont
  {Shiozaki}}\ and\ \bibinfo {author} {\bibfnamefont {M.}~\bibnamefont
  {Sato}},\ }\bibfield  {title} {\bibinfo {title} {Topology of crystalline
  insulators and superconductors},\ }\href
  {https://doi.org/10.1103/PhysRevB.90.165114} {\bibfield  {journal} {\bibinfo
  {journal} {Phys. Rev. B}\ }\textbf {\bibinfo {volume} {90}},\ \bibinfo
  {pages} {165114} (\bibinfo {year} {2014})}\BibitemShut {NoStop}%
\bibitem [{\citenamefont {Shiozaki}\ \emph {et~al.}(2015)\citenamefont
  {Shiozaki}, \citenamefont {Sato},\ and\ \citenamefont {Gomi}}]{Shiozaki15}%
  \BibitemOpen
  \bibfield  {author} {\bibinfo {author} {\bibfnamefont {K.}~\bibnamefont
  {Shiozaki}}, \bibinfo {author} {\bibfnamefont {M.}~\bibnamefont {Sato}},\
  and\ \bibinfo {author} {\bibfnamefont {K.}~\bibnamefont {Gomi}},\ }\bibfield
  {title} {\bibinfo {title} {{${Z}_{2}$} topology in nonsymmorphic crystalline
  insulators: {M\"obius} twist in surface states},\ }\href
  {https://doi.org/10.1103/PhysRevB.91.155120} {\bibfield  {journal} {\bibinfo
  {journal} {Phys. Rev. B}\ }\textbf {\bibinfo {volume} {91}},\ \bibinfo
  {pages} {155120} (\bibinfo {year} {2015})}\BibitemShut {NoStop}%
\bibitem [{\citenamefont {Shiozaki}\ \emph {et~al.}(2016)\citenamefont
  {Shiozaki}, \citenamefont {Sato},\ and\ \citenamefont {Gomi}}]{Shiozaki16}%
  \BibitemOpen
  \bibfield  {author} {\bibinfo {author} {\bibfnamefont {K.}~\bibnamefont
  {Shiozaki}}, \bibinfo {author} {\bibfnamefont {M.}~\bibnamefont {Sato}},\
  and\ \bibinfo {author} {\bibfnamefont {K.}~\bibnamefont {Gomi}},\ }\bibfield
  {title} {\bibinfo {title} {Topology of nonsymmorphic crystalline insulators
  and superconductors},\ }\href {https://doi.org/10.1103/PhysRevB.93.195413}
  {\bibfield  {journal} {\bibinfo  {journal} {Phys. Rev. B}\ }\textbf {\bibinfo
  {volume} {93}},\ \bibinfo {pages} {195413} (\bibinfo {year}
  {2016})}\BibitemShut {NoStop}%
\bibitem [{\citenamefont {Chiu}\ \emph {et~al.}(2016)\citenamefont {Chiu},
  \citenamefont {Teo}, \citenamefont {Schnyder},\ and\ \citenamefont
  {Ryu}}]{Chiu16}%
  \BibitemOpen
  \bibfield  {author} {\bibinfo {author} {\bibfnamefont {C.-K.}\ \bibnamefont
  {Chiu}}, \bibinfo {author} {\bibfnamefont {J.~C.~Y.}\ \bibnamefont {Teo}},
  \bibinfo {author} {\bibfnamefont {A.~P.}\ \bibnamefont {Schnyder}},\ and\
  \bibinfo {author} {\bibfnamefont {S.}~\bibnamefont {Ryu}},\ }\bibfield
  {title} {\bibinfo {title} {Classification of topological quantum matter with
  symmetries},\ }\href {https://doi.org/10.1103/RevModPhys.88.035005}
  {\bibfield  {journal} {\bibinfo  {journal} {Rev. Mod. Phys.}\ }\textbf
  {\bibinfo {volume} {88}},\ \bibinfo {pages} {035005} (\bibinfo {year}
  {2016})}\BibitemShut {NoStop}%
\bibitem [{\citenamefont {Shiozaki}\ \emph {et~al.}(2017)\citenamefont
  {Shiozaki}, \citenamefont {Sato},\ and\ \citenamefont {Gomi}}]{Shiozaki17}%
  \BibitemOpen
  \bibfield  {author} {\bibinfo {author} {\bibfnamefont {K.}~\bibnamefont
  {Shiozaki}}, \bibinfo {author} {\bibfnamefont {M.}~\bibnamefont {Sato}},\
  and\ \bibinfo {author} {\bibfnamefont {K.}~\bibnamefont {Gomi}},\ }\bibfield
  {title} {\bibinfo {title} {Topological crystalline materials: General
  formulation, module structure, and wallpaper groups},\ }\href
  {https://doi.org/10.1103/PhysRevB.95.235425} {\bibfield  {journal} {\bibinfo
  {journal} {Phys. Rev. B}\ }\textbf {\bibinfo {volume} {95}},\ \bibinfo
  {pages} {235425} (\bibinfo {year} {2017})}\BibitemShut {NoStop}%
\bibitem [{\citenamefont {Fang}\ \emph {et~al.}(2017)\citenamefont {Fang},
  \citenamefont {Bernevig},\ and\ \citenamefont {Gilbert}}]{CFang17}%
  \BibitemOpen
  \bibfield  {author} {\bibinfo {author} {\bibfnamefont {C.}~\bibnamefont
  {Fang}}, \bibinfo {author} {\bibfnamefont {B.~A.}\ \bibnamefont {Bernevig}},\
  and\ \bibinfo {author} {\bibfnamefont {M.~J.}\ \bibnamefont {Gilbert}},\
  }\bibfield  {title} {\bibinfo {title} {Topological crystalline
  superconductors with linearly and projectively represented {$C_n$}
  symmetry},\ }\href@noop {} {\bibfield  {journal} {\bibinfo  {journal} {arXiv
  preprint arXiv:1701.01944}\ } (\bibinfo {year} {2017})}\BibitemShut {NoStop}%
\bibitem [{\citenamefont {Shiozaki}\ \emph {et~al.}(2018)\citenamefont
  {Shiozaki}, \citenamefont {Sato},\ and\ \citenamefont
  {Gomi}}]{Shiozaki18atiyah}%
  \BibitemOpen
  \bibfield  {author} {\bibinfo {author} {\bibfnamefont {K.}~\bibnamefont
  {Shiozaki}}, \bibinfo {author} {\bibfnamefont {M.}~\bibnamefont {Sato}},\
  and\ \bibinfo {author} {\bibfnamefont {K.}~\bibnamefont {Gomi}},\ }\bibfield
  {title} {\bibinfo {title} {{Atiyah-Hirzebruch} spectral sequence in band
  topology: General formalism and topological invariants for 230 space
  groups},\ }\href@noop {} {\bibfield  {journal} {\bibinfo  {journal} {arXiv
  preprint arXiv:1802.06694}\ } (\bibinfo {year} {2018})}\BibitemShut {NoStop}%
\bibitem [{\citenamefont {Cornfeld}\ and\ \citenamefont
  {Chapman}(2019)}]{Cornfeld19}%
  \BibitemOpen
  \bibfield  {author} {\bibinfo {author} {\bibfnamefont {E.}~\bibnamefont
  {Cornfeld}}\ and\ \bibinfo {author} {\bibfnamefont {A.}~\bibnamefont
  {Chapman}},\ }\bibfield  {title} {\bibinfo {title} {Classification of
  crystalline topological insulators and superconductors with point group
  symmetries},\ }\href {https://doi.org/10.1103/PhysRevB.99.075105} {\bibfield
  {journal} {\bibinfo  {journal} {Phys. Rev. B}\ }\textbf {\bibinfo {volume}
  {99}},\ \bibinfo {pages} {075105} (\bibinfo {year} {2019})}\BibitemShut
  {NoStop}%
\bibitem [{\citenamefont {Shiozaki}(2019)}]{Shiozaki19classification}%
  \BibitemOpen
  \bibfield  {author} {\bibinfo {author} {\bibfnamefont {K.}~\bibnamefont
  {Shiozaki}},\ }\bibfield  {title} {\bibinfo {title} {The classification of
  surface states of topological insulators and superconductors with magnetic
  point group symmetry},\ }\href@noop {} {\bibfield  {journal} {\bibinfo
  {journal} {arXiv preprint arXiv:1907.09354}\ } (\bibinfo {year}
  {2019})}\BibitemShut {NoStop}%
\bibitem [{\citenamefont {Okuma}\ \emph {et~al.}(2019)\citenamefont {Okuma},
  \citenamefont {Sato},\ and\ \citenamefont {Shiozaki}}]{Okuma19}%
  \BibitemOpen
  \bibfield  {author} {\bibinfo {author} {\bibfnamefont {N.}~\bibnamefont
  {Okuma}}, \bibinfo {author} {\bibfnamefont {M.}~\bibnamefont {Sato}},\ and\
  \bibinfo {author} {\bibfnamefont {K.}~\bibnamefont {Shiozaki}},\ }\bibfield
  {title} {\bibinfo {title} {Topological classification under nonmagnetic and
  magnetic point group symmetry: Application of real-space {Atiyah-Hirzebruch}
  spectral sequence to higher-order topology},\ }\href
  {https://doi.org/10.1103/PhysRevB.99.085127} {\bibfield  {journal} {\bibinfo
  {journal} {Phys. Rev. B}\ }\textbf {\bibinfo {volume} {99}},\ \bibinfo
  {pages} {085127} (\bibinfo {year} {2019})}\BibitemShut {NoStop}%
\bibitem [{\citenamefont {Song}\ \emph
  {et~al.}(2019{\natexlab{a}})\citenamefont {Song}, \citenamefont {Huang},
  \citenamefont {Qi}, \citenamefont {Fang},\ and\ \citenamefont
  {Hermele}}]{Song19topological}%
  \BibitemOpen
  \bibfield  {author} {\bibinfo {author} {\bibfnamefont {Z.}~\bibnamefont
  {Song}}, \bibinfo {author} {\bibfnamefont {S.-J.}\ \bibnamefont {Huang}},
  \bibinfo {author} {\bibfnamefont {Y.}~\bibnamefont {Qi}}, \bibinfo {author}
  {\bibfnamefont {C.}~\bibnamefont {Fang}},\ and\ \bibinfo {author}
  {\bibfnamefont {M.}~\bibnamefont {Hermele}},\ }\bibfield  {title} {\bibinfo
  {title} {Topological states from topological crystals},\ }\href@noop {}
  {\bibfield  {journal} {\bibinfo  {journal} {Science advances}\ }\textbf
  {\bibinfo {volume} {5}},\ \bibinfo {pages} {eaax2007} (\bibinfo {year}
  {2019}{\natexlab{a}})}\BibitemShut {NoStop}%
\bibitem [{\citenamefont {Bradlyn}\ \emph {et~al.}(2017)\citenamefont
  {Bradlyn}, \citenamefont {Elcoro}, \citenamefont {Cano}, \citenamefont
  {Vergniory}, \citenamefont {Wang}, \citenamefont {Felser}, \citenamefont
  {Aroyo},\ and\ \citenamefont {Bernevig}}]{Bradlyn17topological}%
  \BibitemOpen
  \bibfield  {author} {\bibinfo {author} {\bibfnamefont {B.}~\bibnamefont
  {Bradlyn}}, \bibinfo {author} {\bibfnamefont {L.}~\bibnamefont {Elcoro}},
  \bibinfo {author} {\bibfnamefont {J.}~\bibnamefont {Cano}}, \bibinfo {author}
  {\bibfnamefont {M.}~\bibnamefont {Vergniory}}, \bibinfo {author}
  {\bibfnamefont {Z.}~\bibnamefont {Wang}}, \bibinfo {author} {\bibfnamefont
  {C.}~\bibnamefont {Felser}}, \bibinfo {author} {\bibfnamefont
  {M.}~\bibnamefont {Aroyo}},\ and\ \bibinfo {author} {\bibfnamefont {B.~A.}\
  \bibnamefont {Bernevig}},\ }\bibfield  {title} {\bibinfo {title} {Topological
  quantum chemistry},\ }\href@noop {} {\bibfield  {journal} {\bibinfo
  {journal} {Nature}\ }\textbf {\bibinfo {volume} {547}},\ \bibinfo {pages}
  {298} (\bibinfo {year} {2017})}\BibitemShut {NoStop}%
\bibitem [{\citenamefont {Kruthoff}\ \emph {et~al.}(2017)\citenamefont
  {Kruthoff}, \citenamefont {de~Boer}, \citenamefont {van Wezel}, \citenamefont
  {Kane},\ and\ \citenamefont {Slager}}]{Kruthoff17}%
  \BibitemOpen
  \bibfield  {author} {\bibinfo {author} {\bibfnamefont {J.}~\bibnamefont
  {Kruthoff}}, \bibinfo {author} {\bibfnamefont {J.}~\bibnamefont {de~Boer}},
  \bibinfo {author} {\bibfnamefont {J.}~\bibnamefont {van Wezel}}, \bibinfo
  {author} {\bibfnamefont {C.~L.}\ \bibnamefont {Kane}},\ and\ \bibinfo
  {author} {\bibfnamefont {R.-J.}\ \bibnamefont {Slager}},\ }\bibfield  {title}
  {\bibinfo {title} {Topological classification of crystalline insulators
  through band structure combinatorics},\ }\href
  {https://doi.org/10.1103/PhysRevX.7.041069} {\bibfield  {journal} {\bibinfo
  {journal} {Phys. Rev. X}\ }\textbf {\bibinfo {volume} {7}},\ \bibinfo {pages}
  {041069} (\bibinfo {year} {2017})}\BibitemShut {NoStop}%
\bibitem [{\citenamefont {Po}\ \emph {et~al.}(2017)\citenamefont {Po},
  \citenamefont {Vishwanath},\ and\ \citenamefont {Watanabe}}]{Po17}%
  \BibitemOpen
  \bibfield  {author} {\bibinfo {author} {\bibfnamefont {H.~C.}\ \bibnamefont
  {Po}}, \bibinfo {author} {\bibfnamefont {A.}~\bibnamefont {Vishwanath}},\
  and\ \bibinfo {author} {\bibfnamefont {H.}~\bibnamefont {Watanabe}},\
  }\bibfield  {title} {\bibinfo {title} {Symmetry-based indicators of band
  topology in the 230 space groups},\ }\href@noop {} {\bibfield  {journal}
  {\bibinfo  {journal} {Nature communications}\ }\textbf {\bibinfo {volume}
  {8}},\ \bibinfo {pages} {50} (\bibinfo {year} {2017})}\BibitemShut {NoStop}%
\bibitem [{\citenamefont {Song}\ \emph {et~al.}(2018)\citenamefont {Song},
  \citenamefont {Zhang}, \citenamefont {Fang},\ and\ \citenamefont
  {Fang}}]{ZSong18quantitative}%
  \BibitemOpen
  \bibfield  {author} {\bibinfo {author} {\bibfnamefont {Z.}~\bibnamefont
  {Song}}, \bibinfo {author} {\bibfnamefont {T.}~\bibnamefont {Zhang}},
  \bibinfo {author} {\bibfnamefont {Z.}~\bibnamefont {Fang}},\ and\ \bibinfo
  {author} {\bibfnamefont {C.}~\bibnamefont {Fang}},\ }\bibfield  {title}
  {\bibinfo {title} {Quantitative mappings between symmetry and topology in
  solids},\ }\href@noop {} {\bibfield  {journal} {\bibinfo  {journal} {Nature
  communications}\ }\textbf {\bibinfo {volume} {9}},\ \bibinfo {pages} {3530}
  (\bibinfo {year} {2018})}\BibitemShut {NoStop}%
\bibitem [{\citenamefont {Khalaf}\ \emph {et~al.}(2018)\citenamefont {Khalaf},
  \citenamefont {Po}, \citenamefont {Vishwanath},\ and\ \citenamefont
  {Watanabe}}]{Khalaf18}%
  \BibitemOpen
  \bibfield  {author} {\bibinfo {author} {\bibfnamefont {E.}~\bibnamefont
  {Khalaf}}, \bibinfo {author} {\bibfnamefont {H.~C.}\ \bibnamefont {Po}},
  \bibinfo {author} {\bibfnamefont {A.}~\bibnamefont {Vishwanath}},\ and\
  \bibinfo {author} {\bibfnamefont {H.}~\bibnamefont {Watanabe}},\ }\bibfield
  {title} {\bibinfo {title} {Symmetry indicators and anomalous surface states
  of topological crystalline insulators},\ }\href
  {https://doi.org/10.1103/PhysRevX.8.031070} {\bibfield  {journal} {\bibinfo
  {journal} {Phys. Rev. X}\ }\textbf {\bibinfo {volume} {8}},\ \bibinfo {pages}
  {031070} (\bibinfo {year} {2018})}\BibitemShut {NoStop}%
\bibitem [{\citenamefont {Elcoro}\ \emph {et~al.}(2020)\citenamefont {Elcoro},
  \citenamefont {Wieder}, \citenamefont {Song}, \citenamefont {Xu},
  \citenamefont {Bradlyn},\ and\ \citenamefont
  {Bernevig}}]{Elcoro2020magnetic}%
  \BibitemOpen
  \bibfield  {author} {\bibinfo {author} {\bibfnamefont {L.}~\bibnamefont
  {Elcoro}}, \bibinfo {author} {\bibfnamefont {B.~J.}\ \bibnamefont {Wieder}},
  \bibinfo {author} {\bibfnamefont {Z.}~\bibnamefont {Song}}, \bibinfo {author}
  {\bibfnamefont {Y.}~\bibnamefont {Xu}}, \bibinfo {author} {\bibfnamefont
  {B.}~\bibnamefont {Bradlyn}},\ and\ \bibinfo {author} {\bibfnamefont {B.~A.}\
  \bibnamefont {Bernevig}},\ }\bibfield  {title} {\bibinfo {title} {Magnetic
  topological quantum chemistry},\ }\href@noop {} {\bibfield  {journal}
  {\bibinfo  {journal} {arXiv preprint arXiv:2010.00598}\ } (\bibinfo {year}
  {2020})}\BibitemShut {NoStop}%
\bibitem [{\citenamefont {Po}(2020)}]{Po2020symmetry}%
  \BibitemOpen
  \bibfield  {author} {\bibinfo {author} {\bibfnamefont {H.~C.}\ \bibnamefont
  {Po}},\ }\bibfield  {title} {\bibinfo {title} {Symmetry indicators of band
  topology},\ }\href@noop {} {\bibfield  {journal} {\bibinfo  {journal} {J.
  Phys.: Condens. Matter}\ }\textbf {\bibinfo {volume} {32}},\ \bibinfo {pages}
  {263001} (\bibinfo {year} {2020})}\BibitemShut {NoStop}%
\bibitem [{\citenamefont {Zhang}\ \emph {et~al.}(2019)\citenamefont {Zhang},
  \citenamefont {Jiang}, \citenamefont {Song}, \citenamefont {Huang},
  \citenamefont {He}, \citenamefont {Fang}, \citenamefont {Weng},\ and\
  \citenamefont {Fang}}]{Zhang19catalogue}%
  \BibitemOpen
  \bibfield  {author} {\bibinfo {author} {\bibfnamefont {T.}~\bibnamefont
  {Zhang}}, \bibinfo {author} {\bibfnamefont {Y.}~\bibnamefont {Jiang}},
  \bibinfo {author} {\bibfnamefont {Z.}~\bibnamefont {Song}}, \bibinfo {author}
  {\bibfnamefont {H.}~\bibnamefont {Huang}}, \bibinfo {author} {\bibfnamefont
  {Y.}~\bibnamefont {He}}, \bibinfo {author} {\bibfnamefont {Z.}~\bibnamefont
  {Fang}}, \bibinfo {author} {\bibfnamefont {H.}~\bibnamefont {Weng}},\ and\
  \bibinfo {author} {\bibfnamefont {C.}~\bibnamefont {Fang}},\ }\bibfield
  {title} {\bibinfo {title} {Catalogue of topological electronic materials},\
  }\href@noop {} {\bibfield  {journal} {\bibinfo  {journal} {Nature}\ }\textbf
  {\bibinfo {volume} {566}},\ \bibinfo {pages} {475} (\bibinfo {year}
  {2019})}\BibitemShut {NoStop}%
\bibitem [{\citenamefont {Vergniory}\ \emph {et~al.}(2019)\citenamefont
  {Vergniory}, \citenamefont {Elcoro}, \citenamefont {Felser}, \citenamefont
  {Regnault}, \citenamefont {Bernevig},\ and\ \citenamefont
  {Wang}}]{Vergniory19complete}%
  \BibitemOpen
  \bibfield  {author} {\bibinfo {author} {\bibfnamefont {M.}~\bibnamefont
  {Vergniory}}, \bibinfo {author} {\bibfnamefont {L.}~\bibnamefont {Elcoro}},
  \bibinfo {author} {\bibfnamefont {C.}~\bibnamefont {Felser}}, \bibinfo
  {author} {\bibfnamefont {N.}~\bibnamefont {Regnault}}, \bibinfo {author}
  {\bibfnamefont {B.~A.}\ \bibnamefont {Bernevig}},\ and\ \bibinfo {author}
  {\bibfnamefont {Z.}~\bibnamefont {Wang}},\ }\bibfield  {title} {\bibinfo
  {title} {A complete catalogue of high-quality topological materials},\
  }\href@noop {} {\bibfield  {journal} {\bibinfo  {journal} {Nature}\ }\textbf
  {\bibinfo {volume} {566}},\ \bibinfo {pages} {480} (\bibinfo {year}
  {2019})}\BibitemShut {NoStop}%
\bibitem [{\citenamefont {Tang}\ \emph
  {et~al.}(2019{\natexlab{a}})\citenamefont {Tang}, \citenamefont {Po},
  \citenamefont {Vishwanath},\ and\ \citenamefont {Wan}}]{Tang19sciadv}%
  \BibitemOpen
  \bibfield  {author} {\bibinfo {author} {\bibfnamefont {F.}~\bibnamefont
  {Tang}}, \bibinfo {author} {\bibfnamefont {H.~C.}\ \bibnamefont {Po}},
  \bibinfo {author} {\bibfnamefont {A.}~\bibnamefont {Vishwanath}},\ and\
  \bibinfo {author} {\bibfnamefont {X.}~\bibnamefont {Wan}},\ }\bibfield
  {title} {\bibinfo {title} {Topological materials discovery by large-order
  symmetry indicators},\ }\href@noop {} {\bibfield  {journal} {\bibinfo
  {journal} {Science Advances}\ }\textbf {\bibinfo {volume} {5}},\ \bibinfo
  {pages} {eaau8725} (\bibinfo {year} {2019}{\natexlab{a}})}\BibitemShut
  {NoStop}%
\bibitem [{\citenamefont {Tang}\ \emph
  {et~al.}(2019{\natexlab{b}})\citenamefont {Tang}, \citenamefont {Po},
  \citenamefont {Vishwanath},\ and\ \citenamefont {Wan}}]{Tang2019efficient}%
  \BibitemOpen
  \bibfield  {author} {\bibinfo {author} {\bibfnamefont {F.}~\bibnamefont
  {Tang}}, \bibinfo {author} {\bibfnamefont {H.~C.}\ \bibnamefont {Po}},
  \bibinfo {author} {\bibfnamefont {A.}~\bibnamefont {Vishwanath}},\ and\
  \bibinfo {author} {\bibfnamefont {X.}~\bibnamefont {Wan}},\ }\bibfield
  {title} {\bibinfo {title} {Efficient topological materials discovery using
  symmetry indicators},\ }\href@noop {} {\bibfield  {journal} {\bibinfo
  {journal} {Nature Physics}\ }\textbf {\bibinfo {volume} {15}},\ \bibinfo
  {pages} {470} (\bibinfo {year} {2019}{\natexlab{b}})}\BibitemShut {NoStop}%
\bibitem [{\citenamefont {Wang}\ \emph {et~al.}(2019)\citenamefont {Wang},
  \citenamefont {Tang}, \citenamefont {Ji}, \citenamefont {Zhang},
  \citenamefont {Vishwanath}, \citenamefont {Po},\ and\ \citenamefont
  {Wan}}]{DWang19}%
  \BibitemOpen
  \bibfield  {author} {\bibinfo {author} {\bibfnamefont {D.}~\bibnamefont
  {Wang}}, \bibinfo {author} {\bibfnamefont {F.}~\bibnamefont {Tang}}, \bibinfo
  {author} {\bibfnamefont {J.}~\bibnamefont {Ji}}, \bibinfo {author}
  {\bibfnamefont {W.}~\bibnamefont {Zhang}}, \bibinfo {author} {\bibfnamefont
  {A.}~\bibnamefont {Vishwanath}}, \bibinfo {author} {\bibfnamefont {H.~C.}\
  \bibnamefont {Po}},\ and\ \bibinfo {author} {\bibfnamefont {X.}~\bibnamefont
  {Wan}},\ }\bibfield  {title} {\bibinfo {title} {Two-dimensional topological
  materials discovery by symmetry-indicator method},\ }\href
  {https://doi.org/10.1103/PhysRevB.100.195108} {\bibfield  {journal} {\bibinfo
   {journal} {Phys. Rev. B}\ }\textbf {\bibinfo {volume} {100}},\ \bibinfo
  {pages} {195108} (\bibinfo {year} {2019})}\BibitemShut {NoStop}%
\bibitem [{\citenamefont {Xu}\ \emph {et~al.}(2020)\citenamefont {Xu},
  \citenamefont {Elcoro}, \citenamefont {Song}, \citenamefont {Wieder},
  \citenamefont {Vergniory}, \citenamefont {Regnault}, \citenamefont {Chen},
  \citenamefont {Felser},\ and\ \citenamefont {Bernevig}}]{Xu2020high}%
  \BibitemOpen
  \bibfield  {author} {\bibinfo {author} {\bibfnamefont {Y.}~\bibnamefont
  {Xu}}, \bibinfo {author} {\bibfnamefont {L.}~\bibnamefont {Elcoro}}, \bibinfo
  {author} {\bibfnamefont {Z.-D.}\ \bibnamefont {Song}}, \bibinfo {author}
  {\bibfnamefont {B.~J.}\ \bibnamefont {Wieder}}, \bibinfo {author}
  {\bibfnamefont {M.}~\bibnamefont {Vergniory}}, \bibinfo {author}
  {\bibfnamefont {N.}~\bibnamefont {Regnault}}, \bibinfo {author}
  {\bibfnamefont {Y.}~\bibnamefont {Chen}}, \bibinfo {author} {\bibfnamefont
  {C.}~\bibnamefont {Felser}},\ and\ \bibinfo {author} {\bibfnamefont {B.~A.}\
  \bibnamefont {Bernevig}},\ }\bibfield  {title} {\bibinfo {title}
  {High-throughput calculations of magnetic topological materials},\
  }\href@noop {} {\bibfield  {journal} {\bibinfo  {journal} {Nature}\ }\textbf
  {\bibinfo {volume} {586}},\ \bibinfo {pages} {702} (\bibinfo {year}
  {2020})}\BibitemShut {NoStop}%
\bibitem [{\citenamefont {Moore}\ \emph {et~al.}(2008)\citenamefont {Moore},
  \citenamefont {Ran},\ and\ \citenamefont {Wen}}]{Moore2008}%
  \BibitemOpen
  \bibfield  {author} {\bibinfo {author} {\bibfnamefont {J.~E.}\ \bibnamefont
  {Moore}}, \bibinfo {author} {\bibfnamefont {Y.}~\bibnamefont {Ran}},\ and\
  \bibinfo {author} {\bibfnamefont {X.-G.}\ \bibnamefont {Wen}},\ }\bibfield
  {title} {\bibinfo {title} {Topological surface states in three-dimensional
  magnetic insulators},\ }\href
  {https://doi.org/10.1103/PhysRevLett.101.186805} {\bibfield  {journal}
  {\bibinfo  {journal} {Phys. Rev. Lett.}\ }\textbf {\bibinfo {volume} {101}},\
  \bibinfo {pages} {186805} (\bibinfo {year} {2008})}\BibitemShut {NoStop}%
\bibitem [{\citenamefont {Deng}\ \emph {et~al.}(2013)\citenamefont {Deng},
  \citenamefont {Wang}, \citenamefont {Shen},\ and\ \citenamefont
  {Duan}}]{Deng2013}%
  \BibitemOpen
  \bibfield  {author} {\bibinfo {author} {\bibfnamefont {D.-L.}\ \bibnamefont
  {Deng}}, \bibinfo {author} {\bibfnamefont {S.-T.}\ \bibnamefont {Wang}},
  \bibinfo {author} {\bibfnamefont {C.}~\bibnamefont {Shen}},\ and\ \bibinfo
  {author} {\bibfnamefont {L.-M.}\ \bibnamefont {Duan}},\ }\bibfield  {title}
  {\bibinfo {title} {Hopf insulators and their topologically protected surface
  states},\ }\href {https://doi.org/10.1103/PhysRevB.88.201105} {\bibfield
  {journal} {\bibinfo  {journal} {Phys. Rev. B}\ }\textbf {\bibinfo {volume}
  {88}},\ \bibinfo {pages} {201105} (\bibinfo {year} {2013})}\BibitemShut
  {NoStop}%
\bibitem [{\citenamefont {Kennedy}(2016)}]{Kennedy2016}%
  \BibitemOpen
  \bibfield  {author} {\bibinfo {author} {\bibfnamefont {R.}~\bibnamefont
  {Kennedy}},\ }\bibfield  {title} {\bibinfo {title} {Topological {Hopf-Chern}
  insulators and the {Hopf} superconductor},\ }\href
  {https://doi.org/10.1103/PhysRevB.94.035137} {\bibfield  {journal} {\bibinfo
  {journal} {Phys. Rev. B}\ }\textbf {\bibinfo {volume} {94}},\ \bibinfo
  {pages} {035137} (\bibinfo {year} {2016})}\BibitemShut {NoStop}%
\bibitem [{\citenamefont {Liu}\ \emph {et~al.}(2017)\citenamefont {Liu},
  \citenamefont {Vafa},\ and\ \citenamefont {Xu}}]{Liu2017}%
  \BibitemOpen
  \bibfield  {author} {\bibinfo {author} {\bibfnamefont {C.}~\bibnamefont
  {Liu}}, \bibinfo {author} {\bibfnamefont {F.}~\bibnamefont {Vafa}},\ and\
  \bibinfo {author} {\bibfnamefont {C.}~\bibnamefont {Xu}},\ }\bibfield
  {title} {\bibinfo {title} {Symmetry-protected topological {Hopf} insulator
  and its generalizations},\ }\href
  {https://doi.org/10.1103/PhysRevB.95.161116} {\bibfield  {journal} {\bibinfo
  {journal} {Phys. Rev. B}\ }\textbf {\bibinfo {volume} {95}},\ \bibinfo
  {pages} {161116} (\bibinfo {year} {2017})}\BibitemShut {NoStop}%
\bibitem [{\citenamefont {Schuster}\ \emph {et~al.}(2019)\citenamefont
  {Schuster}, \citenamefont {Gazit}, \citenamefont {Moore},\ and\ \citenamefont
  {Yao}}]{Schuster2019}%
  \BibitemOpen
  \bibfield  {author} {\bibinfo {author} {\bibfnamefont {T.}~\bibnamefont
  {Schuster}}, \bibinfo {author} {\bibfnamefont {S.}~\bibnamefont {Gazit}},
  \bibinfo {author} {\bibfnamefont {J.~E.}\ \bibnamefont {Moore}},\ and\
  \bibinfo {author} {\bibfnamefont {N.~Y.}\ \bibnamefont {Yao}},\ }\bibfield
  {title} {\bibinfo {title} {Floquet {Hopf} insulators},\ }\href
  {https://doi.org/10.1103/PhysRevLett.123.266803} {\bibfield  {journal}
  {\bibinfo  {journal} {Phys. Rev. Lett.}\ }\textbf {\bibinfo {volume} {123}},\
  \bibinfo {pages} {266803} (\bibinfo {year} {2019})}\BibitemShut {NoStop}%
\bibitem [{\citenamefont {Schuster}\ \emph {et~al.}(2021)\citenamefont
  {Schuster}, \citenamefont {Flicker}, \citenamefont {Li}, \citenamefont
  {Kotochigova}, \citenamefont {Moore}, \citenamefont {Ye},\ and\ \citenamefont
  {Yao}}]{Schuster2021}%
  \BibitemOpen
  \bibfield  {author} {\bibinfo {author} {\bibfnamefont {T.}~\bibnamefont
  {Schuster}}, \bibinfo {author} {\bibfnamefont {F.}~\bibnamefont {Flicker}},
  \bibinfo {author} {\bibfnamefont {M.}~\bibnamefont {Li}}, \bibinfo {author}
  {\bibfnamefont {S.}~\bibnamefont {Kotochigova}}, \bibinfo {author}
  {\bibfnamefont {J.~E.}\ \bibnamefont {Moore}}, \bibinfo {author}
  {\bibfnamefont {J.}~\bibnamefont {Ye}},\ and\ \bibinfo {author}
  {\bibfnamefont {N.~Y.}\ \bibnamefont {Yao}},\ }\bibfield  {title} {\bibinfo
  {title} {Realizing {Hopf} insulators in dipolar spin systems},\ }\href
  {https://doi.org/10.1103/PhysRevLett.127.015301} {\bibfield  {journal}
  {\bibinfo  {journal} {Phys. Rev. Lett.}\ }\textbf {\bibinfo {volume} {127}},\
  \bibinfo {pages} {015301} (\bibinfo {year} {2021})}\BibitemShut {NoStop}%
\bibitem [{\citenamefont {Po}\ \emph {et~al.}(2018)\citenamefont {Po},
  \citenamefont {Watanabe},\ and\ \citenamefont {Vishwanath}}]{Po2018fragile}%
  \BibitemOpen
  \bibfield  {author} {\bibinfo {author} {\bibfnamefont {H.~C.}\ \bibnamefont
  {Po}}, \bibinfo {author} {\bibfnamefont {H.}~\bibnamefont {Watanabe}},\ and\
  \bibinfo {author} {\bibfnamefont {A.}~\bibnamefont {Vishwanath}},\ }\bibfield
   {title} {\bibinfo {title} {Fragile topology and {Wannier} obstructions},\
  }\href {https://doi.org/10.1103/PhysRevLett.121.126402} {\bibfield  {journal}
  {\bibinfo  {journal} {Phys. Rev. Lett.}\ }\textbf {\bibinfo {volume} {121}},\
  \bibinfo {pages} {126402} (\bibinfo {year} {2018})}\BibitemShut {NoStop}%
\bibitem [{\citenamefont {Wieder}\ and\ \citenamefont
  {Bernevig}(2018)}]{Wieder2018axion}%
  \BibitemOpen
  \bibfield  {author} {\bibinfo {author} {\bibfnamefont {B.~J.}\ \bibnamefont
  {Wieder}}\ and\ \bibinfo {author} {\bibfnamefont {B.~A.}\ \bibnamefont
  {Bernevig}},\ }\bibfield  {title} {\bibinfo {title} {The axion insulator as a
  pump of fragile topology},\ }\href@noop {} {\bibfield  {journal} {\bibinfo
  {journal} {arXiv preprint arXiv:1810.02373}\ } (\bibinfo {year}
  {2018})}\BibitemShut {NoStop}%
\bibitem [{\citenamefont {Kooi}\ \emph {et~al.}(2019)\citenamefont {Kooi},
  \citenamefont {van Miert},\ and\ \citenamefont {Ortix}}]{Kooi2019}%
  \BibitemOpen
  \bibfield  {author} {\bibinfo {author} {\bibfnamefont {S.~H.}\ \bibnamefont
  {Kooi}}, \bibinfo {author} {\bibfnamefont {G.}~\bibnamefont {van Miert}},\
  and\ \bibinfo {author} {\bibfnamefont {C.}~\bibnamefont {Ortix}},\ }\bibfield
   {title} {\bibinfo {title} {Classification of crystalline insulators without
  symmetry indicators: Atomic and fragile topological phases in twofold
  rotation symmetric systems},\ }\href
  {https://doi.org/10.1103/PhysRevB.100.115160} {\bibfield  {journal} {\bibinfo
   {journal} {Phys. Rev. B}\ }\textbf {\bibinfo {volume} {100}},\ \bibinfo
  {pages} {115160} (\bibinfo {year} {2019})}\BibitemShut {NoStop}%
\bibitem [{\citenamefont {Hwang}\ \emph {et~al.}(2019)\citenamefont {Hwang},
  \citenamefont {Ahn},\ and\ \citenamefont {Yang}}]{Hwang2019}%
  \BibitemOpen
  \bibfield  {author} {\bibinfo {author} {\bibfnamefont {Y.}~\bibnamefont
  {Hwang}}, \bibinfo {author} {\bibfnamefont {J.}~\bibnamefont {Ahn}},\ and\
  \bibinfo {author} {\bibfnamefont {B.-J.}\ \bibnamefont {Yang}},\ }\bibfield
  {title} {\bibinfo {title} {Fragile topology protected by inversion symmetry:
  Diagnosis, bulk-boundary correspondence, and {Wilson} loop},\ }\href
  {https://doi.org/10.1103/PhysRevB.100.205126} {\bibfield  {journal} {\bibinfo
   {journal} {Phys. Rev. B}\ }\textbf {\bibinfo {volume} {100}},\ \bibinfo
  {pages} {205126} (\bibinfo {year} {2019})}\BibitemShut {NoStop}%
\bibitem [{\citenamefont {Song}\ \emph {et~al.}(2020)\citenamefont {Song},
  \citenamefont {Elcoro}, \citenamefont {Xu}, \citenamefont {Regnault},\ and\
  \citenamefont {Bernevig}}]{Song2020fragile}%
  \BibitemOpen
  \bibfield  {author} {\bibinfo {author} {\bibfnamefont {Z.-D.}\ \bibnamefont
  {Song}}, \bibinfo {author} {\bibfnamefont {L.}~\bibnamefont {Elcoro}},
  \bibinfo {author} {\bibfnamefont {Y.-F.}\ \bibnamefont {Xu}}, \bibinfo
  {author} {\bibfnamefont {N.}~\bibnamefont {Regnault}},\ and\ \bibinfo
  {author} {\bibfnamefont {B.~A.}\ \bibnamefont {Bernevig}},\ }\bibfield
  {title} {\bibinfo {title} {Fragile phases as {Affine} monoids: Classification
  and material examples},\ }\href {https://doi.org/10.1103/PhysRevX.10.031001}
  {\bibfield  {journal} {\bibinfo  {journal} {Phys. Rev. X}\ }\textbf {\bibinfo
  {volume} {10}},\ \bibinfo {pages} {031001} (\bibinfo {year}
  {2020})}\BibitemShut {NoStop}%
\bibitem [{\citenamefont {Nelson}\ \emph {et~al.}(2021)\citenamefont {Nelson},
  \citenamefont {Neupert}, \citenamefont {Bzdu\ifmmode~\check{s}\else
  \v{s}\fi{}ek},\ and\ \citenamefont {Alexandradinata}}]{Nelson2021}%
  \BibitemOpen
  \bibfield  {author} {\bibinfo {author} {\bibfnamefont {A.}~\bibnamefont
  {Nelson}}, \bibinfo {author} {\bibfnamefont {T.}~\bibnamefont {Neupert}},
  \bibinfo {author} {\bibfnamefont {T.}~\bibnamefont
  {Bzdu\ifmmode~\check{s}\else \v{s}\fi{}ek}},\ and\ \bibinfo {author}
  {\bibfnamefont {A.}~\bibnamefont {Alexandradinata}},\ }\bibfield  {title}
  {\bibinfo {title} {Multicellularity of delicate topological insulators},\
  }\href {https://doi.org/10.1103/PhysRevLett.126.216404} {\bibfield  {journal}
  {\bibinfo  {journal} {Phys. Rev. Lett.}\ }\textbf {\bibinfo {volume} {126}},\
  \bibinfo {pages} {216404} (\bibinfo {year} {2021})}\BibitemShut {NoStop}%
\bibitem [{Note1()}]{Note1}%
  \BibitemOpen
  \bibinfo {note} {There are various definitions of unstable/fragile
  topological insulators in the literature. An initial example of unstable
  topological insulators is a Hopf insulator~\cite {Schnyder08,Moore2008} of a
  three-dimensional two-band insulator. The term {"fragile topology"} was
  introduced in the analysis using the symmetry-based indicators~\cite
  {Po2018fragile} for energy bands that can be mathematically written as a
  \protect \textit {difference} between trivial (atomic) bands. Since atomic
  insulators do not host gapless surface states, the same holds true for these
  fragile topological bands. In contrast, recent studies~\cite
  {Bouhon2020,Alexandradinata2020} have found fragile topological insulators
  that are not captured by the symmetry-based indicators but often have gapless
  surface states. In this paper, we adopt the broad definition of fragile
  topological insulators that includes the latter cases.}\BibitemShut {Stop}%
\bibitem [{\citenamefont {Fu}(2011)}]{Fu2011}%
  \BibitemOpen
  \bibfield  {author} {\bibinfo {author} {\bibfnamefont {L.}~\bibnamefont
  {Fu}},\ }\bibfield  {title} {\bibinfo {title} {Topological crystalline
  insulators},\ }\href {https://doi.org/10.1103/PhysRevLett.106.106802}
  {\bibfield  {journal} {\bibinfo  {journal} {Phys. Rev. Lett.}\ }\textbf
  {\bibinfo {volume} {106}},\ \bibinfo {pages} {106802} (\bibinfo {year}
  {2011})}\BibitemShut {NoStop}%
\bibitem [{\citenamefont {Alexandradinata}\ \emph
  {et~al.}(2014{\natexlab{a}})\citenamefont {Alexandradinata}, \citenamefont
  {Fang}, \citenamefont {Gilbert},\ and\ \citenamefont
  {Bernevig}}]{Alexandradinata2014spin}%
  \BibitemOpen
  \bibfield  {author} {\bibinfo {author} {\bibfnamefont {A.}~\bibnamefont
  {Alexandradinata}}, \bibinfo {author} {\bibfnamefont {C.}~\bibnamefont
  {Fang}}, \bibinfo {author} {\bibfnamefont {M.~J.}\ \bibnamefont {Gilbert}},\
  and\ \bibinfo {author} {\bibfnamefont {B.~A.}\ \bibnamefont {Bernevig}},\
  }\bibfield  {title} {\bibinfo {title} {Spin-orbit-free topological insulators
  without time-reversal symmetry},\ }\href
  {https://doi.org/10.1103/PhysRevLett.113.116403} {\bibfield  {journal}
  {\bibinfo  {journal} {Phys. Rev. Lett.}\ }\textbf {\bibinfo {volume} {113}},\
  \bibinfo {pages} {116403} (\bibinfo {year} {2014}{\natexlab{a}})}\BibitemShut
  {NoStop}%
\bibitem [{\citenamefont {Alexandradinata}\ \emph {et~al.}(2020)\citenamefont
  {Alexandradinata}, \citenamefont {H\"oller}, \citenamefont {Wang},
  \citenamefont {Cheng},\ and\ \citenamefont {Lu}}]{Alexandradinata2020}%
  \BibitemOpen
  \bibfield  {author} {\bibinfo {author} {\bibfnamefont {A.}~\bibnamefont
  {Alexandradinata}}, \bibinfo {author} {\bibfnamefont {J.}~\bibnamefont
  {H\"oller}}, \bibinfo {author} {\bibfnamefont {C.}~\bibnamefont {Wang}},
  \bibinfo {author} {\bibfnamefont {H.}~\bibnamefont {Cheng}},\ and\ \bibinfo
  {author} {\bibfnamefont {L.}~\bibnamefont {Lu}},\ }\bibfield  {title}
  {\bibinfo {title} {Crystallographic splitting theorem for band
  representations and fragile topological photonic crystals},\ }\href
  {https://doi.org/10.1103/PhysRevB.102.115117} {\bibfield  {journal} {\bibinfo
   {journal} {Phys. Rev. B}\ }\textbf {\bibinfo {volume} {102}},\ \bibinfo
  {pages} {115117} (\bibinfo {year} {2020})}\BibitemShut {NoStop}%
\bibitem [{\citenamefont {Fang}\ and\ \citenamefont {Fu}(2015)}]{CFang15}%
  \BibitemOpen
  \bibfield  {author} {\bibinfo {author} {\bibfnamefont {C.}~\bibnamefont
  {Fang}}\ and\ \bibinfo {author} {\bibfnamefont {L.}~\bibnamefont {Fu}},\
  }\bibfield  {title} {\bibinfo {title} {New classes of three-dimensional
  topological crystalline insulators: Nonsymmorphic and magnetic},\ }\href
  {https://doi.org/10.1103/PhysRevB.91.161105} {\bibfield  {journal} {\bibinfo
  {journal} {Phys. Rev. B}\ }\textbf {\bibinfo {volume} {91}},\ \bibinfo
  {pages} {161105(R)} (\bibinfo {year} {2015})}\BibitemShut {NoStop}%
\bibitem [{\citenamefont {Alexandradinata}\ and\ \citenamefont
  {Bernevig}(2016)}]{Alexandradinata2016Berry}%
  \BibitemOpen
  \bibfield  {author} {\bibinfo {author} {\bibfnamefont {A.}~\bibnamefont
  {Alexandradinata}}\ and\ \bibinfo {author} {\bibfnamefont {B.~A.}\
  \bibnamefont {Bernevig}},\ }\bibfield  {title} {\bibinfo {title} {Berry-phase
  description of topological crystalline insulators},\ }\href
  {https://doi.org/10.1103/PhysRevB.93.205104} {\bibfield  {journal} {\bibinfo
  {journal} {Phys. Rev. B}\ }\textbf {\bibinfo {volume} {93}},\ \bibinfo
  {pages} {205104} (\bibinfo {year} {2016})}\BibitemShut {NoStop}%
\bibitem [{\citenamefont {Ahn}\ \emph {et~al.}(2018)\citenamefont {Ahn},
  \citenamefont {Kim}, \citenamefont {Kim},\ and\ \citenamefont
  {Yang}}]{Ahn2018band}%
  \BibitemOpen
  \bibfield  {author} {\bibinfo {author} {\bibfnamefont {J.}~\bibnamefont
  {Ahn}}, \bibinfo {author} {\bibfnamefont {D.}~\bibnamefont {Kim}}, \bibinfo
  {author} {\bibfnamefont {Y.}~\bibnamefont {Kim}},\ and\ \bibinfo {author}
  {\bibfnamefont {B.-J.}\ \bibnamefont {Yang}},\ }\bibfield  {title} {\bibinfo
  {title} {Band topology and linking structure of nodal line semimetals with
  {${Z}_{2}$} monopole charges},\ }\href
  {https://doi.org/10.1103/PhysRevLett.121.106403} {\bibfield  {journal}
  {\bibinfo  {journal} {Phys. Rev. Lett.}\ }\textbf {\bibinfo {volume} {121}},\
  \bibinfo {pages} {106403} (\bibinfo {year} {2018})}\BibitemShut {NoStop}%
\bibitem [{\citenamefont {Bouhon}\ \emph {et~al.}(2019)\citenamefont {Bouhon},
  \citenamefont {Black-Schaffer},\ and\ \citenamefont {Slager}}]{Bouhon2019}%
  \BibitemOpen
  \bibfield  {author} {\bibinfo {author} {\bibfnamefont {A.}~\bibnamefont
  {Bouhon}}, \bibinfo {author} {\bibfnamefont {A.~M.}\ \bibnamefont
  {Black-Schaffer}},\ and\ \bibinfo {author} {\bibfnamefont {R.-J.}\
  \bibnamefont {Slager}},\ }\bibfield  {title} {\bibinfo {title} {Wilson loop
  approach to fragile topology of split elementary band representations and
  topological crystalline insulators with time-reversal symmetry},\ }\href
  {https://doi.org/10.1103/PhysRevB.100.195135} {\bibfield  {journal} {\bibinfo
   {journal} {Phys. Rev. B}\ }\textbf {\bibinfo {volume} {100}},\ \bibinfo
  {pages} {195135} (\bibinfo {year} {2019})}\BibitemShut {NoStop}%
\bibitem [{\citenamefont {Ahn}\ and\ \citenamefont {Yang}(2019)}]{Ahn2019}%
  \BibitemOpen
  \bibfield  {author} {\bibinfo {author} {\bibfnamefont {J.}~\bibnamefont
  {Ahn}}\ and\ \bibinfo {author} {\bibfnamefont {B.-J.}\ \bibnamefont {Yang}},\
  }\bibfield  {title} {\bibinfo {title} {Symmetry representation approach to
  topological invariants in {${C}_{2z}T$}-symmetric systems},\ }\href
  {https://doi.org/10.1103/PhysRevB.99.235125} {\bibfield  {journal} {\bibinfo
  {journal} {Phys. Rev. B}\ }\textbf {\bibinfo {volume} {99}},\ \bibinfo
  {pages} {235125} (\bibinfo {year} {2019})}\BibitemShut {NoStop}%
\bibitem [{\citenamefont {Song}\ \emph
  {et~al.}(2019{\natexlab{b}})\citenamefont {Song}, \citenamefont {Wang},
  \citenamefont {Shi}, \citenamefont {Li}, \citenamefont {Fang},\ and\
  \citenamefont {Bernevig}}]{Song2019all}%
  \BibitemOpen
  \bibfield  {author} {\bibinfo {author} {\bibfnamefont {Z.}~\bibnamefont
  {Song}}, \bibinfo {author} {\bibfnamefont {Z.}~\bibnamefont {Wang}}, \bibinfo
  {author} {\bibfnamefont {W.}~\bibnamefont {Shi}}, \bibinfo {author}
  {\bibfnamefont {G.}~\bibnamefont {Li}}, \bibinfo {author} {\bibfnamefont
  {C.}~\bibnamefont {Fang}},\ and\ \bibinfo {author} {\bibfnamefont {B.~A.}\
  \bibnamefont {Bernevig}},\ }\bibfield  {title} {\bibinfo {title} {All magic
  angles in twisted bilayer graphene are topological},\ }\href
  {https://doi.org/10.1103/PhysRevLett.123.036401} {\bibfield  {journal}
  {\bibinfo  {journal} {Phys. Rev. Lett.}\ }\textbf {\bibinfo {volume} {123}},\
  \bibinfo {pages} {036401} (\bibinfo {year} {2019}{\natexlab{b}})}\BibitemShut
  {NoStop}%
\bibitem [{\citenamefont {Ahn}\ \emph {et~al.}(2019)\citenamefont {Ahn},
  \citenamefont {Park},\ and\ \citenamefont {Yang}}]{Ahn2019failure}%
  \BibitemOpen
  \bibfield  {author} {\bibinfo {author} {\bibfnamefont {J.}~\bibnamefont
  {Ahn}}, \bibinfo {author} {\bibfnamefont {S.}~\bibnamefont {Park}},\ and\
  \bibinfo {author} {\bibfnamefont {B.-J.}\ \bibnamefont {Yang}},\ }\bibfield
  {title} {\bibinfo {title} {Failure of {Nielsen-Ninomiya} theorem and fragile
  topology in two-dimensional systems with space-time inversion symmetry:
  Application to twisted bilayer graphene at magic angle},\ }\href
  {https://doi.org/10.1103/PhysRevX.9.021013} {\bibfield  {journal} {\bibinfo
  {journal} {Phys. Rev. X}\ }\textbf {\bibinfo {volume} {9}},\ \bibinfo {pages}
  {021013} (\bibinfo {year} {2019})}\BibitemShut {NoStop}%
\bibitem [{\citenamefont {Bouhon}\ \emph {et~al.}(2020)\citenamefont {Bouhon},
  \citenamefont {Bzdu\ifmmode~\check{s}\else \v{s}\fi{}ek},\ and\ \citenamefont
  {Slager}}]{Bouhon2020}%
  \BibitemOpen
  \bibfield  {author} {\bibinfo {author} {\bibfnamefont {A.}~\bibnamefont
  {Bouhon}}, \bibinfo {author} {\bibfnamefont {T.}~\bibnamefont
  {Bzdu\ifmmode~\check{s}\else \v{s}\fi{}ek}},\ and\ \bibinfo {author}
  {\bibfnamefont {R.-J.}\ \bibnamefont {Slager}},\ }\bibfield  {title}
  {\bibinfo {title} {Geometric approach to fragile topology beyond symmetry
  indicators},\ }\href {https://doi.org/10.1103/PhysRevB.102.115135} {\bibfield
   {journal} {\bibinfo  {journal} {Phys. Rev. B}\ }\textbf {\bibinfo {volume}
  {102}},\ \bibinfo {pages} {115135} (\bibinfo {year} {2020})}\BibitemShut
  {NoStop}%
\bibitem [{\citenamefont {\"Unal}\ \emph {et~al.}(2020)\citenamefont {\"Unal},
  \citenamefont {Bouhon},\ and\ \citenamefont {Slager}}]{Unal2020}%
  \BibitemOpen
  \bibfield  {author} {\bibinfo {author} {\bibfnamefont {F.~N.}\ \bibnamefont
  {\"Unal}}, \bibinfo {author} {\bibfnamefont {A.}~\bibnamefont {Bouhon}},\
  and\ \bibinfo {author} {\bibfnamefont {R.-J.}\ \bibnamefont {Slager}},\
  }\bibfield  {title} {\bibinfo {title} {Topological {Euler} class as a
  dynamical observable in optical lattices},\ }\href
  {https://doi.org/10.1103/PhysRevLett.125.053601} {\bibfield  {journal}
  {\bibinfo  {journal} {Phys. Rev. Lett.}\ }\textbf {\bibinfo {volume} {125}},\
  \bibinfo {pages} {053601} (\bibinfo {year} {2020})}\BibitemShut {NoStop}%
\bibitem [{\citenamefont {Umerski}(1997)}]{Umerski1997}%
  \BibitemOpen
  \bibfield  {author} {\bibinfo {author} {\bibfnamefont {A.}~\bibnamefont
  {Umerski}},\ }\bibfield  {title} {\bibinfo {title} {Closed-form solutions to
  surface {Green's} functions},\ }\href
  {https://doi.org/10.1103/PhysRevB.55.5266} {\bibfield  {journal} {\bibinfo
  {journal} {Phys. Rev. B}\ }\textbf {\bibinfo {volume} {55}},\ \bibinfo
  {pages} {5266} (\bibinfo {year} {1997})}\BibitemShut {NoStop}%
\bibitem [{Note2()}]{Note2}%
  \BibitemOpen
  \bibinfo {note} {It should be noted that $C_3$ symmetry allows another
  possibility that $f$ does not have a constant term and $g \propto k_+^3 +
  \protect \mathrm {H.c.}$, which leaves a quadratic band touching intact.
  However, the surface quadratic band touching cannot be split into multiple
  Dirac cones due to the absence of the $C_2T$ symmetry.}\BibitemShut {Stop}%
\bibitem [{\citenamefont {Fu}\ \emph {et~al.}(2007)\citenamefont {Fu},
  \citenamefont {Kane},\ and\ \citenamefont {Mele}}]{Fu2007}%
  \BibitemOpen
  \bibfield  {author} {\bibinfo {author} {\bibfnamefont {L.}~\bibnamefont
  {Fu}}, \bibinfo {author} {\bibfnamefont {C.~L.}\ \bibnamefont {Kane}},\ and\
  \bibinfo {author} {\bibfnamefont {E.~J.}\ \bibnamefont {Mele}},\ }\bibfield
  {title} {\bibinfo {title} {Topological insulators in three dimensions},\
  }\href {https://doi.org/10.1103/PhysRevLett.98.106803} {\bibfield  {journal}
  {\bibinfo  {journal} {Phys. Rev. Lett.}\ }\textbf {\bibinfo {volume} {98}},\
  \bibinfo {pages} {106803} (\bibinfo {year} {2007})}\BibitemShut {NoStop}%
\bibitem [{\citenamefont {Nielsen}\ and\ \citenamefont
  {Ninomiya}(1981{\natexlab{a}})}]{Nielsen1981absenceI}%
  \BibitemOpen
  \bibfield  {author} {\bibinfo {author} {\bibfnamefont {H.}~\bibnamefont
  {Nielsen}}\ and\ \bibinfo {author} {\bibfnamefont {M.}~\bibnamefont
  {Ninomiya}},\ }\bibfield  {title} {\bibinfo {title} {Absence of neutrinos on
  a lattice: {(I)}. proof by homotopy theory},\ }\href
  {https://doi.org/https://doi.org/10.1016/0550-3213(81)90361-8} {\bibfield
  {journal} {\bibinfo  {journal} {Nuclear Physics B}\ }\textbf {\bibinfo
  {volume} {185}},\ \bibinfo {pages} {20} (\bibinfo {year}
  {1981}{\natexlab{a}})}\BibitemShut {NoStop}%
\bibitem [{\citenamefont {Nielsen}\ and\ \citenamefont
  {Ninomiya}(1981{\natexlab{b}})}]{Nielsen1981absenceII}%
  \BibitemOpen
  \bibfield  {author} {\bibinfo {author} {\bibfnamefont {H.}~\bibnamefont
  {Nielsen}}\ and\ \bibinfo {author} {\bibfnamefont {M.}~\bibnamefont
  {Ninomiya}},\ }\bibfield  {title} {\bibinfo {title} {Absence of neutrinos on
  a lattice: {(II)}. intuitive topological proof},\ }\href
  {https://doi.org/https://doi.org/10.1016/0550-3213(81)90524-1} {\bibfield
  {journal} {\bibinfo  {journal} {Nuclear Physics B}\ }\textbf {\bibinfo
  {volume} {193}},\ \bibinfo {pages} {173} (\bibinfo {year}
  {1981}{\natexlab{b}})}\BibitemShut {NoStop}%
\bibitem [{\citenamefont {Nielsen}\ and\ \citenamefont
  {Ninomiya}(1981{\natexlab{c}})}]{Nielsen1981no-go}%
  \BibitemOpen
  \bibfield  {author} {\bibinfo {author} {\bibfnamefont {H.}~\bibnamefont
  {Nielsen}}\ and\ \bibinfo {author} {\bibfnamefont {M.}~\bibnamefont
  {Ninomiya}},\ }\bibfield  {title} {\bibinfo {title} {A no-go theorem for
  regularizing chiral fermions},\ }\href
  {https://doi.org/https://doi.org/10.1016/0370-2693(81)91026-1} {\bibfield
  {journal} {\bibinfo  {journal} {Physics Letters B}\ }\textbf {\bibinfo
  {volume} {105}},\ \bibinfo {pages} {219} (\bibinfo {year}
  {1981}{\natexlab{c}})}\BibitemShut {NoStop}%
\bibitem [{\citenamefont {Fang}\ and\ \citenamefont {Fu}(2019)}]{CFang19}%
  \BibitemOpen
  \bibfield  {author} {\bibinfo {author} {\bibfnamefont {C.}~\bibnamefont
  {Fang}}\ and\ \bibinfo {author} {\bibfnamefont {L.}~\bibnamefont {Fu}},\
  }\bibfield  {title} {\bibinfo {title} {New classes of topological crystalline
  insulators having surface rotation anomaly},\ }\href
  {https://doi.org/10.1126/sciadv.aat2374} {\bibfield  {journal} {\bibinfo
  {journal} {Sci. Adv.}\ }\textbf {\bibinfo {volume} {5}},\ \bibinfo {pages}
  {eaat2374} (\bibinfo {year} {2019})}\BibitemShut {NoStop}%
\bibitem [{\citenamefont {Fang}\ \emph {et~al.}(2012)\citenamefont {Fang},
  \citenamefont {Gilbert},\ and\ \citenamefont {Bernevig}}]{CFang2012}%
  \BibitemOpen
  \bibfield  {author} {\bibinfo {author} {\bibfnamefont {C.}~\bibnamefont
  {Fang}}, \bibinfo {author} {\bibfnamefont {M.~J.}\ \bibnamefont {Gilbert}},\
  and\ \bibinfo {author} {\bibfnamefont {B.~A.}\ \bibnamefont {Bernevig}},\
  }\bibfield  {title} {\bibinfo {title} {Bulk topological invariants in
  noninteracting point group symmetric insulators},\ }\href
  {https://doi.org/10.1103/PhysRevB.86.115112} {\bibfield  {journal} {\bibinfo
  {journal} {Phys. Rev. B}\ }\textbf {\bibinfo {volume} {86}},\ \bibinfo
  {pages} {115112} (\bibinfo {year} {2012})}\BibitemShut {NoStop}%
\bibitem [{\citenamefont {Morimoto}\ and\ \citenamefont
  {Furusaki}(2014)}]{Morimoto2014}%
  \BibitemOpen
  \bibfield  {author} {\bibinfo {author} {\bibfnamefont {T.}~\bibnamefont
  {Morimoto}}\ and\ \bibinfo {author} {\bibfnamefont {A.}~\bibnamefont
  {Furusaki}},\ }\bibfield  {title} {\bibinfo {title} {Weyl and {Dirac}
  semimetals with {$\mathbb{Z}_{2}$} topological charge},\ }\href
  {https://doi.org/10.1103/PhysRevB.89.235127} {\bibfield  {journal} {\bibinfo
  {journal} {Phys. Rev. B}\ }\textbf {\bibinfo {volume} {89}},\ \bibinfo
  {pages} {235127} (\bibinfo {year} {2014})}\BibitemShut {NoStop}%
\bibitem [{\citenamefont {Fang}\ \emph {et~al.}(2015)\citenamefont {Fang},
  \citenamefont {Chen}, \citenamefont {Kee},\ and\ \citenamefont
  {Fu}}]{CFang2015}%
  \BibitemOpen
  \bibfield  {author} {\bibinfo {author} {\bibfnamefont {C.}~\bibnamefont
  {Fang}}, \bibinfo {author} {\bibfnamefont {Y.}~\bibnamefont {Chen}}, \bibinfo
  {author} {\bibfnamefont {H.-Y.}\ \bibnamefont {Kee}},\ and\ \bibinfo {author}
  {\bibfnamefont {L.}~\bibnamefont {Fu}},\ }\bibfield  {title} {\bibinfo
  {title} {Topological nodal line semimetals with and without spin-orbital
  coupling},\ }\href {https://doi.org/10.1103/PhysRevB.92.081201} {\bibfield
  {journal} {\bibinfo  {journal} {Phys. Rev. B}\ }\textbf {\bibinfo {volume}
  {92}},\ \bibinfo {pages} {081201(R)} (\bibinfo {year} {2015})}\BibitemShut
  {NoStop}%
\bibitem [{\citenamefont {Zhao}\ and\ \citenamefont {Lu}(2017)}]{Zhao2017}%
  \BibitemOpen
  \bibfield  {author} {\bibinfo {author} {\bibfnamefont {Y.~X.}\ \bibnamefont
  {Zhao}}\ and\ \bibinfo {author} {\bibfnamefont {Y.}~\bibnamefont {Lu}},\
  }\bibfield  {title} {\bibinfo {title} {$pt$-symmetric real {Dirac} fermions
  and semimetals},\ }\href {https://doi.org/10.1103/PhysRevLett.118.056401}
  {\bibfield  {journal} {\bibinfo  {journal} {Phys. Rev. Lett.}\ }\textbf
  {\bibinfo {volume} {118}},\ \bibinfo {pages} {056401} (\bibinfo {year}
  {2017})}\BibitemShut {NoStop}%
\bibitem [{\citenamefont {Hatcher}(2002)}]{Hatcher2002}%
  \BibitemOpen
  \bibfield  {author} {\bibinfo {author} {\bibfnamefont {A.}~\bibnamefont
  {Hatcher}},\ }\href@noop {} {\emph {\bibinfo {title} {Algebraic Topology}}}\
  (\bibinfo  {publisher} {Cambridge University Press},\ \bibinfo {address}
  {Cambridge},\ \bibinfo {year} {2002})\BibitemShut {NoStop}%
\bibitem [{\citenamefont {Yu}\ \emph {et~al.}(2011)\citenamefont {Yu},
  \citenamefont {Qi}, \citenamefont {Bernevig}, \citenamefont {Fang},\ and\
  \citenamefont {Dai}}]{Yu2011}%
  \BibitemOpen
  \bibfield  {author} {\bibinfo {author} {\bibfnamefont {R.}~\bibnamefont
  {Yu}}, \bibinfo {author} {\bibfnamefont {X.~L.}\ \bibnamefont {Qi}}, \bibinfo
  {author} {\bibfnamefont {A.}~\bibnamefont {Bernevig}}, \bibinfo {author}
  {\bibfnamefont {Z.}~\bibnamefont {Fang}},\ and\ \bibinfo {author}
  {\bibfnamefont {X.}~\bibnamefont {Dai}},\ }\bibfield  {title} {\bibinfo
  {title} {Equivalent expression of {$\mathbb{Z}_{2}$} topological invariant
  for band insulators using the non-abelian berry connection},\ }\href
  {https://doi.org/10.1103/PhysRevB.84.075119} {\bibfield  {journal} {\bibinfo
  {journal} {Phys. Rev. B}\ }\textbf {\bibinfo {volume} {84}},\ \bibinfo
  {pages} {075119} (\bibinfo {year} {2011})}\BibitemShut {NoStop}%
\bibitem [{\citenamefont {Alexandradinata}\ \emph
  {et~al.}(2014{\natexlab{b}})\citenamefont {Alexandradinata}, \citenamefont
  {Dai},\ and\ \citenamefont {Bernevig}}]{Alexandradinata2014}%
  \BibitemOpen
  \bibfield  {author} {\bibinfo {author} {\bibfnamefont {A.}~\bibnamefont
  {Alexandradinata}}, \bibinfo {author} {\bibfnamefont {X.}~\bibnamefont
  {Dai}},\ and\ \bibinfo {author} {\bibfnamefont {B.~A.}\ \bibnamefont
  {Bernevig}},\ }\bibfield  {title} {\bibinfo {title} {Wilson-loop
  characterization of inversion-symmetric topological insulators},\ }\href
  {https://doi.org/10.1103/PhysRevB.89.155114} {\bibfield  {journal} {\bibinfo
  {journal} {Phys. Rev. B}\ }\textbf {\bibinfo {volume} {89}},\ \bibinfo
  {pages} {155114} (\bibinfo {year} {2014}{\natexlab{b}})}\BibitemShut
  {NoStop}%
\bibitem [{\citenamefont {Alexandradinata}\ \emph {et~al.}(2016)\citenamefont
  {Alexandradinata}, \citenamefont {Wang},\ and\ \citenamefont
  {Bernevig}}]{Alexandradinata2016}%
  \BibitemOpen
  \bibfield  {author} {\bibinfo {author} {\bibfnamefont {A.}~\bibnamefont
  {Alexandradinata}}, \bibinfo {author} {\bibfnamefont {Z.}~\bibnamefont
  {Wang}},\ and\ \bibinfo {author} {\bibfnamefont {B.~A.}\ \bibnamefont
  {Bernevig}},\ }\bibfield  {title} {\bibinfo {title} {Topological insulators
  from group cohomology},\ }\href {https://doi.org/10.1103/PhysRevX.6.021008}
  {\bibfield  {journal} {\bibinfo  {journal} {Phys. Rev. X}\ }\textbf {\bibinfo
  {volume} {6}},\ \bibinfo {pages} {021008} (\bibinfo {year}
  {2016})}\BibitemShut {NoStop}%
\bibitem [{\citenamefont {Cano}\ \emph {et~al.}(2018)\citenamefont {Cano},
  \citenamefont {Bradlyn}, \citenamefont {Wang}, \citenamefont {Elcoro},
  \citenamefont {Vergniory}, \citenamefont {Felser}, \citenamefont {Aroyo},\
  and\ \citenamefont {Bernevig}}]{Cano2018}%
  \BibitemOpen
  \bibfield  {author} {\bibinfo {author} {\bibfnamefont {J.}~\bibnamefont
  {Cano}}, \bibinfo {author} {\bibfnamefont {B.}~\bibnamefont {Bradlyn}},
  \bibinfo {author} {\bibfnamefont {Z.}~\bibnamefont {Wang}}, \bibinfo {author}
  {\bibfnamefont {L.}~\bibnamefont {Elcoro}}, \bibinfo {author} {\bibfnamefont
  {M.~G.}\ \bibnamefont {Vergniory}}, \bibinfo {author} {\bibfnamefont
  {C.}~\bibnamefont {Felser}}, \bibinfo {author} {\bibfnamefont {M.~I.}\
  \bibnamefont {Aroyo}},\ and\ \bibinfo {author} {\bibfnamefont {B.~A.}\
  \bibnamefont {Bernevig}},\ }\bibfield  {title} {\bibinfo {title} {Topology of
  disconnected elementary band representations},\ }\href
  {https://doi.org/10.1103/PhysRevLett.120.266401} {\bibfield  {journal}
  {\bibinfo  {journal} {Phys. Rev. Lett.}\ }\textbf {\bibinfo {volume} {120}},\
  \bibinfo {pages} {266401} (\bibinfo {year} {2018})}\BibitemShut {NoStop}%
\bibitem [{\citenamefont {Bradlyn}\ \emph {et~al.}(2019)\citenamefont
  {Bradlyn}, \citenamefont {Wang}, \citenamefont {Cano},\ and\ \citenamefont
  {Bernevig}}]{Bradlyn2019}%
  \BibitemOpen
  \bibfield  {author} {\bibinfo {author} {\bibfnamefont {B.}~\bibnamefont
  {Bradlyn}}, \bibinfo {author} {\bibfnamefont {Z.}~\bibnamefont {Wang}},
  \bibinfo {author} {\bibfnamefont {J.}~\bibnamefont {Cano}},\ and\ \bibinfo
  {author} {\bibfnamefont {B.~A.}\ \bibnamefont {Bernevig}},\ }\bibfield
  {title} {\bibinfo {title} {Disconnected elementary band representations,
  fragile topology, and wilson loops as topological indices: An example on the
  triangular lattice},\ }\href {https://doi.org/10.1103/PhysRevB.99.045140}
  {\bibfield  {journal} {\bibinfo  {journal} {Phys. Rev. B}\ }\textbf {\bibinfo
  {volume} {99}},\ \bibinfo {pages} {045140} (\bibinfo {year}
  {2019})}\BibitemShut {NoStop}%
\bibitem [{\citenamefont {Su}\ \emph {et~al.}(1979)\citenamefont {Su},
  \citenamefont {Schrieffer},\ and\ \citenamefont {Heeger}}]{Su1979}%
  \BibitemOpen
  \bibfield  {author} {\bibinfo {author} {\bibfnamefont {W.~P.}\ \bibnamefont
  {Su}}, \bibinfo {author} {\bibfnamefont {J.~R.}\ \bibnamefont {Schrieffer}},\
  and\ \bibinfo {author} {\bibfnamefont {A.~J.}\ \bibnamefont {Heeger}},\
  }\bibfield  {title} {\bibinfo {title} {Solitons in {Polyacetylene}},\ }\href
  {https://doi.org/10.1103/PhysRevLett.42.1698} {\bibfield  {journal} {\bibinfo
   {journal} {Phys. Rev. Lett.}\ }\textbf {\bibinfo {volume} {42}},\ \bibinfo
  {pages} {1698} (\bibinfo {year} {1979})}\BibitemShut {NoStop}%
\bibitem [{\citenamefont {Turner}\ \emph {et~al.}(2021)\citenamefont {Turner},
  \citenamefont {Berg},\ and\ \citenamefont {Stern}}]{turner2021}%
  \BibitemOpen
  \bibfield  {author} {\bibinfo {author} {\bibfnamefont {A.~M.}\ \bibnamefont
  {Turner}}, \bibinfo {author} {\bibfnamefont {E.}~\bibnamefont {Berg}},\ and\
  \bibinfo {author} {\bibfnamefont {A.}~\bibnamefont {Stern}},\ }\bibfield
  {title} {\bibinfo {title} {Gapping fragile topological bands by
  interactions},\ }\href@noop {} {\bibfield  {journal} {\bibinfo  {journal}
  {arXiv preprint arXiv:2104.09528}\ } (\bibinfo {year} {2021})}\BibitemShut
  {NoStop}%
\bibitem [{\citenamefont {Yannopapas}(2011)}]{Yannopapas2011}%
  \BibitemOpen
  \bibfield  {author} {\bibinfo {author} {\bibfnamefont {V.}~\bibnamefont
  {Yannopapas}},\ }\bibfield  {title} {\bibinfo {title} {Gapless surface states
  in a lattice of coupled cavities: A photonic analog of topological
  crystalline insulators},\ }\href {https://doi.org/10.1103/PhysRevB.84.195126}
  {\bibfield  {journal} {\bibinfo  {journal} {Phys. Rev. B}\ }\textbf {\bibinfo
  {volume} {84}},\ \bibinfo {pages} {195126} (\bibinfo {year}
  {2011})}\BibitemShut {NoStop}%
\bibitem [{\citenamefont {Lu}\ \emph {et~al.}(2014)\citenamefont {Lu},
  \citenamefont {Joannopoulos},\ and\ \citenamefont
  {Solja{\v{c}}i{\'c}}}]{Lu2014topological}%
  \BibitemOpen
  \bibfield  {author} {\bibinfo {author} {\bibfnamefont {L.}~\bibnamefont
  {Lu}}, \bibinfo {author} {\bibfnamefont {J.~D.}\ \bibnamefont
  {Joannopoulos}},\ and\ \bibinfo {author} {\bibfnamefont {M.}~\bibnamefont
  {Solja{\v{c}}i{\'c}}},\ }\bibfield  {title} {\bibinfo {title} {Topological
  photonics},\ }\href@noop {} {\bibfield  {journal} {\bibinfo  {journal}
  {Nature photonics}\ }\textbf {\bibinfo {volume} {8}},\ \bibinfo {pages} {821}
  (\bibinfo {year} {2014})}\BibitemShut {NoStop}%
\bibitem [{\citenamefont {Slobozhanyuk}\ \emph {et~al.}(2017)\citenamefont
  {Slobozhanyuk}, \citenamefont {Mousavi}, \citenamefont {Ni}, \citenamefont
  {Smirnova}, \citenamefont {Kivshar},\ and\ \citenamefont
  {Khanikaev}}]{Slobozhanyuk2017three}%
  \BibitemOpen
  \bibfield  {author} {\bibinfo {author} {\bibfnamefont {A.}~\bibnamefont
  {Slobozhanyuk}}, \bibinfo {author} {\bibfnamefont {S.~H.}\ \bibnamefont
  {Mousavi}}, \bibinfo {author} {\bibfnamefont {X.}~\bibnamefont {Ni}},
  \bibinfo {author} {\bibfnamefont {D.}~\bibnamefont {Smirnova}}, \bibinfo
  {author} {\bibfnamefont {Y.~S.}\ \bibnamefont {Kivshar}},\ and\ \bibinfo
  {author} {\bibfnamefont {A.~B.}\ \bibnamefont {Khanikaev}},\ }\bibfield
  {title} {\bibinfo {title} {Three-dimensional all-dielectric photonic
  topological insulator},\ }\href@noop {} {\bibfield  {journal} {\bibinfo
  {journal} {Nature Photonics}\ }\textbf {\bibinfo {volume} {11}},\ \bibinfo
  {pages} {130} (\bibinfo {year} {2017})}\BibitemShut {NoStop}%
\bibitem [{\citenamefont {Ochiai}(2017)}]{Ochiai2017}%
  \BibitemOpen
  \bibfield  {author} {\bibinfo {author} {\bibfnamefont {T.}~\bibnamefont
  {Ochiai}},\ }\bibfield  {title} {\bibinfo {title} {Gapless surface states
  originating from accidentally degenerate quadratic band touching in a
  three-dimensional tetragonal photonic crystal},\ }\href
  {https://doi.org/10.1103/PhysRevA.96.043842} {\bibfield  {journal} {\bibinfo
  {journal} {Phys. Rev. A}\ }\textbf {\bibinfo {volume} {96}},\ \bibinfo
  {pages} {043842} (\bibinfo {year} {2017})}\BibitemShut {NoStop}%
\bibitem [{\citenamefont {Ozawa}\ \emph {et~al.}(2019)\citenamefont {Ozawa},
  \citenamefont {Price}, \citenamefont {Amo}, \citenamefont {Goldman},
  \citenamefont {Hafezi}, \citenamefont {Lu}, \citenamefont {Rechtsman},
  \citenamefont {Schuster}, \citenamefont {Simon}, \citenamefont {Zilberberg},\
  and\ \citenamefont {Carusotto}}]{Ozawa2019}%
  \BibitemOpen
  \bibfield  {author} {\bibinfo {author} {\bibfnamefont {T.}~\bibnamefont
  {Ozawa}}, \bibinfo {author} {\bibfnamefont {H.~M.}\ \bibnamefont {Price}},
  \bibinfo {author} {\bibfnamefont {A.}~\bibnamefont {Amo}}, \bibinfo {author}
  {\bibfnamefont {N.}~\bibnamefont {Goldman}}, \bibinfo {author} {\bibfnamefont
  {M.}~\bibnamefont {Hafezi}}, \bibinfo {author} {\bibfnamefont
  {L.}~\bibnamefont {Lu}}, \bibinfo {author} {\bibfnamefont {M.~C.}\
  \bibnamefont {Rechtsman}}, \bibinfo {author} {\bibfnamefont {D.}~\bibnamefont
  {Schuster}}, \bibinfo {author} {\bibfnamefont {J.}~\bibnamefont {Simon}},
  \bibinfo {author} {\bibfnamefont {O.}~\bibnamefont {Zilberberg}},\ and\
  \bibinfo {author} {\bibfnamefont {I.}~\bibnamefont {Carusotto}},\ }\bibfield
  {title} {\bibinfo {title} {Topological photonics},\ }\href
  {https://doi.org/10.1103/RevModPhys.91.015006} {\bibfield  {journal}
  {\bibinfo  {journal} {Rev. Mod. Phys.}\ }\textbf {\bibinfo {volume} {91}},\
  \bibinfo {pages} {015006} (\bibinfo {year} {2019})}\BibitemShut {NoStop}%
\bibitem [{\citenamefont {Yang}\ \emph {et~al.}(2019)\citenamefont {Yang},
  \citenamefont {Gao}, \citenamefont {Xue}, \citenamefont {Zhang},
  \citenamefont {He}, \citenamefont {Yang}, \citenamefont {Singh},
  \citenamefont {Chong}, \citenamefont {Zhang},\ and\ \citenamefont
  {Chen}}]{Yang2019realization}%
  \BibitemOpen
  \bibfield  {author} {\bibinfo {author} {\bibfnamefont {Y.}~\bibnamefont
  {Yang}}, \bibinfo {author} {\bibfnamefont {Z.}~\bibnamefont {Gao}}, \bibinfo
  {author} {\bibfnamefont {H.}~\bibnamefont {Xue}}, \bibinfo {author}
  {\bibfnamefont {L.}~\bibnamefont {Zhang}}, \bibinfo {author} {\bibfnamefont
  {M.}~\bibnamefont {He}}, \bibinfo {author} {\bibfnamefont {Z.}~\bibnamefont
  {Yang}}, \bibinfo {author} {\bibfnamefont {R.}~\bibnamefont {Singh}},
  \bibinfo {author} {\bibfnamefont {Y.}~\bibnamefont {Chong}}, \bibinfo
  {author} {\bibfnamefont {B.}~\bibnamefont {Zhang}},\ and\ \bibinfo {author}
  {\bibfnamefont {H.}~\bibnamefont {Chen}},\ }\bibfield  {title} {\bibinfo
  {title} {Realization of a three-dimensional photonic topological insulator},\
  }\href@noop {} {\bibfield  {journal} {\bibinfo  {journal} {Nature}\ }\textbf
  {\bibinfo {volume} {565}},\ \bibinfo {pages} {622} (\bibinfo {year}
  {2019})}\BibitemShut {NoStop}%
\bibitem [{\citenamefont {Kim}\ \emph {et~al.}(2020)\citenamefont {Kim},
  \citenamefont {Jacob},\ and\ \citenamefont {Rho}}]{Kim2020recent}%
  \BibitemOpen
  \bibfield  {author} {\bibinfo {author} {\bibfnamefont {M.}~\bibnamefont
  {Kim}}, \bibinfo {author} {\bibfnamefont {Z.}~\bibnamefont {Jacob}},\ and\
  \bibinfo {author} {\bibfnamefont {J.}~\bibnamefont {Rho}},\ }\bibfield
  {title} {\bibinfo {title} {Recent advances in {2D}, {3D} and higher-order
  topological photonics},\ }\href@noop {} {\bibfield  {journal} {\bibinfo
  {journal} {Light: Science \& Applications}\ }\textbf {\bibinfo {volume}
  {9}},\ \bibinfo {pages} {130} (\bibinfo {year} {2020})}\BibitemShut {NoStop}%
\bibitem [{\citenamefont {He}\ \emph {et~al.}(2020)\citenamefont {He},
  \citenamefont {Lai}, \citenamefont {He}, \citenamefont {Yu}, \citenamefont
  {Xu}, \citenamefont {Lu},\ and\ \citenamefont {Chen}}]{He2020acoustic}%
  \BibitemOpen
  \bibfield  {author} {\bibinfo {author} {\bibfnamefont {C.}~\bibnamefont
  {He}}, \bibinfo {author} {\bibfnamefont {H.-S.}\ \bibnamefont {Lai}},
  \bibinfo {author} {\bibfnamefont {B.}~\bibnamefont {He}}, \bibinfo {author}
  {\bibfnamefont {S.-Y.}\ \bibnamefont {Yu}}, \bibinfo {author} {\bibfnamefont
  {X.}~\bibnamefont {Xu}}, \bibinfo {author} {\bibfnamefont {M.-H.}\
  \bibnamefont {Lu}},\ and\ \bibinfo {author} {\bibfnamefont {Y.-F.}\
  \bibnamefont {Chen}},\ }\bibfield  {title} {\bibinfo {title} {Acoustic
  analogues of three-dimensional topological insulators},\ }\href@noop {}
  {\bibfield  {journal} {\bibinfo  {journal} {Nature communications}\ }\textbf
  {\bibinfo {volume} {11}},\ \bibinfo {pages} {1} (\bibinfo {year}
  {2020})}\BibitemShut {NoStop}%
\bibitem [{Note3()}]{Note3}%
  \BibitemOpen
  \bibinfo {note} {Exactly speaking, when the first homotopy group is
  nontrivial, it also affects the band topology through the Whitney sum
  formula~\cite {Ahn2018band}. However, in the pseudo-spin basis, the first
  homotopy group becomes trivial.}\BibitemShut {Stop}%
\bibitem [{\citenamefont {Jackiw}\ and\ \citenamefont
  {Rebbi}(1976)}]{Jackiw-Rebbi}%
  \BibitemOpen
  \bibfield  {author} {\bibinfo {author} {\bibfnamefont {R.}~\bibnamefont
  {Jackiw}}\ and\ \bibinfo {author} {\bibfnamefont {C.}~\bibnamefont {Rebbi}},\
  }\bibfield  {title} {\bibinfo {title} {Solitons with fermion number
  \textonehalf{}},\ }\href {https://doi.org/10.1103/PhysRevD.13.3398}
  {\bibfield  {journal} {\bibinfo  {journal} {Phys. Rev. D}\ }\textbf {\bibinfo
  {volume} {13}},\ \bibinfo {pages} {3398} (\bibinfo {year}
  {1976})}\BibitemShut {NoStop}%
\bibitem [{\citenamefont {Khalaf}(2018)}]{Khalaf2018high}%
  \BibitemOpen
  \bibfield  {author} {\bibinfo {author} {\bibfnamefont {E.}~\bibnamefont
  {Khalaf}},\ }\bibfield  {title} {\bibinfo {title} {Higher-order topological
  insulators and superconductors protected by inversion symmetry},\ }\href
  {https://doi.org/10.1103/PhysRevB.97.205136} {\bibfield  {journal} {\bibinfo
  {journal} {Phys. Rev. B}\ }\textbf {\bibinfo {volume} {97}},\ \bibinfo
  {pages} {205136} (\bibinfo {year} {2018})}\BibitemShut {NoStop}%
\bibitem [{\citenamefont {Geier}\ \emph {et~al.}(2018)\citenamefont {Geier},
  \citenamefont {Trifunovic}, \citenamefont {Hoskam},\ and\ \citenamefont
  {Brouwer}}]{Geier2018}%
  \BibitemOpen
  \bibfield  {author} {\bibinfo {author} {\bibfnamefont {M.}~\bibnamefont
  {Geier}}, \bibinfo {author} {\bibfnamefont {L.}~\bibnamefont {Trifunovic}},
  \bibinfo {author} {\bibfnamefont {M.}~\bibnamefont {Hoskam}},\ and\ \bibinfo
  {author} {\bibfnamefont {P.~W.}\ \bibnamefont {Brouwer}},\ }\bibfield
  {title} {\bibinfo {title} {Second-order topological insulators and
  superconductors with an order-two crystalline symmetry},\ }\href
  {https://doi.org/10.1103/PhysRevB.97.205135} {\bibfield  {journal} {\bibinfo
  {journal} {Phys. Rev. B}\ }\textbf {\bibinfo {volume} {97}},\ \bibinfo
  {pages} {205135} (\bibinfo {year} {2018})}\BibitemShut {NoStop}%
\bibitem [{\citenamefont {Trifunovic}\ and\ \citenamefont
  {Brouwer}(2019)}]{Trifunovic2019}%
  \BibitemOpen
  \bibfield  {author} {\bibinfo {author} {\bibfnamefont {L.}~\bibnamefont
  {Trifunovic}}\ and\ \bibinfo {author} {\bibfnamefont {P.~W.}\ \bibnamefont
  {Brouwer}},\ }\bibfield  {title} {\bibinfo {title} {Higher-order
  bulk-boundary correspondence for topological crystalline phases},\ }\href
  {https://doi.org/10.1103/PhysRevX.9.011012} {\bibfield  {journal} {\bibinfo
  {journal} {Phys. Rev. X}\ }\textbf {\bibinfo {volume} {9}},\ \bibinfo {pages}
  {011012} (\bibinfo {year} {2019})}\BibitemShut {NoStop}%
\end{thebibliography}%

\end{document}